\documentclass[preprint,12pt,authoryear]{elsarticle}

\usepackage{epsfig}

\usepackage{amsmath}
\usepackage{amssymb}
\usepackage{mathtools}
\usepackage{subfigure}
\usepackage{color}
\usepackage{multirow}
\usepackage{tabularx}
\usepackage{booktabs}
\usepackage{pdflscape}

\biboptions{round,semicolon,numbers}

\journal{International Journal of Heat and Fluid Flow}

\begin{document}

\begin{frontmatter}


\title{Pressure Drop and Flow development in the Entrance Region of Micro-Channels with Second Order Velocity Slip Condition and the Requirement for Development Length}
\author[label1]{Baibhab Ray\corref{cor1}}
\cortext[cor1]{Corresponding Authors}
\ead{baibhab.ray@mailbox.tu-dresden.de \& baibhab\_ray@yahoo.in}
\address[label1]{Department of Physics, Technische Universit\"{a}t Dresden} 

\author[label2]{Franz Durst}
\ead{f.durst@fmp-technology.com}
\address[label2]{FMP Technology GmbH, Am Weichselgarten 34, 91058 Erlangen, Germany}

\author[label3]{Subhashis Ray\corref{cor1}}
\ead{ray@iwtt.tu-freiberg.de \& juhp\_sray@yahoo.co.in}
\address[label3]{Institute of Thermal Engineering, Technische Universit\"{a}t Bergakademie Freiberg\\Gustav-Zeuner Stra{\ss}e 7, 09596 Freiberg, Germany}


\begin{abstract}

In the present investigation, the development of axial velocity profile, the requirement for development length ($L^{*}_{fd} = L/D_{h}$) and the pressure drop in the entrance region of circular and parallel plate micro-channels have been critically analysed for a large range of operating conditions ($10^{-2} \le Re \le 10^{4}$, $10^{-4} \le Kn \le 0.2$ and $0 \le C_{2} \le 0.5$). For this purpose, the conventional Navier-Stokes equations have been numerically solved using the finite volume method on non-staggered grid, while employing the second-order velocity slip condition at the wall with $C_{1} = 1$. The results indicate that although the magnitude of local velocity slip at the wall is always greater than that for the fully-developed section, the local wall shear stress, particularly for higher $Kn$ and $C_2$, could be considerably lower than its fully-developed value. This effect, which is more prominent for lower $Re$, significantly affects the local and the fully-developed incremental pressure drop number $K(x)$ and $K_{fd}$, respectively. As a result, depending upon the operating condition, $K_{fd}$, as well as $K(x)$, could assume negative values. This never reported observation implies that in the presence of enhanced velocity slip at the wall, the pressure gradient in the developing region could even be less than that in the fully-developed section. From simulated data, it has been observed that both $L^{*}_{fd}$ and $K_{fd}$ are characterised by the low and the high $Re$ asymptotes, using which, extremely accurate correlations for them have been proposed for both geometries. Although owing to the complex nature, no correlation could be derived for $K(x)$ and an exact knowledge of $K(x)$ is necessary for evaluating the actual pressure drop for a duct length $L^{*} < L^{*}_{fd}$, a method has been proposed that provides a conservative estimate of the pressure drop for both $K_{fd} > 0$ and $K_{fd} \le 0$. \\

\end{abstract}

\begin{keyword}
Pressure drop \sep Flow Development \sep Development length \sep Pipe and channel flows \sep Second-order slip boundary condition \sep Incremental pressure drop number 


\end{keyword}

\end{frontmatter}




\tableofcontents

\newpage
\addcontentsline{toc}{section}{Nomenclature}

\section*{Nomenclature}

\begin{tabbing}
123456789 \= \kill
$A_{c}$ \> Cross-sectional area (m$^2$) \\
$b$ \> Half gap between two parallel plates (m) \\
$C_{1}$, $C_{2}$ \> First and second order coefficients for wall velocity slip condition \\
$D$ \> Pipe diameter (m) \\
$D_{h}$ \> Hydraulic diameter $=D$ for pipe and $=4b$ for parallel plate channel (m) \\
$f$ \> Fanning friction factor, $2 \tau_{w} / \rho u_{av}^{2}$ \\
$H$ \> Gap between two parallel plates = $2b$ (m) \\
$k_{ij}$ \> Coefficients in the correlation for $K_{i}$ \\
$K$ \> Incremental pressure drop number \\
$K_{i}$ \> Coefficients in the correlation for $K_{fd}$ \\
$Kn$ \> Knudsen number \\
$l_{ij}$ \> Coefficients in the correlation for $L_{i}$ \\
$L$ \> Axial length (m) \\
$L_{i}$ \> Coefficients in the correlation for $L^{*}_{fd}$ \\
$n$ \> Outward normal coordinate from the computational domain (m) \\
$p$ \> Effective pressure (Pa) \\
$P_{w}$ \> Wetted perimeter (m) \\
$q$ \> Exponent in the correlation for $L^{*}_{fd}$ \\
$r$ \> Identifier for the coordinate system, also the radial coordinate (m) \\
$R$ \> Radius of the pipe $=D/2$ (m) \\
$Re$ \> Reynolds number, $\rho u_{av} D_{h} / \mu$ \\
$S_{axi}$ \> Special source term for axi-symmetric coordinates \\
$u$ \> Axial velocity (m/s) \\
$v$ \> Radial (for pipe) or transverse (for channel) velocity (m/s) \\
$x$ \> Axial coordinate (m) \\
$y$ \> Radial (for pipe) or transverse (for channel) coordinate (m)
\end{tabbing}

\subsection*{Greek Letters}

\begin{tabbing}
123456789 \= \kill
$\lambda$ \> Mean free path (m) \\
$\mu$ \> Dynamic viscosity (Pa~s) \\
$\rho$ \> Density (kg/m$^3$) \\
$\sigma$ \> Tangential momentum accommodation coefficient \\
$\tau$ \> Shear stress (N/m$^2$)
\end{tabbing}
 
\subsection*{Subscripts}

\begin{tabbing}
123456789 \= \kill
$app$ \> Apparent \\
$av$ \> Average \\
$c$ \> Centre-line \\
$f$ \> Due to friction \\
$fd$ \> Fully-developed \\
$m$ \> Due to change in momentum \\
$s$ \> Slip \\
$t$ \> Tangential direction \\
$w$ \> Pertaining to wall \\
$x$ \> Pertaining to axial coordinate
\end{tabbing}

\subsection*{Superscripts}

\begin{tabbing}
123456789 \= \kill
$*$ \> Dimensionless quantity
\end{tabbing}

\section{Introduction and Aim of Work}\label{Sec:IntroAndAim}

With the increasing miniaturisation, considerable technological developments have recently been taken place in order to manufacture fluidic systems with channel dimensions in the micro-metre scale, where the overall system dimensions could vary between a few $\mu$m to 1~mm \citep{HoAndTai_1998, Gad-el-Hak_1999, Gad-el-Hak_2001, KarniadakisAndBeskok_2002, MicroflowsNanoflows}. As a result, one can easily find numerous examples of micro-channel flows in a broad range of scientific applications and also in everyday use scenarios, such as, the cooling of micro-electronic components and integrated circuits, the gas lubrication in micro-bearings, the active control of aerodynamic flows, the liquid ink flow through the print-heads of ink-jet printers, the extraction of biological samples and the development of micro analysis platforms dubbed ``Lab-on-a-chip'', etc., to mention a few \citep{BarberAndEmerson_2006, Tang_etal_2008}. \\

Owing to its importance in the present context, several articles providing reviews on general and specific topics, monographs and books have been published over the past few decades, emphasising on different aspects of the micro-channel flows. Other than those presenting the broad-based general information, mentioned before, these documents may be classified into the ones specifically dealing with:

\begin{enumerate}
\item Predictions of flows using the conventional Navier-Stokes (NS) equations along with different velocity slip and temperature jump conditions at the wall \citep{Akrilic_etal_1997, KarniadakisAndBeskok_2002, Colin_2005, MicroflowsNanoflows, BarberAndEmerson_2006, Dongari2007, Tang_etal_2007a, Tang_etal_2007b, WengAndChen_2008, Cao_etal_2009, ChenAndBogy_2010, Colin_2012, Zhang_etal_2012},
\item Modelling of flows employing higher order equations than the conventional NS equations \citep{Hadjiconstantinou_2000, JinAndSlemrod_2001, StruchtrupAndTorrilhon_2003, Hadjiconstantinou_2006, Shan_etal_2006, Ansumali_etal_2007, LilleyAndSader_2008, StruchtrupAndTorrilhon_2008, WengAndChen_2008, Dongari_etal_2009, RoohiAndDarbandi_2009, Dongari_etal_2010},
\item Use of i) kinetic theory of gases \citep{Loyalka_1971, Cercignani_1990}, ii) Molecular Dynamic (MD) simulations \citep{Bird_1994, ZhangHW_etal_2012}, iii) Discrete Simulations Monte Carlo (DSMC) methods \citep{Pan_etal_1999, Hadjiconstantinou_2000}, iv) Non-linear and linearised Boltzmann Equation (BE) \citep{Cercignani_1975, Cercignani_1988, LiAndKwok_2003} and v) Lattice Boltzmann Method (LBM) \citep{Cornubert_etal_1991, SbragagliaAndSucci_2005, Zheng_etal_2006, Tang_etal_2008, AidunAndClausen_2010, Zhang_2011} for predictions,
\item Effects of compressibility on the flow behaviour and employment of rarefied gas dynamics \citep{BeskokAndKarniadakis_1996, SharipovAndSeleznev_1998, Cercignani_2000, SiewertAndSharipov_2002, Sharipov_2003, Colin_2005, BarberAndEmerson_2006, Dongari_etal_2011_1, Dongari_etal_2011_2} for predicting micro-channel flows and
\item Heat transfer enhancement and its characteristics \citep{Gad-el-Hak_2006, Colin_2012, Kandlikar_etal_2013} associated with micro-channel flows.
\end{enumerate}

Micro-channel flows are often characterised by the higher mean free path ($\lambda$) of the gas molecules that is comparable with the typical system dimension. In this respect, the Knudsen number may be defined as $Kn = \lambda / L_{ref}$, where $L_{ref}$ is the reference or the characteristic length. For the present investigation, $L_{ref}$ for defining  $Kn$ has been chosen as the diameter of the capillary $D$ for pipe flows and the gap between two parallel plates $H = 2b$ for channel flows. It appears that \citet{SchaafAndChambre_1961} first proposed the classifications of gas flow regimes based on $Kn$ as:
\begin{enumerate}
\item The continuum regime holds for $Kn \le 10^{-2}$, where both continuum and local thermodynamic equilibrium assumptions are valid and hence the flow is governed by the conventional NS equations with the traditional no-slip condition at the wall.
\item The slip flow regime is identified by $10^{-2} < Kn \le 10^{-1}$, where the non-equilibrium effects, particularly close to the walls, start dominating the flow and hence the conventional no-slip boundary condition becomes invalid. It is, however, well established that the bulk flow outside the Knudsen layer, whose thickness is estimated to be of the order of $\lambda$ \citep{ZhangR_etal_2006, ZhangYH_etal_2006, Zhang_etal_2012}, is still governed by the traditional NS equations \citep{Dongari2007, MicroflowsNanoflows} and hence the gaseous micro-channel flows in this regime may still be predicted by employing the velocity slip and the temperature jump conditions at the walls, while describing the fluid motion by the continuous NS equations. \\

Alternatively, as Durst and his co-workers \citep{SumanAndDurst_2007, Dongari_etal_2009, Dongari_etal_2010} suggested, the extended NS equations may be invoked for predictions that takes the self diffusion of mass into account. In this formulation, the mass velocity of the fluid is divided into its diffusive and convective parts, where the former explicitly depends on the local gradients of pressure and temperature and produces velocity slip at the wall, while the no-slip boundary condition applies for the convective velocity \citep{Sambasivam_PhD_2013}. \\

The present investigation has been carried out for $Kn \le 0.2$ and hence it primarily falls into the slip flow regime. Therefore, the conventional NS equations, along with the velocity slip condition at the wall, have been employed for modelling.
\item The transition regime is characterised by $10^{-1} < Kn \le 10$, where the rarefaction effects dominate and hence both continuum and local thermodynamic equilibrium assumptions tend to fail. For the early transition regime, however, the predictions obtained using the conventional NS equations require the employment of higher order velocity slip condition at the wall in order to compensate for the non-linear stress-strain relationship and the variation in effective viscosity within the Knudsen layer \citep{BarberAndEmerson_2006, Dongari2007}. Alternatively, costlier methods, like MD simulations, DSMC, or methods derived from the kinetic theory and the BE should be adopted for reliable predictions.
\item The regime beyond $Kn = 10$ is recognised as the free molecular regime, where, owing to the large separation between the gas molecules, the inter-molecular interactions are negligible as compared to the collisions of molecules with the confining walls \citep{BeskokAndKarniadakis_1999, KarniadakisAndBeskok_2002}.
\end{enumerate}

While analysing ducted flows, however, it is essential to differentiate the hydro-dynamically fully-developed region\footnote{Where analytical solutions for velocity distributions could be obtained for most cases.} from the developing region that is observed close to the entrance. It is, therefore, important to determine the development length $L_{fd}$, in order to ascertain whether simple analytical solutions could be applied for predicting the flow characteristics. As summarised by \citet{Durstetal2005}, for the developing region, semi-analytical, experimental and numerical methods were adopted in the past. \\

Semi-analytical solutions in the developing region can be obtained only with considerable simplifying assumptions. By neglecting the axial diffusion of momentum, the boundary layer-type assumptions\footnote{With parabolic axial velocity profile that depends on the local centre-line velocity.} are typically invoked for this purpose \citep{ShahandLondon1978}. However, the experimental and the numerical investigations showed the existence of {\em velocity overshoots} close to the duct wall, particularly near the inlet, which proves incompatible with the concept of a boundary layer. Moreover, as \citet{Durstetal2005} pointed out, the axial diffusion plays an extremely important role in the momentum transfer for low Reynolds number flows, pertinent specifically for micro-channels, and hence it cannot be neglected in this regime. Nevertheless, employing such semi-analytical treatments applicable only for the high Reynolds number regime, $L_{fd}$ could be obtained as:
\begin{equation}
 {L_{fd}}/D_{h} = C \cdot Re \label{Eq:LbyD_HighReLimit}
\end{equation}
where $C$ is a constant and $Re$ is the Reynolds number.\footnote{Most often defined on the basis of average axial velocity $u_{av}$ and hydraulic diameter $D_{h}$.} Based on several investigations, $C \approx 0.05$ is often cited in the standard text books \citep[see for example,][]{White,FoxMcDonald} for pipe flows. \\

Experimental measurements of $L_{fd}$ are similarly confronted with considerable difficulties. Present levels of uncertainty in the measurement of small difference in the centre-line velocity can produce large errors in determining $L_{fd}$. This was already reported by \citet{Durstetal2005}, which is also evident from the relatively high scattering of the experimentally determined values of $C$. \\

Since the semi-analytical and the experimental methods fail to deliver accurate results for $L_{fd}$, it is obvious that only a numerical approach would be meaningful. \citet{Durstetal2005} obtained the following correlations for $L_{fd}$ from their numerical investigations, covering a wide range of Reynolds number ($0.01 \leq Re \leq 4000$):
\begin{subequations}
\begin{eqnarray}
L_{fd} / D_{h} & = & {\left[ {\left( 0.619 \right)}^{1.6} + {\left(0.0567~Re \right)}^{1.6} \right]}^{{1}/{1.6}} ~~~~~ \mbox{for pipe} \label{Eq:LbyD_Pipe_Durst} \\
L_{fd} / D_{h} & = & {\left[ {\left( 0.3155 \right)}^{1.6} + {\left(0.01105~Re \right)}^{1.6} \right]}^{{1}/{1.6}} ~~~~~ \mbox{for channel} \label{Eq:LbyD_Channel_Durst}
\end{eqnarray}
\label{Eq:LbyD_Durst}
\end{subequations}
\noindent For channel flows, \citet{Durstetal2005} presented their correlation for $L_{fd} / H$ as a function of $Re_{H}$. In Eq.~(\ref{Eq:LbyD_Channel_Durst}), however, $Re = \rho u_{av} D_{h} / \mu$ is used, which is consistent with the present definition and hence the constants differ from that of the original article. \\

As far as the developing flow through micro-channels are concerned, in spite of a thorough literature review, the present authors are aware about only two articles by \citet{BarberAndEmerson_2001} and \citet{Ferrasetal2012}, where the authors numerically determined $L_{fd}$, employing the simplified first order velocity slip condition at the wall. While \citet{BarberAndEmerson_2001} considered flows through both pipes and parallel plate channels for $Re \le 400$, \citet{Ferrasetal2012} dealt with only the latter geometry and restricted their study for $Re \le 100$. In order to define the $R$ and $Kn$, the hydraulic diameter $D_{h}$ and the gap between parallel plates $H=2b$ were chosen as the length scales by \citet{BarberAndEmerson_2001} and \citet{Ferrasetal2012}, respectively.\\

It may be noted that for pipe flows, in spite of observing the dependence of $L_{fd}$ on both $Re$ and $Kn$, \citet{BarberAndEmerson_2001} recommended the use of correlations from \citet{Chen_1973} and \citet{Dombrowski_etal_1993} irrespective of $Kn$, although they were obtained for $Kn = 0$ in the continuum regime. Therefore, the validity of the proposed predictive equations for $L_{fd}$ of pipe flows is questionable, particularly for higher $Kn$. \\

For parallel plate channels, \citet{BarberAndEmerson_2001} and \citet{Ferrasetal2012} proposed different correlations for $L_{fd}$, presented in Eqs.~(\ref{Eq:BarberAndEmerson_2001}) and (\ref{Eq:Ferrasetal2012}), respectively:
\begin{subequations}
\begin{eqnarray}
 \frac {L_{fd}} {4b} & = & 
 \frac {0.332} {1 + 0.0271~Re} + \left( \frac {1 + 14.78~C_{1} Kn_{Dh}} {1 + 9.78~C_{1} Kn_{Dh}} \right) 0.011~Re ~~~~~ \label{Eq:BarberAndEmerson_2001} \\
 \frac {L_{fd}} {2b} & = & 
 \frac {1 + 3.15~{\left( C_{1} Kn \right)}^{1.2} + 0.28~\left( C_{1} Kn \right) Re^{0.5}_{H}} {1 + 3.82~{\left( C_{1} Kn \right)}^{1.5} + 0.018~\left( C_{1} Kn \right) Re_{H}} 
 {\left[ {\left( 0.631 \right)}^{1.8} + {\left(0.047~Re_{H} \right)}^{1.8} \right]}^{{1}/{1.8}} ~~~~~ \label{Eq:Ferrasetal2012}
\end{eqnarray}
\label{Eq:LbyD_Slip_Old}
\end{subequations}
where the symbols $Kn_{Dh} = \lambda / D_{h}$ and $Re_{H} = \rho u_{av} H / \mu$ are used, highlighting the difference between the former and the present definitions. In Eq.~(\ref{Eq:BarberAndEmerson_2001}), the first order velocity slip coefficient is defined as $C_{1} = \left( 2 - \sigma \right) / \sigma$ \citep{BarberAndEmerson_2001}, where $\sigma$ is the tangential momentum accommodation coefficient that varies between $0$ and $1$ for specular and diffusive reflections, respectively \citep{WuAndBogy_2001, Lockerby_etal_2004, LockerbyAndReese_2008}, while for Eq.~(\ref{Eq:Ferrasetal2012}), $\overline{k}_{l} = C_{1} Kn$ is substituted for consistent representation. Comparison of these correlations clearly reveals the apparent contradictions. Equation~(\ref{Eq:BarberAndEmerson_2001}) suggests that the low $Re$ ($Re \rightarrow 0$) asymptote is independent of $Kn_{Dh}$,\footnote{$L^{*}_{fd} = 0.332$ is obtained from the first term in Eq.~(\ref{Eq:BarberAndEmerson_2001}) in this limit.} whereas the high $Re$ asymptote ($Re \rightarrow \infty$) is a linear function of $Re$ that explicitly depends on $Kn_{Dh}$. On the other hand, Eq.~(\ref{Eq:Ferrasetal2012}) shows that the high $Re$ asymptote is given by $L_{fd} / 2b \simeq 0.731 {Re}^{1/2}_{H}$ and hence is independent of $Kn$,\footnote{In this regime, $L_{fd} / 2H$ is expected to be a linear function of $Re_{H}$ that depends on $Kn$.} although the low $Re$ asymptote clearly depends on $Kn$. It is, therefore, evident that the discrepancy in $L_{fd}$ for flows through parallel plate channel must be resolved through systematic investigation. \\

Although the aforementioned contradicting correlations from \citet{BarberAndEmerson_2001} in Eq.~(\ref{Eq:BarberAndEmerson_2001}) and \citet{Ferrasetal2012} in Eq.~(\ref{Eq:Ferrasetal2012}) were obtained employing the first order velocity slip boundary condition at the wall,\footnote{Where the effects of $C_{1}$ and $Kn$ cannot be separately distinguished.} \citet{Dongari2007} clearly demonstrated that such a simplified approximation fails to explain certain unexpected behaviours, such as the \citet{Knudsen1909} paradox. As a viable alternative, they suggested employing the more general second order velocity slip condition at the wall as:
\begin{equation}
 u_{t,w} - u_{w} = - C_{1} \lambda {\left. \frac {\partial u_{t}} {\partial n} \right|}_{w} - C_{2} \lambda^{2} {\left. \frac {\partial^{2} u_{t}} {\partial n^{2}} \right|}_{w}  \label{Eq:General-Second-Order-Slip-Condition}
\end{equation}
where $u_{w}$ is the velocity of the solid wall, $u_{t}$ is the fluid velocity tangential to the wall and the suffix $w$ refers to the wall. Further, $C_{1}$ and $C_{2}$ are first and second order velocity slip coefficients, respectively, while $n$ is the local spatial coordinate, pointing outward from the domain. It is also evident that $C_{2}=0$ was explicitly set by both \citet{BarberAndEmerson_2001} and \citet{Ferrasetal2012} and hence their predicted results must be questionable. \\

Another important observation is that although both \citet{BarberAndEmerson_2001} and \citet{Ferrasetal2012} numerically solved the developing flow through circular pipes and parallel plate channels, neither of them reported the variations of pressure drop in the entrance region. However, these data, in the form of incremental pressure drop number $K \left(x \right)$ \citep{ShahandLondon1978}, should be considered extremely important. \\

In view of the brief literature review, presented so far, few comments are now in order:
\begin{enumerate}
\item In the slip flow and the early transition regimes, the conventional NS equations, originally derived for the continuum regime, could be employed for predictions of micro-channels flows, as long as the second order velocity slip condition is applied at the wall. Similar to the previous studies \citep{BarberAndEmerson_2001, Ferrasetal2012}, this modelling approach has been adopted also for the present investigation, which has been conducted for $Kn \le 0.2$.
\item Extremely insufficient data and no reliable correlation are available for $L_{fd}$ of flow through circular micro-channels. For the parallel plate micro-channels, on the other hand, the available correlations for $L_{fd}$ contradict each other and do not respect either the low or the high $Re$ asymptotes for all $Kn$.
\item Both previous investigations on $L_{fd}$ were carried out by employing the simplified first order velocity slip condition at the wall that fails to explain the \citet{Knudsen1909} paradox \citep{Dongari2007}. It is, therefore, evident that the more general second order velocity slip condition in Eq.~(\ref{Eq:General-Second-Order-Slip-Condition}) should be adopted, from which, the results for the first order velocity slip condition could be retrieved by setting $C_{2}=0$.
\item The investigation for micro-channel flows should also be accompanied by the associated pressure drop data in the developing region. This information was missing in the previous studies and hence demands for a thorough investigation.
\end{enumerate}

Based on the aforementioned observations, the present investigation has been carried out in order to study the pressure drop and the development of axial velocity profile in the entrance regions of circular and parallel plate micro-channels and to determine $L_{fd}$ for such flows. For this purpose, the conventional NS equations have been employed, along with the second order velocity slip condition at the wall for $10^{-4} \le Kn \le 0.2$, while assuming the flow to be steady as well as laminar. Since most of the micro-channel flows operate at low Mach numbers, the compressibility effects have been neglected and hence the fluid has been assumed to be incompressible \citep{BarberAndEmerson_2001}. In addition, the fluid has been considered to be Newtonian and the flow as isothermal. Since both density and temperature of the fluid have been assumed to be constants and the viscosity of a Newtonian fluid is a weak function of pressure, its variation in space has been neglected. \\

The present article has been organised as follows. After this section, presenting a brief introduction, the literature review and the motivation, section~\ref{mathform} deals with the governing equations, their scaling and the employed boundary conditions. This section also provides analytical solutions for the fully-developed flow through both circular and parallel plate micro-channels with second order velocity slip condition at the wall along with some other relevant characteristics. In addition, a brief description of the numerical method is also outlined for the sake of completeness, along with the post-processing of relevant data. The main results, in the form of development of the axial velocity profile, the variations in $L_{fd}$, their correlations and the variations in $K \left(x \right)$ as well as $K_{fd}$ along with the correlations for the latter are presented in section~\ref{results}. Finally, the conclusions are reported and the final remarks are made in section~\ref{concls}.

\section{Mathematical Formulation}\label{mathform}

\subsection{Governing Equations and Boundary Conditions}\label{GovEqn}

The conventional NS equations have been used for modelling the developing flow through micro-channels. In order to express the conservation equations in their non-dimensional forms, all coordinates and lengths have been made dimensionless with respect to $D_{h}$, while the velocity components have been normalised using $u_{av}$. On the other hand, the effective pressure, that also includes the hydrostatic pressure variation, has been scaled with respect to $\rho u^{2}_{av}$. As a consequence, the governing mass, axial and transverse (for parallel plate channel) or radial (for pipe) momentum conservation equations are obtained as presented in Eqs.~(\ref{eq:mass}), (\ref{eq:x-mom}) and (\ref{eq:y-mom}), respectively \citep{TransportPhenomena, White, FoxMcDonald}:
\begin{equation}
\frac {\partial u^{*}} {\partial x^{*}} + 
\frac {1} {r^{*}} ~\frac {\partial} {\partial y^{*}} \left( r^{*} v^{*} \right) = 0 \label{eq:mass}
\end{equation}
\begin{equation}
\frac {\partial} {\partial x^{*}} \left( u^{*} u^{*} \right) + 
\frac {1} {r^{*}} ~\frac {\partial} {\partial y^{*}} \left( r^{*} v^{*} u^{*} \right) = 
- ~\frac {\partial p^{*}} {\partial x^{*}} + 
\frac {\partial} {\partial x^{*}} \left( \mu^{*} \frac {\partial u^{*}} {\partial x^{*}} \right) + 
\frac {1} {r^{*}} ~\frac {\partial} {\partial y^{*}} \left( \mu^{*} r^{*} \frac{ \partial u^{*}} {\partial y^{*}} \right) \label{eq:x-mom}
\end{equation}
\begin{equation}
\frac {\partial} {\partial x^{*}} \left( u^{*} v^{*} \right) + 
\frac {1} {r^{*}} ~\frac {\partial} {\partial y^{*}} \left( r^{*} v^{*} v^{*} \right) = 
- ~\frac {\partial p^{*}} {\partial y^{*}} + 
\frac {\partial} {\partial x^{*}} \left( \mu^{*} \frac {\partial v^{*}} {\partial x^{*}} \right) + 
\frac {1} {r^{*}} ~\frac {\partial} {\partial y^{*}} \left( \mu^{*} r^{*} \frac{ \partial v^{*}} {\partial y^{*}} \right) +
S^{*}_{axi} \label{eq:y-mom}
\end{equation} 
where $r^{*} = r /D_{h}$ is the identifier for the coordinate system that is set to unity and $y^{*}$ for the Cartesian and the cylindrical axi-symmetric coordinates, respectively. Further in Eqs.~(\ref{eq:x-mom}) and (\ref{eq:y-mom}), $\mu^{*} = 1/Re$ is the dimensionless viscosity, where $Re = \rho u_{av} D_{h} / \mu$ is the Reynolds number defined on the basis of the average axial velocity $u_{av}$ and the hydraulic diameter $D_{h}$ and $S^{*}_{axi} = \mu^{*} v^{*} / {r^{*}}^{2}$ in Eq.~(\ref{eq:y-mom}) is the special source term that appears only for the cylindrical axi-symmetric coordinates (pipe flows) and is set to zero for the Cartesian coordinates (flows through parallel plate channels). For further definitions, the nomenclature section may be referred. \\

In order to solve Eqs.~(\ref{eq:mass}) -- (\ref{eq:y-mom}), the following boundary conditions have been applied:
\begin{enumerate}
\item At the inlet, i.e., at $x^{*} = 0$, the flow has been assumed to be uniform and hence $u^{*} = u^{*}_{av} = 1$ and $v^{*} = 0$ have been set for $0 \le y^{*} \le y^{*}_{\max}$, where $y^{*}_{\max} = R^{*} = 1/2$ for pipes and $y^{*}_{\max} = b^{*} = 1/4$ (since $D_{h} = 4b$) for parallel plate channels.
\item At the exit, i.e., at $x^{*} = x^{*}_{\max}$, the axial diffusion terms in Eqs.~(\ref{eq:x-mom}) and (\ref{eq:y-mom}) have been neglected, i.e., ${\partial^{2} \left( u^{*}, v^{*} \right)} / {\partial {x^{*}}^{2}} = 0$ have been set for $0 \le y^{*} \le y^{*}_{\max}$. This condition is considered less stringent as compared to setting ${\partial \left( u^{*}, v^{*} \right)} / {\partial x^{*}} = 0$, which may be regarded as a special case. 
\item On the line of symmetry (centre-line), i.e., at $y^{*} = 0$, ${\partial u^{*}} / {\partial y^{*}} = 0$ and $v^{*} = 0$ have been set for $0 \le x^{*} \le x^{*}_{\max}$.
\item On the wall, i.e., at $y^{*} = 1/2$ for pipes and $y^{*} = 1/4$ for parallel plate channels, the impermeable condition $v^{*} = 0$ and the second order velocity slip condition in Eq.~(\ref{Eq:General-Second-Order-Slip-Condition}) have been applied for $0 \le x^{*} \le x^{*}_{\max}$. Using the dimensionless variables and defining $Kn$ as $\lambda / D$ and $\lambda / H = \lambda / 2b$ for pipes and parallel plate channels, respectively, Eq.~(\ref{Eq:General-Second-Order-Slip-Condition}) may be written for $u^{*}_{w} = 0$ in its dimensionless form as:
\begin{equation}
 u^{*} = - C_{1} ~\widetilde{Kn} {\left. \frac {\partial u^{*}} {\partial y^{*}} \right|}_{y^{*} = y^{*}_{\max}} - 
 C_{2} ~{\widetilde{Kn}}^{2} {\left. \frac {\partial^{2} u^{*}} {\partial {y^{*}}^{2}} \right|}_{y^{*} = y^{*}_{\max}} \label{Eq:General-Second-Order-Slip-Condition-Dimless}
\end{equation}
where, as the definitions suggest, $\widetilde{Kn} = Kn$ and $\widetilde{Kn} = Kn / 2$ have been set for pipes and parallel plate channels, respectively. Fitting a second degree polynomial for $u^{*}$ close to the wall using the boundary and two interior nodes, located at same $x^{*}$, the boundary velocity has been iteratively updated using Eq.~(\ref{Eq:General-Second-Order-Slip-Condition-Dimless}).
\end{enumerate}

\subsection{Analytical Solution for Fully-developed Flow}\label{AnSoln}

\begin{figure}[htbp]
\begin{center}
\subfigure[$u^{*}_{s,fd}$ as function of $Kn$]{
\includegraphics[width=0.45\textwidth]{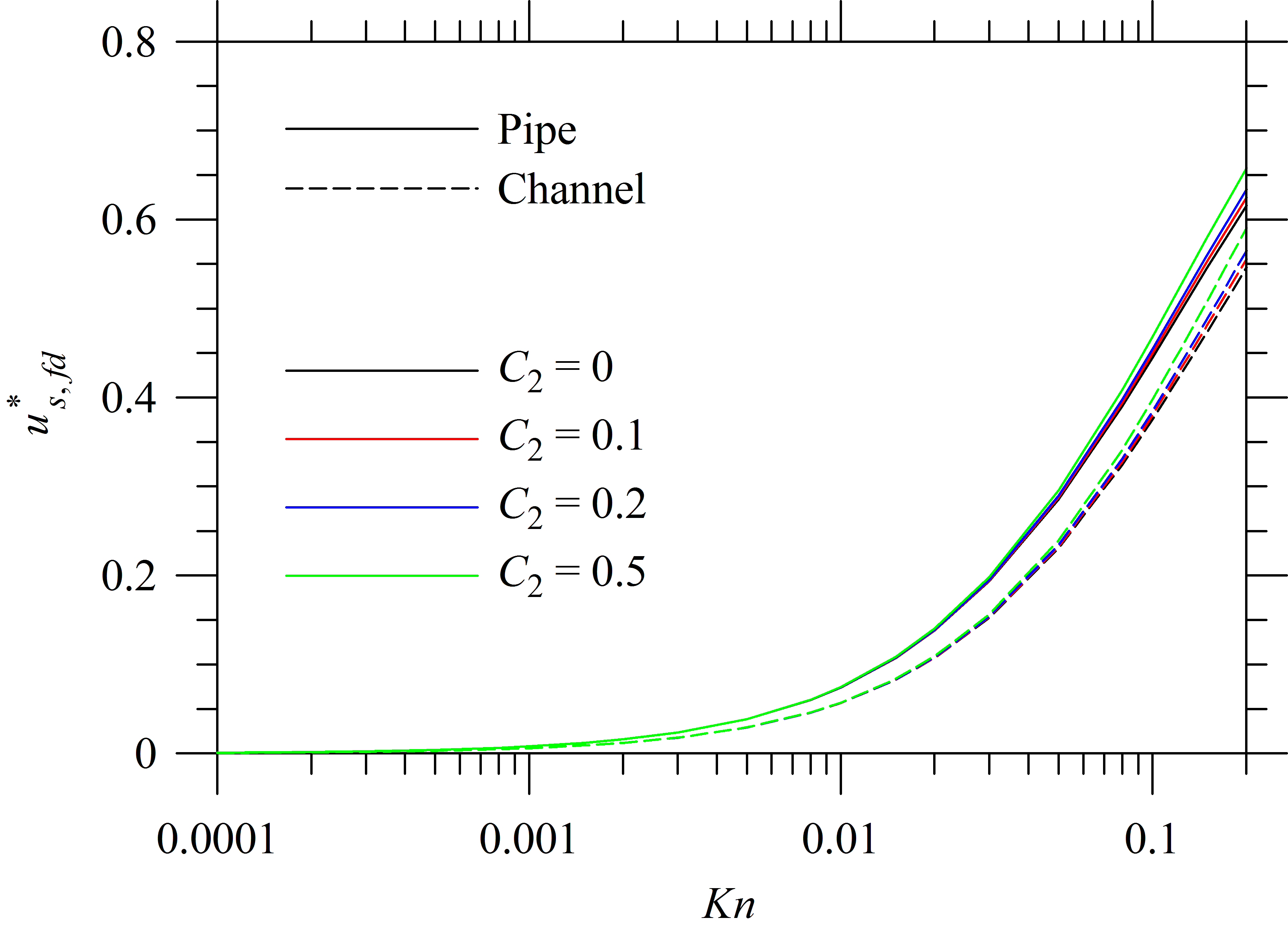}
} \label{fig:Us_PipeChannel}
\subfigure[$u^{*}_{c,fd}$ as function of $Kn$]{
\includegraphics[width=0.45\textwidth]{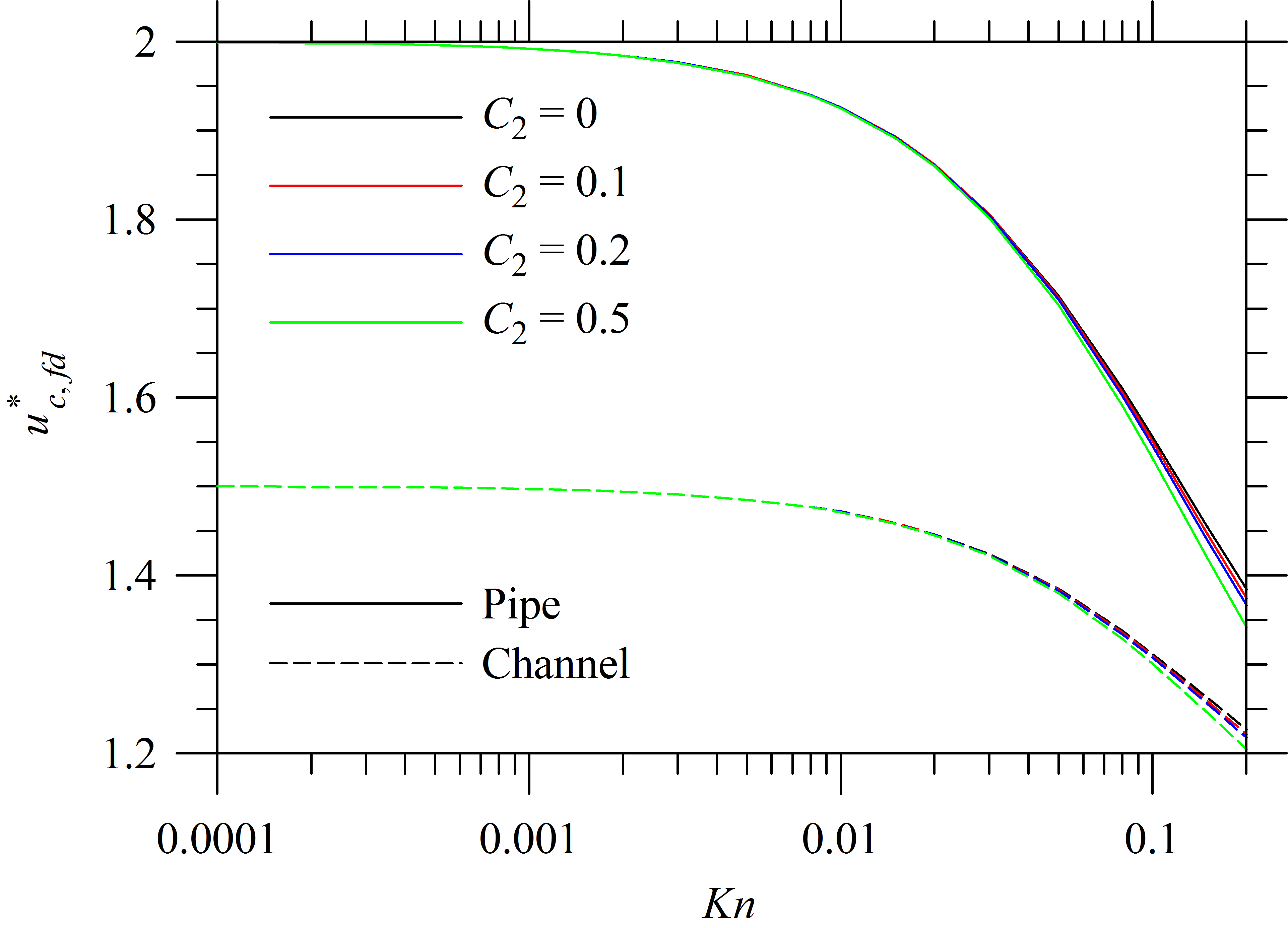}
} \label{fig:Uc_PipeChannel}
\end{center}
\caption{$u^{*}_{s,fd}$ and $u^{*}_{c,fd}$ for pipe and channel flows as functions of $Kn$ for different $C_{2}$.}
\label{fig:UsUc}
\end{figure}

For fully-developed flows, $\partial u^{*} / \partial x^{*} = 0$ and $v^{*} = 0$ and hence Eqs.~(\ref{eq:mass}) and (\ref{eq:y-mom}) are automatically satisfied. Solving the further simplified form of Eq.~(\ref{eq:x-mom}) using the boundary conditions at $y^{*} = 0$ and $y^{*} = y^{*}_{\max}$, the axial velocity profiles may be obtained as:
\begin{subequations}
\begin{eqnarray}
 u^{*}_{fd} = \frac {u_{fd}} {u_{av}} & = & 2 \left[ \frac {1 - {\left( r / R \right)}^{2} + 4 C_{1} Kn + 8 C_{2} Kn^{2}} {1 + 8 C_{1} Kn + 16 C_{2} Kn^{2}} \right] ~~~~~ \mbox{for pipe} \label{Eq:ufd_pipe} \\
 & = & \frac {3} {2} \left[ \frac {1 - {\left( y / b \right)}^{2} + 4 C_{1} Kn + 8 C_{2} Kn^{2}} {1 + 6 C_{1} Kn + 12 C_{2} Kn^{2}} \right] ~~~~~ \mbox{for channel} \label{Eq:ufd_ppchannel}
\end{eqnarray}
\label{Eq:ufd}
\end{subequations}
The slip velocity at the wall $u^{*}_{s,fd}$ and the centre-line velocity $u^{*}_{c,fd}$ for fully-developed flows may now be easily retrieved from Eq.~(\ref{Eq:ufd}) by setting $r=R$ or $y=b$ and $r=0$ or $y=0$ for pipe or channel flows, respectively. The variations in $u^{*}_{s,fd}$ and $u^{*}_{c,fd}$ as functions of $Kn$ are presented in Fig.~\ref{fig:UsUc} for different $C_{2}$. The figure shows that $u^{*}_{s,fd}$ increases and in order to maintain the same dimensionless mass flow rate ($u^{*}_{av} = 1$), $u^{*}_{c,fd}$ decreases with the increase in both $Kn$ and $C_{2}$. These variations for different $C_{2}$ are also more prominent for $Kn \ge 0.01$, which indicates that the difference between the first and the second order velocity slip conditions at the wall are expected to be more significant for higher $Kn$ and justifies the present investigation. For developing flows, on the other hand, owing to the higher axial velocity gradients at the wall, particularly close to the entrance, the effects of $C_{2}$ are expected to be more prominent even at lower $Kn$. \\

The Fanning friction factor is defined as $f = 2 \tau_{w} / \rho u^{2}_{av}$ \citep{ShahandLondon1978}, where $\tau_{w}$ as the wall shear stress, and $f_{fd} Re$ may be obtained from Eq.~(\ref{Eq:ufd}) as:
\begin{subequations}
\begin{eqnarray}
 f_{fd} Re & = & \frac {16} {1 + 8 C_{1} Kn + 16 C_{2} Kn^{2}} ~~~~~ \mbox{for pipe} \label{Eq:ffdRe_pipe} \\
 & = & \frac {24} {1 + 6 C_{1} Kn + 12 C_{2} Kn^{2}} ~~~~~ \mbox{for channel} \label{Eq:ffdRe_ppchannel}
\end{eqnarray}
\label{Eq:ffdRe}
\end{subequations}
In the fully-developed section, owing to the absence of inertia (convection) and axial diffusion, the integral force balance could be obtained as $\Delta p_{fd} A_{c} = \tau_{w,fd} P_{w} x$, where $\Delta p_{fd} = \left( - dp / dx \right) x$ is the pressure drop in the fully-developed section over an axial length $x$. This relation may be expressed in dimensionless form as $\Delta p^{*}_{fd} = 2 f_{fd} x^{*}$.

\subsection{Numerical Simulations and Post Processing of Data}\label{NumSim}

The numerical code, used by \citet{Durstetal2005}, has been employed also for the present study by modifying the wall boundary condition according to Eq.~(\ref{Eq:General-Second-Order-Slip-Condition-Dimless}). For this purpose, Eqs.~(\ref{eq:mass}) -- (\ref{eq:y-mom}) have been discretised for a given non-staggered control volume (CV) using the finite volume approach \citep{peric}, where the cell-face velocities have been evaluated using the momentum interpolation method \citep{RhieChow1983}. The central differencing scheme has been used for both convective and diffusive terms, where the deferred correction approach has been used for the former \citep{khosla-rubin}. \\
 
The set of discretised equations for a given equation have been solved iteratively by employing the Strongly Implicit Procedure \citep{stone}, while the SIMPLE algorithm \citep{patankar-spalding, patankar} has been used in order to ensure the pressure-velocity coupling. After each iteration, the L2 norms of the residues for all conservation equations have been calculated and the solution has been accepted as converged when all these norms have been found to be less than $10^{-7}$. \\

From the converged solutions, the local friction factors have been obtained as $f_{x} = 2 \tau_{w,x} / \rho u^{2}_{av}$, where $\tau_{w,x} = \mu {\left( \partial u / \partial y \right)}_{y=y_{\max}}$ is the wall shear stress at $x$. Using the dimensionless variables and the expression for $\tau_{w,x}$, one obtains:
\begin{equation}
f_{x} Re = 2 ~{\left. \frac {\partial u^{*}} {\partial y^{*}} \right|}_{y^{*}=y^{*}_{\max}}
\label{Eq:fxRe}
\end{equation} 
\citet{Durstetal2005} pointed out that the flow develops much faster (i.e., at least distance from the entrance) close to the wall as compared to the centre-line. This observation has been found to be true also for the present investigation, irrespective of $C_{2}$, $Kn$ and $Re$. Therefore, $L_{fd}$ has been always determined by the linear interpolation using two successive nodal velocities between which the centre-line velocity equals 99\% of the analytical fully-developed value in Eq.~(\ref{Eq:ufd}). \\

Alternatively, the length required for $f_{x} Re$ in  Eq.~(\ref{Eq:fxRe}) to differ by 1\% from $f_{fd} Re$ in Eq.~(\ref{Eq:ffdRe}) could also be considered as a measure of $L_{fd}$. As expected, however, since the development of velocity gradient at the wall depends directly on the flow development close to the wall, $f_{x} Re$ also develops much faster than the centre-line velocity. Therefore, $L_{fd}$, calculated on the basis of centre-line velocity, provides the most conservative estimate and hence has been adopted for the present investigation. \\

\begin{figure}[htbp]
\begin{center}
\includegraphics[width=0.6\textwidth]{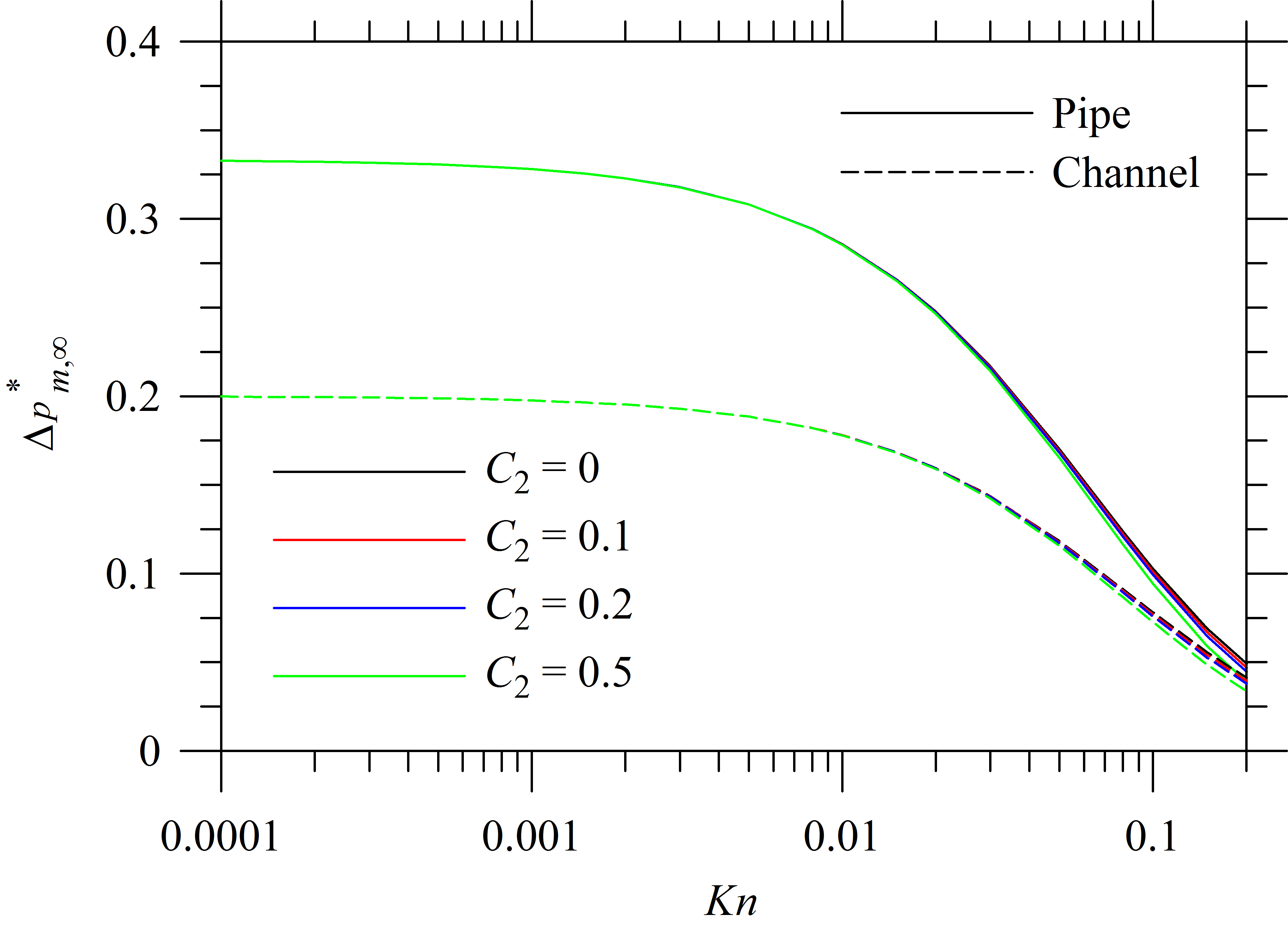}
\end{center}
\caption{Variations in $\Delta p^{*}_{m,\infty}$ as functions of $Kn$ for pipe and channel flows.}
\label{fig:DelPmInf}
\end{figure}

In order to quantify the pressure drop in the developing region, $K(x)$, according to \citep{ShahandLondon1978}, has been evaluated. Assuming a uniform velocity profile at the inlet ($u=u_{av}$) and integrating the dimensional form of Eq.~(\ref{eq:x-mom}), one obtains:
\begin{equation}
\Delta p_{x} A_{c} = \left( p_{0} - p_{x} \right) A_{c} = \int_{0}^{x} \tau_{w,x} P_{w} dx + \int_{A_{c}} \rho u^{2} d A_{c} - \rho u^{2}_{av} A_{c}
\label{Eq:ForceBalanceDimensional}
\end{equation}
where $p_{0}$ and $p_{x}$ are the cross-sectional averaged pressure at the inlet and $x$, respectively:
\begin{equation}
p_{0} = \frac {1} {A_{c}} \int_{A_{c}} p \left( 0, y \right) d A_{c}; ~~~~~ p_{x} = \frac {1} {A_{c}} \int_{A_{c}} p \left( x, y \right) d A_{c}; ~~~~~ d A_{c} = r dy
\label{Eq:px}
\end{equation}
In the expression for $dA_{c}$ in Eq.~(\ref{Eq:px}), $r$ has the similar meaning as in Eqs.~(\ref{eq:mass}) -- (\ref{eq:y-mom}) with respect to the coordinate system. Dividing Eq.~(\ref{Eq:ForceBalanceDimensional}) by the inlet momentum $\rho u^{2}_{av} A_{c}$, one obtains $\Delta p^{*}_{x}$ as:
\begin{equation}
\Delta p^{*}_{x} = 2 \int_{0}^{x^{*}} f_{x} dx^{*} + \frac {1} {A^{*}_{c}} \int_{A^{*}_{c}} {\left( u^{*} \right)}^{2} d A^{*}_{c} - 1 = \Delta p^{*}_{f,x} + \Delta p^{*}_{m,x}
\label{Eq:Deltapx}
\end{equation}
Equation~(\ref{Eq:Deltapx}) clearly shows that the pressure drop occurs in the axial direction in order to 1) overcome the frictional resistance at the wall (the first term, $\Delta p^{*}_{f,x}$) and 2) increase the total momentum of the fluid (the last two terms, $\Delta p^{*}_{m,x}$). For the fully-developed flows ($x \rightarrow \infty$), since the velocity distribution is given by Eq.~(\ref{Eq:ufd}), $\Delta p^{*}_{m,\infty}$ could be analytically obtained as presented in Eqs.~(\ref{Eq:DelPmInf_pipe}) and (\ref{Eq:DelPmInf_ppchannel}) for pipe and channel flows, respectively:
\begin{subequations}
\begin{eqnarray}
 \Delta p^{*}_{m,\infty} & = & \frac {4 \left[ 1/3 + \left( 1 + 4 C_{1} Kn + 8 C_{2} Kn^{2} \right) \left( 4 C_{1} Kn + 8 C_{2} Kn^{2} \right) \right]} {{\left( 1 + 8 C_{1} Kn + 16 C_{2} Kn^{2} \right)}^{2}} - 1 ~~~~~ \label{Eq:DelPmInf_pipe} \\
 & = & \frac {9 \left[ 1/5 + \left( 1 + 4 C_{1} Kn + 8 C_{2} Kn^{2} \right) \left( 1/3 + 4 C_{1} Kn + 8 C_{2} Kn^{2} \right) \right]} {4 {\left( 1 + 6 C_{1} Kn + 12 C_{2} Kn^{2} \right)}^{2}} - 1 ~~~~~  \label{Eq:DelPmInf_ppchannel}
\end{eqnarray}
\label{Eq:DelPmInf}
\end{subequations}
Like the fully-developed velocity profiles in Eq.~(\ref{Eq:ufd}), $\Delta p^{*}_{m,\infty}$ is also independent of $Re$. For $Kn = 0$ (no-slip condition), $\Delta p^{*}_{m,\infty}$ is obtained as $1/3$ and $1/5$ for pipe and channel flows, respectively. The variations in $\Delta p^{*}_{m,\infty}$ as functions of $Kn$ for different $C_{2}$ are presented in Fig.~\ref{fig:DelPmInf}, which shows $\Delta p^{*}_{m,\infty}$ decreases rapidly with the increase in $Kn$, particularly for $Kn \ge 0.01$. Like $u^{*}_{s}$ and $u^{*}_{c}$ in Fig.~\ref{fig:UsUc}, the variations are clearly more sensitive on $Kn$ than on $C_{2}$ for the investigated ranges. \\

Similar to $f_{fd}$ in the fully-developed region, the apparent friction factor $f_{app,x}$ may be defined according to $\Delta p^{*}_{x} = 2 f_{app,x} x^{*}$. Using this relation and Eq.~(\ref{Eq:Deltapx}), one obtains:
\begin{equation}
f_{app,x} = \frac {\Delta p^{*}_{x}} {2x^{*}} = \frac {1} {x^{*}} \left[ \int_{0}^{x^{*}} f_{x} dx^{*} + \frac {1} {2} \left( \frac {1} {A^{*}_{c}} \int_{A^{*}_{c}} {\left( u^{*} \right)}^{2} d A^{*}_{c} - 1 \right) \right]
\label{Eq:f_app}
\end{equation}
If the flow is assumed to be fully-developed even in the developing region, the pressure drop would be obtained as $\Delta p^{*}_{fd} = 2 f_{fd} x^{*}$. The incremental pressure drop number $K (x)$ is defined as the difference between the true and the expected fully-developed pressure drops, normalised with respect to $\rho u^{2}_{av} / 2$:
\begin{eqnarray}
K \left( x \right) & = & \frac {2} {\rho u^{2}_{av}} \left( \Delta p_{x} - \Delta p_{fd} \right) = 2 \left( \Delta p^{*}_{x} - \Delta p^{*}_{fd} \right) \nonumber \\
& = & 2 \left( \Delta p^{*}_{f,x} + \Delta p^{*}_{m,x} - 2 f_{fd} x^{*} \right) 
\label{Eq:Kx}
\end{eqnarray}
Since $\tau_{w,x}$ is expected to be higher than $\tau_{w,fd}$, $\Delta p_{x}$ is also expected to be higher than $\Delta p_{fd}$ for the same axial distance and hence $K (x)$, irrespective of the operating condition, is expected to be always positive. The present authors are unaware about any case where $K (x)$ has been reported to be negative. Nevertheless, once $K (x)$ is known, $\Delta p_{x}$ may be evaluated as:
\begin{equation}
\Delta p_{x} = \frac {1} {2} \rho u^{2}_{av} \left[ K (x) + 4 f_{fd} x^{*} \right]
\label{Eq:Deltapx_From_Kx}
\end{equation}
Unlike $f_{app,x}$, as $x \rightarrow \infty$, $K (x)$ asymptotically assumes a constant value that strongly depends on $Re$, $Kn$ and $C_{2}$ and is denoted as $K_{fd}$. It is evident that when the length of the duct $L \geq L_{fd}$, $K_{fd}$ plays a direct as well as important role in the calculation of true pressure drop $\Delta p_{L}$ and hence the variation in $K (x)$ is considered more important than $f_{app,x}$ in order to characterise the pressure drop in the developing region.

\section{Results and Discussion}\label{results}

In the present investigation, numerical simulations have been performed for both pipe and channel flows by varying $Re$ over a wide range from $10^{-2}$ (diffusion dominated regime) to $10^{4}$ (convective regime), although the micro-channel flows may never encounter a case with extremely high Reynolds number. However, as will be shortly apparent, accurate low and high $Re$ asymptotes are required in order to obtain reliable correlations for $L^{*}_{fd}$ even in the moderate $Re$ range of practical interest $1 \le Re \le 100$ \citep{BarberAndEmerson_2001, Ferrasetal2012}. For each $Re$, $Kn$ has been varied from $10^{-4}$ (continuum regime) to $0.2$ (early transition regime). As indicated by \citet{Dongari2007} and \citet{Zhang_etal_2012}, most theoretical and experimental studies reported $C_{1} \approx 1$, while $C_{2}$ varies over a wide range. As a result, for the present investigation, $C_{1} = 1$ (special case for $\sigma = 1$) has been chosen, while $C_{2}$ has been varied from 0 (first order) to 0.5. The results of \citet{Durstetal2005} have been reproduced by setting $Kn = 0$. \\

Prior to obtaining the results, however, a careful grid independence study has been carried out and $x^{*}_{\max}$ has been determined in order to ensure that the fully-developed flow condition for a given $Re$ is always satisfied well inside the chosen duct length, irrespective of $Kn$ and $C_{2}$. A preliminary investigation has revealed that other than pipe flows for high $Re$, $L^{*}_{fd}$ always increases with the increase in $Kn$ and $C_{2}$. For the most critical case with $C_{2} = 0.5$ and $Kn = 0.2$, $L^{*}_{fd}$ has been observed to be approximately $1.5$ to $2$ times of that calculated from Eq.~(\ref{Eq:LbyD_Durst}) for the continuum regime. Therefore, $x^{*}_{\max}$ for all investigated cases has been specified as $3$ times of $L^{*}_{fd}$ for $Kn = 0$. \\

The computational domain has been discretised using non-uniform CVs, expanding in the axial direction and contracting in the transverse or the radial direction in geometric progression with common ratios $1.02$ and $1 / 1.02$, respectively. The grid independence study showed $300$ CVs are required in order to resolve the axial direction, while for the transverse or the radial direction, $80$ CVs have been found to be sufficient as far as the evaluation of $L^{*}_{fd}$ is concerned. The detailed results of the grid independence study, however, are not presented here for brevity. \\

Unless absolutely required, the results for pipe flows are only presented in this article in order to demonstrate and explain various physical aspects of micro-channel flows, while similar data for parallel plate channels are not presented here for the sake of brevity.

\subsection{Development of Velocity Profile} \label{development_velocity_profile}

The fully-developed velocity profile in Eq.~(\ref{Eq:ufd}) still remains parabolic, even in the presence of velocity slip at the wall. For no-slip condition at the wall, \citet{Durstetal2005} already reported that owing to the higher velocity gradient, velocity overshoots occur close to the wall, particularly near the entrance, which subsequently decays and finally disappears. \citet{BarberAndEmerson_2001} and \citet{Ferrasetal2012} also observed similar velocity overshoots with the first order velocity slip condition and hence it would be worthwhile exploring if (and how) this behaviour changes for the second order velocity slip condition. \\

\begin{figure}[htbp]
\begin{center}
\subfigure[$Re = 0.1, Kn = 0.01, C_{2} = 0$]{
\includegraphics[width=0.45\textwidth]{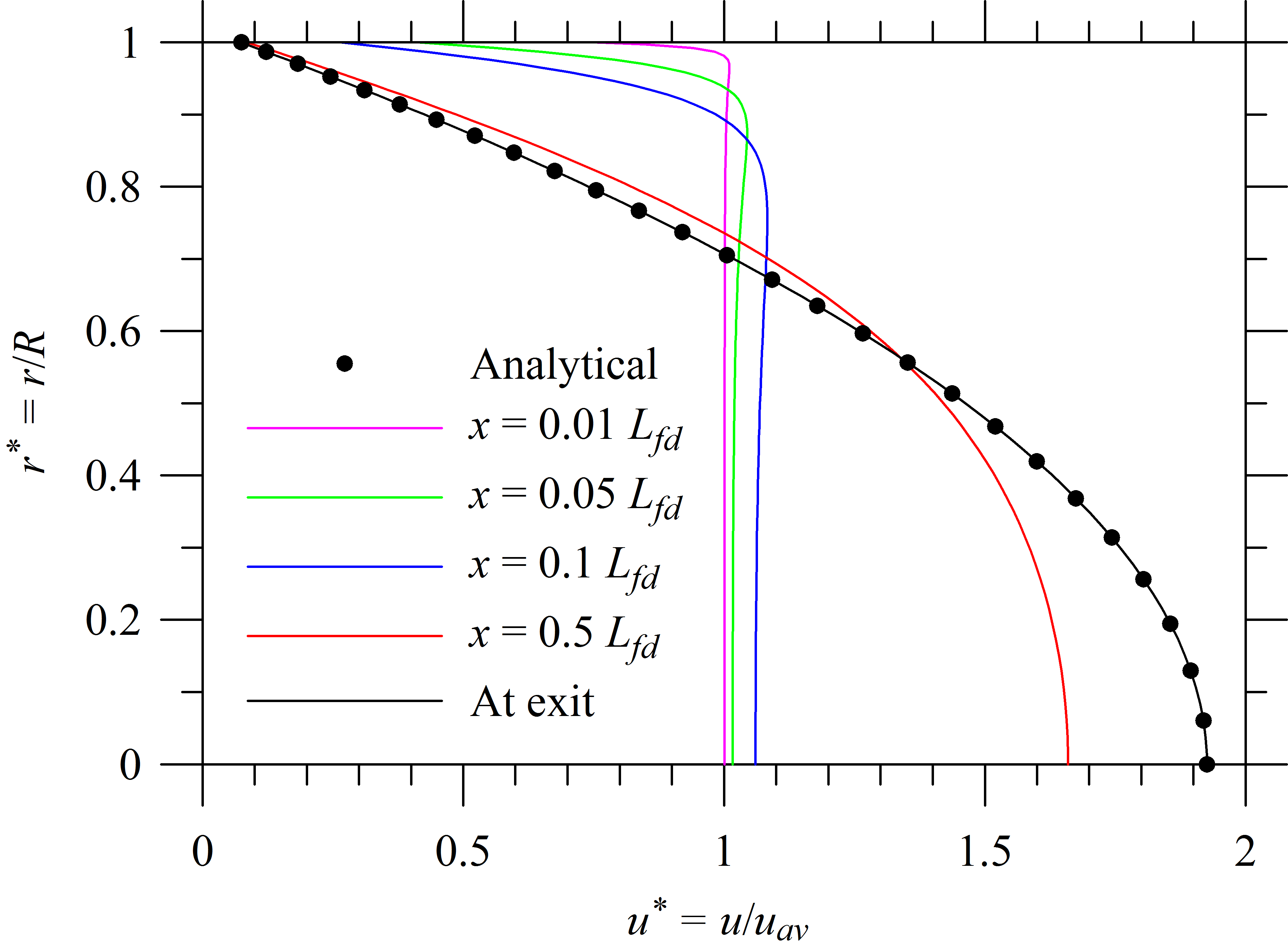} \label{fig:Re_01Kn_001C2_0}
} 
\subfigure[$Re = 0.1, Kn = 0.1, C_{2} = 0.5$]{
\includegraphics[width=0.45\textwidth]{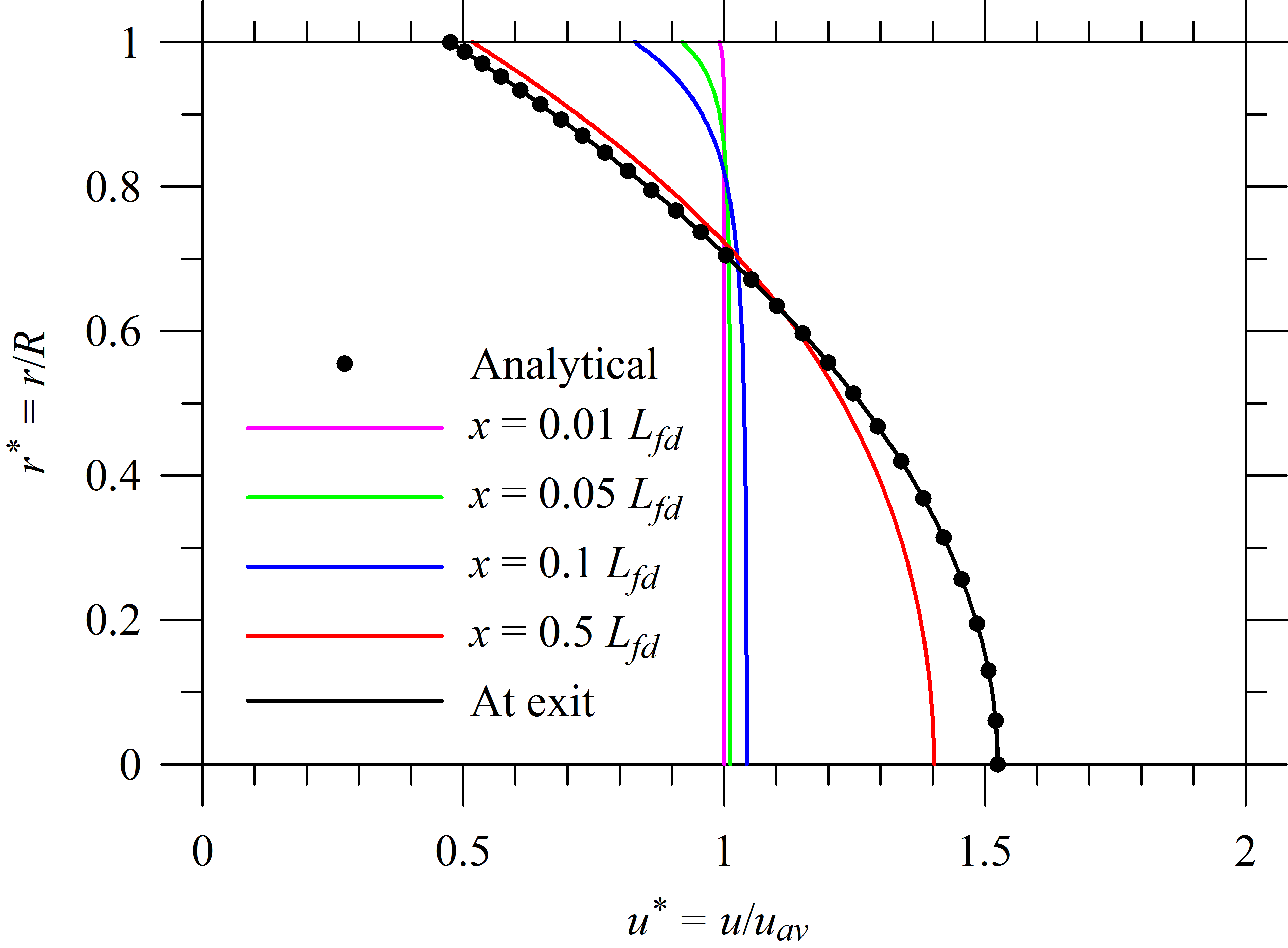} \label{fig:Re01_Kn_01C2_05}
} 
\subfigure[$Re = 100, Kn = 0.01, C_{2} = 0$]{
\includegraphics[width=0.45\textwidth]{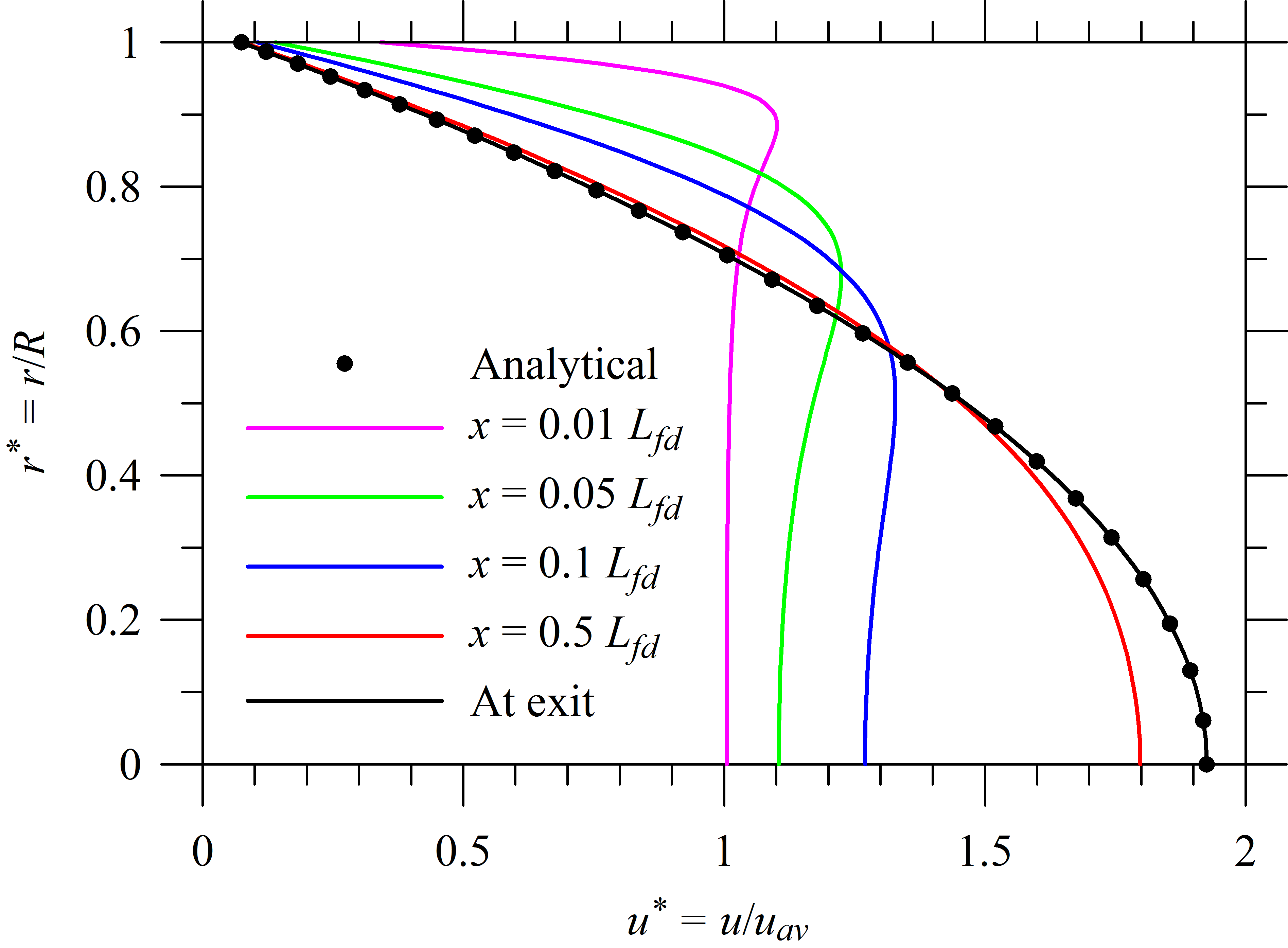} \label{fig:Re100Kn_001C2_00}
} 
\subfigure[$Re = 100, Kn = 0.1, C_{2} = 0.5$]{
\includegraphics[width=0.45\textwidth]{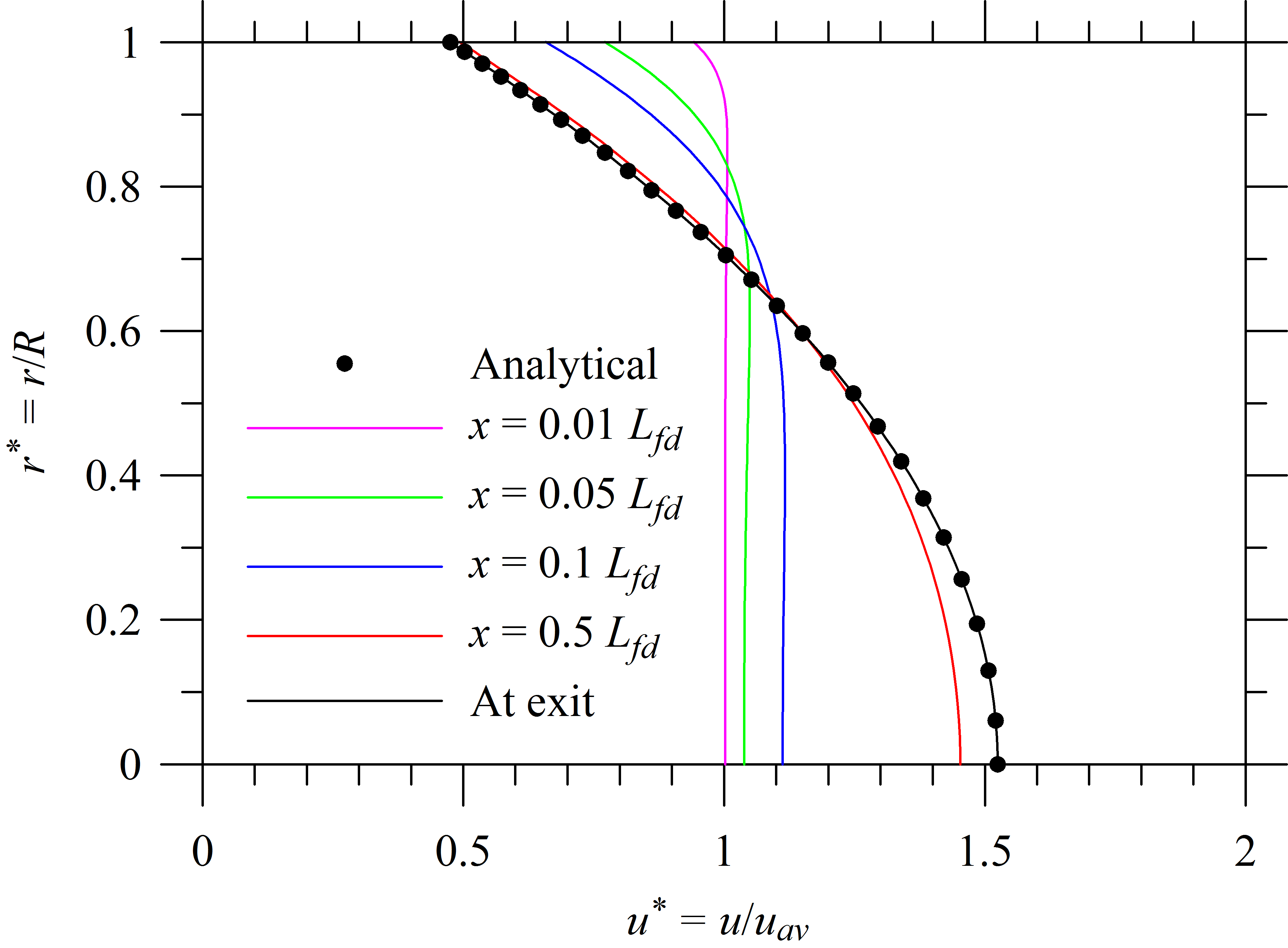} \label{fig:Re100Kn_01C2_00}
} 
\end{center}
\caption{Development of $u / u_{av}$ as functions of $r /R $ at different axial locations for pipe flows.}
\label{fig:overshoots}
\end{figure}

The velocity profiles at different axial locations are presented in Fig.~\ref{fig:overshoots} for pipe flows with $Re = 0.1$ and $100$ for $Kn = 0.01$ with $C_{2} = 0$ and $Kn = 0.1$ with $C_{2} = 0.5$. The figure shows that irrespective of the chosen parameters, the analytical solutions for the fully-developed flows have always been achieved at the exit that justify the present choice of $x^{*}_{\max}$. In addition, the velocity overshoot is observed to be more pronounced for higher $Re$, although even in this regime, it decreases considerably with the increase in both $Kn$ and $C_{2}$. Accordingly, the largest velocity overshoot occurs for the highest $Re$ with $C_2 = 0$ and $Kn = 0$, although the results are not presented in here for brevity. For higher $Kn$ with $C_{2} = 0.5$, the velocity overshoots could be so minute that it may hardly be noticeable. \\

From the variations of axial velocity profiles in Fig~\ref{fig:overshoots}, it may appear that the parabolic (quadratic) velocity profile, without any velocity overshoot, could be assumed at least for higher $Kn$ and $C_{2}$ and the boundary layer theory could be applied for the prediction. However, similar to $Kn = 0$, such an apprehension could be true only for $Re \rightarrow \infty$ since the effects of axial diffusion are always neglected in such analysis and hence relations similar to Eq.~(\ref{Eq:LbyD_HighReLimit}) could be retrieved only in the convection dominated regime. Nevertheless, as \citet{Durstetal2005} demonstrated, the constant in such relation still remains a function of $Re$, even in the apparently convection dominated regime ($Re \ge 100$) and hence numerical simulations are inevitable for all investigated ranges of parameters. \\

Another important observation is that irrespective of the operating condition, $u^{*}_{s}$ in the developing region is always higher than $u^{*}_{s,fd}$. However, for higher $Kn$ and $C_{2}$, the velocity gradients at the wall are observed to be less that in the fully-developed section, which may be attributed to the presence of second term in Eq.~(\ref{Eq:General-Second-Order-Slip-Condition-Dimless}) that becomes important with the increase in both $Kn$ and $C_{2}$. As a consequence, for such cases, the wall shear stresses in the developing region are found to be less than that in the fully-developed section, which is expected to significantly affect the variations in $K (x)$, as shall be discussed later.

\subsection{Development Lengths and Correlations} \label{sub:development_length}

\begin{figure}[htbp]
\begin{center}
\includegraphics[width=0.6\textwidth]{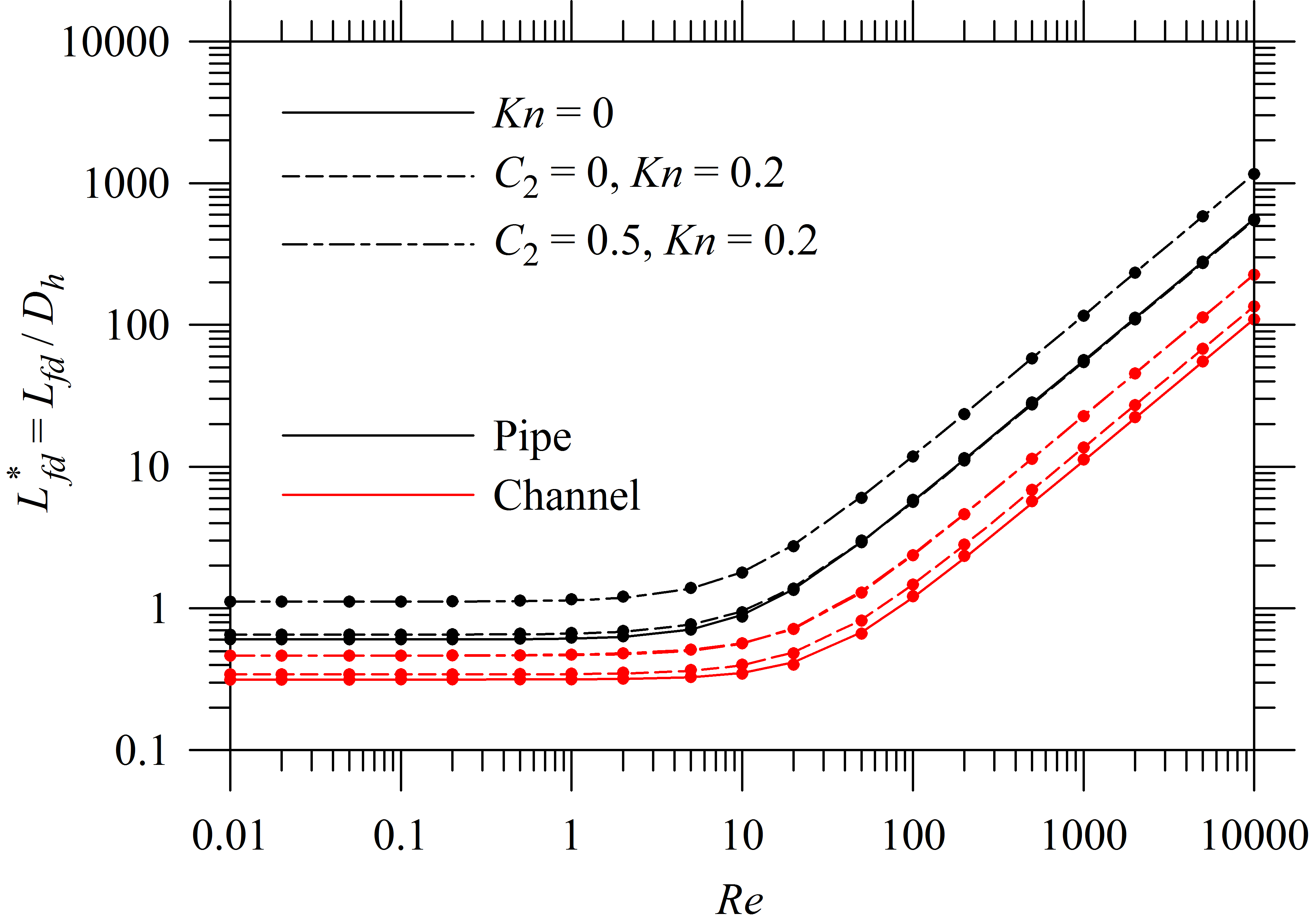}
\end{center}
\caption{$L^{*}_{fd} = L_{fd} / D_{h}$ as functions of $Re$. The symbols represent the raw computational data and the lines are obtained according to Eq.~(\ref{eq:corr_general}).}
\label{fig:twoAsymptotes}
\end{figure}

The dimensionless development lengths $L^{*}_{fd} = L_{fd} / D_{h}$ for both pipe and channel flows are still characterised by the low and the high $Re$ asymptotes, similar to that reported by \citet{Durstetal2005} for $Kn = 0$. The results for $Kn = 0.2$ with $C_{2} = 0$ and $C_{2} = 0.5$, along with that for $Kn = 0$, are presented in Fig.~\ref{fig:twoAsymptotes}. A careful examination of the results, which will be shortly evident from the data in Table~\ref{tab:L0L1_PipeCHannel}, shows that except for pipe flows in the high $Re$ regime and for both pipe and channel flows in the low $Re$ regime only with $C_{2} = 0$ and high $Kn$, $L^{*}_{fd}$ increases with the increase in both $Kn$ and $C_{2}$. In the high $Re$ regime, however, irrespective of $C_{2}$, $L^{*}_{fd}$ for pipe flows first decreases with the initial increase in $Kn$, where both reduction in $L^{*}_{fd}$ and value of $Kn$, up to which $L^{*}_{fd}$ decreases, depend on $C_{2}$. With the subsequent increase in $Kn$, $L^{*}_{fd}$ increases irrespective of $C_{2}$, although only for the first order velocity slip condition ($C_{2} = 0$), $L^{*}_{fd}$ decreases once again with the further increase in $Kn$. Similar behaviour, however, could not be detected for channel flows, irrespective of $C_{2}$. In the low $Re$ regime, on the other hand, $L^{*}_{fd}$ for both pipe and channel flows only with $C_{2} = 0$ has been found to decrease marginally for higher values of $Kn$. For all other $C_{2}$, $L^{*}_{fd}$ in the diffusion dominated regime consistently increases with the increase in $kn$. Nevertheless, the intermediate data are not presented in Fig.~\ref{fig:twoAsymptotes} for brevity, since they will be shortly apparent from the proposed correlations and the asymptotic behaviours for $Re \rightarrow 0$ and $Re \rightarrow \infty$ in Table~\ref{tab:L0L1_PipeCHannel}. \\

Comparing the variations in $L^{*}_{fd}$ for $Kn= 0$ and $0.2$, it may be concluded that earlier correlations, developed specifically for the continuum regime \citep{Chen_1973, Dombrowski_etal_1993, Durstetal2005}, cannot be used for predicting $L^{*}_{fd}$ of flows in presence of velocity slip at the wall, without causing substantial error. Nevertheless, the dependence of $L^{*}_{fd}$ on $Re$ could still be represented in its general form as:
\begin{equation}
L^{*}_{fd} = \left[ \left( L_{0} \right)^{q} + \left( L_{1} Re \right)^{q} \right]^{1/q}
\label{eq:corr_general}
\end{equation}
where $L_{0}$ and $L_{1}$ are functions of $Kn$ and $C_{2}$. They could also be functions of $C_{1}$. However, since $C_{1}$ has been kept fixed to unity for all investigated cases, its effect could not be ascertained from the present study. Nevertheless, $L_{0}$ and $L_{1}$ have been directly obtained from the simulated data for $L^{*}_{fd}$ at $Re = 10^{-2}$ and $L^{*}_{fd} / Re$ at $Re = 10^{4}$, respectively. The results for different combinations of $C_{2}$ and $Kn$ are presented in Table~\ref{tab:L0L1_PipeCHannel} for completeness, where the minimum $L_{1}$ for all $C_{2}$ and the local maximum $L_{1}$ for $C_{2} = 0$ are highlighted for easy identification of the features described earlier. \\

As compared to the correlations proposed by \citet{Durstetal2005} for $Kn = 0$, certain values have been further adjusted in order to improve the predictability. The exponent $q$ has been found to be $1.5975$ and $1.6002$ for pipe and channel flows, respectively, instead of $1.6$ for both geometries in Eq.~(\ref{Eq:LbyD_Durst}).\footnote{Corrected up to the first place of decimal, both exponents are, however, equal to $1.6$.} For pipe flows, $L_{0} = 0.6044$ (instead of $0.619$ with $2.42$ \% deviation) and $L_{1} = 5.5935 \times 10^{-2}$ (rather than $5.67 \times 10^{-2}$ with $1.37$ \% deviation) have been determined with a maximum absolute error of $2.47$ \%, in contrast to $4.38$ \%, obtained from the correlation of \citet{Durstetal2005}. Similarly, for channel flows, $L_{0} = 0.3152$ (as opposed to $0.3155$, with $0.1$ \% deviation) and $L_{1} = 1.0984 \times 10^{-2}$ (in lieu of $1.105 \times 10^{-2}$ with $0.6$ \% deviation) have been evaluated with a maximum absolute error of $3.86$ \%, as compared to $4.16$ \% achieved according to Eq.~(\ref{Eq:LbyD_Durst}). \\

\begin{figure}[htbp]
	\begin{center}
		\subfigure[Pipe]{
			\includegraphics[width=0.45\textwidth]{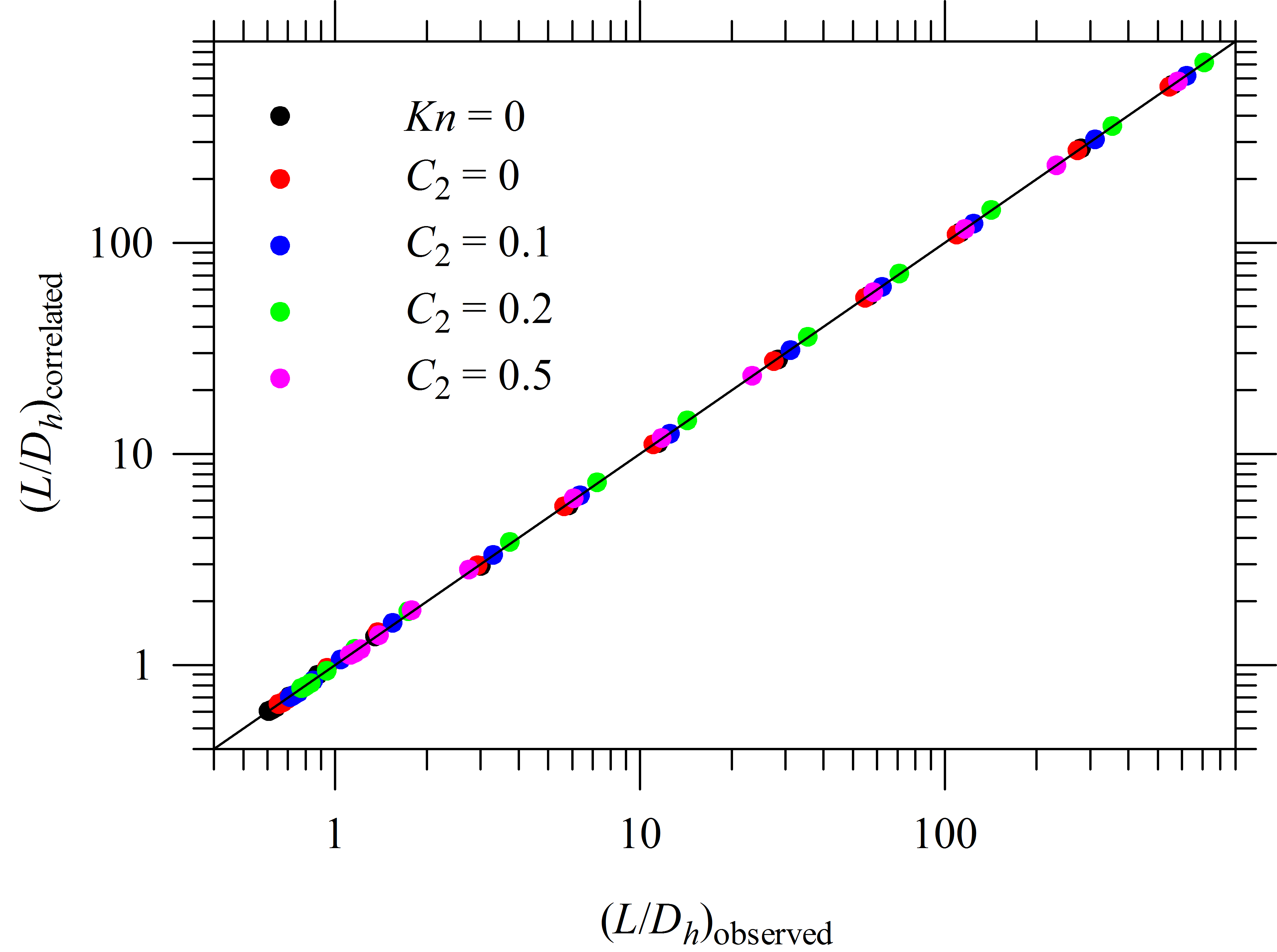}
		} \label{fig:LbyDCorr_Pipe}
		\subfigure[Parallel Plate Channel]{
			\includegraphics[width=0.45\textwidth]{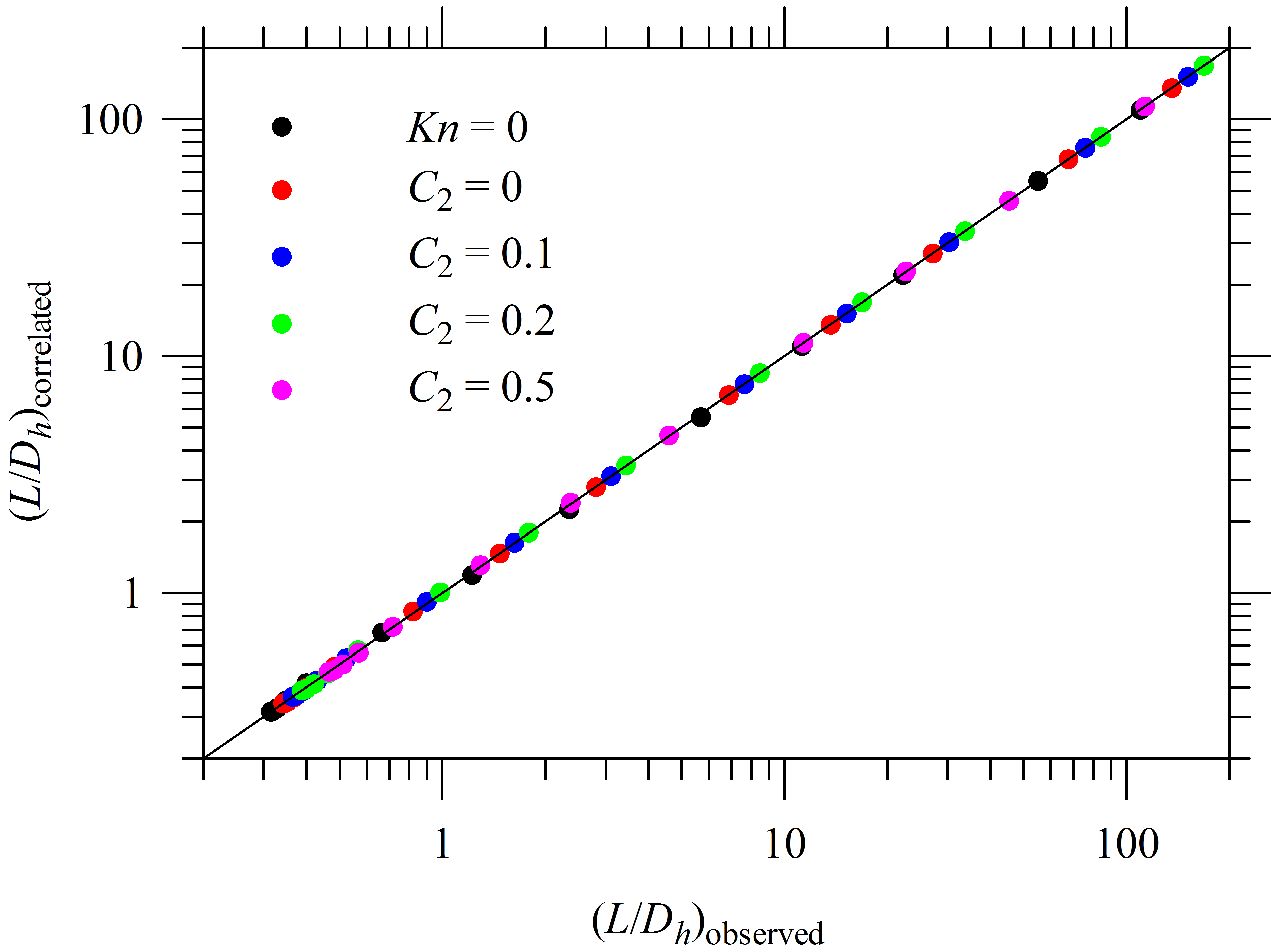}
		}\label{fig:LbyDCorr_PP}
	\end{center}
	\caption{Performance of the correlations for $L^{*}_{fd}$. The solid lines represent 100\% accuracy.}
	\label{fig:LbyDCorr}
\end{figure}

Insignificant deviations in the exponent $q$ has been observed since the earlier correlations \citep{Durstetal2005} were obtained by allowing $q$ to vary only up to the first place of decimal, while minimising the maximum absolute relative error in $L^{*}_{fd}$ using a search method. The marginal variations in $L_{1}$ may be explained by the fact that the earlier high $Re$ asymptotes were obtained for $Re = 4 \times 10^{3}$, whereas the present computations have been extended up to  $Re = 10^{4}$. On the other hand, small differences in $L^{*}_{fd}$ for $Re \rightarrow 0$, given by $L_{0}$, may be attributed to the use of non-uniform grid in the radial (or the transverse) direction that better resolves the velocity gradients close to the wall than on uniform grid, employed by \citet{Durstetal2005}. Nevertheless, the dependence of $L_{0}$ and $L_{1}$ on $Kn$ for a fixed $C_{2}$ could be expressed as:
\begin{equation}
L_{i} = \sum_{j=0}^{2}~l_{ij}\ Kn^{j} = l_{i0} + l_{i1} Kn + l_{i2} Kn^{2} ~~~~~ \mbox{for}~ i=1,2 \label{Eq:Corr_Ai}
\end{equation}
where $l_{00} = 0.6044$ and $0.3152$, while $l_{10} = 5.5935 \times 10^{-2}$ and $1.0984 \times 10^{-2}$ are constants, obtained directly from the numerical simulations with $Kn = 0$ for pipe and channel flows, respectively. Other $l_{0j}$ and $l_{1j}$ for $j = 1$ and $2$ have been found to be best represented by the quadratic functions of $C_{2}$. For pipe flows, they have been obtained as:
\begin{subequations}
\begin{eqnarray}
l_{01} & = & 0.7937 + 1.652~C_{2} - 2.1152~C_{2}^{2} \label{Eq:a01_Pipe} \\
l_{02} & = & - 2.7519 + 2.2478~C_{2} + 35.8177~C_{2}^{2} \label{Eq:a02_Pipe} \\
l_{11} & = & 7.691 \times 10^{-4} + 0.2212~C_{2} - 0.2954~C_{2}^{2} \label{Eq:a11_Pipe} \\
l_{12} & = & - 3.3061 \times 10^{-2} + 0.3009~C_{2} + 4.8206~C_{2}^{2} \label{Eq:a12_Pipe}
\end{eqnarray}
\label{Eq:aij_Pipe}
\end{subequations}
Similarly, for channel flows, these coefficients have been correlated as:
\begin{subequations}
\begin{eqnarray}
l_{01} & = & 0.4189 + 0.4671~C_{2} + 1.8294 \times 10^{-2}~C_{2}^{2} \label{Eq:a01_Channel} \\
l_{02} & = & - 1.4249 + 3.1608~C_{2} + 1.2335~C_{2}^{2} \label{Eq:a02_Channel} \\
l_{11} & = & 2.1484 \times 10^{-2} + 2.2458 \times 10^{-2}~C_{2} - 6.637 \times 10^{-3}~C_{2}^{2} \label{Eq:a11_Channel} \\
l_{12} & = & - 4.312 \times 10^{-2} + 0.2681~C_{2} + 0.1863~C_{2}^{2} \label{Eq:a12_Channel}
\end{eqnarray}
\label{Eq:aij_Channel}
\end{subequations}
The exponent $q$ in Eq.~(\ref{eq:corr_general}) has been found to vary between $1.5975$ and $1.4992$ for pipe flows and $1.6002$ and $1.4923$ for channel flows. Similar to \citet{Durstetal2005}, these values have been obtained by minimising the maximum absolute relative error in $L^{*}_{fd}$ for different combinations of $Kn$ and $C_{2}$. Since the predicted $L^{*}_{fd}$ is less sensitive to the variations in $q$ and since $q$ varies over a relatively small range, it has been correlated as:
\begin{subequations}
\begin{eqnarray}
q & = & 1.5975 - \left( 0.4956 - 0.6511~C_{2} + 0.7115~C_{2}^{2} \right)~Kn ~~~~~ \mbox{for pipe} \label{Eq:n_Pipe} \\
  & = & 1.6002 - \left( 0.5743 - 0.9495~C_{2} + 0.7444~C_{2}^{2} \right)~Kn ~~~~~ \mbox{for channel} \label{Eq:n_Channel}
\end{eqnarray}
\label{Eq:n_PipeChannel}
\end{subequations}
The performance of the correlations is presented in Fig.~\ref{fig:LbyDCorr} for the most stringent case with $Kn = 0.2$ for different $C_{2}$ and also for $Kn = 0$, similar to that obtained by \citet{Durstetal2005} in the continuum regime. It is evident from the figure that irrespective of $Re$, $Kn$ and $C_{2}$, the present correlations perform extremely well for both geometries and produce maximum absolute errors of $2.47$ \% and $3.86$ \% for pipe and channel flows, respectively. Comparing these deviations with that obtained earlier for the continuum regime, it is obvious that the maximum errors in prediction occur for $Kn = 0$ for both pipe and channel flows and hence the exponent $q$ may be further adjusted.\\

At this point, it may be mentioned that instead of evaluating $q$ from Eq.~(\ref{Eq:n_PipeChannel}), if constant values of $q = 1.5975$ and $1.6002$ are used irrespective of $C_{2}$ and $Kn$, the resultant correlations produce maximum errors of $3.65$~\% (for $C_{2} = 0.1$ and $Kn = 0.2$) and $3.86$~\% (for $Kn = 0$) for pipe and channel flows, respectively. On the other hand the use of $q = 1.6$ for both pipe and channel flows, as proposed by \citet{Durstetal2005} for the continuum regime, yields maximum absolute errors of $3.70$ \% and $3.87$ \% for pipe and channel flows, respectively.\footnote{The maximum absolute error for pipe flows is still less than that for channel flows.} Therefore, if some additional error ($\approx 1.23$ \%) is considered acceptable for pipe flows, $q = 1.6$ may be recommended for both geometries in order to correlate $L^{*}_{fd}$ even in the presence of substantial velocity slip at the wall, provided $L_{0}$ and $L_{1}$ are evaluated according to Eq.~(\ref{Eq:Corr_Ai}), while calculating $l_{ij}$ from Eqs.~(\ref{Eq:aij_Pipe}) and (\ref{Eq:aij_Channel}) for pipe and channel flows, respectively. \\

\begin{figure}[htbp]
\begin{center}
\includegraphics[width=0.6\textwidth]{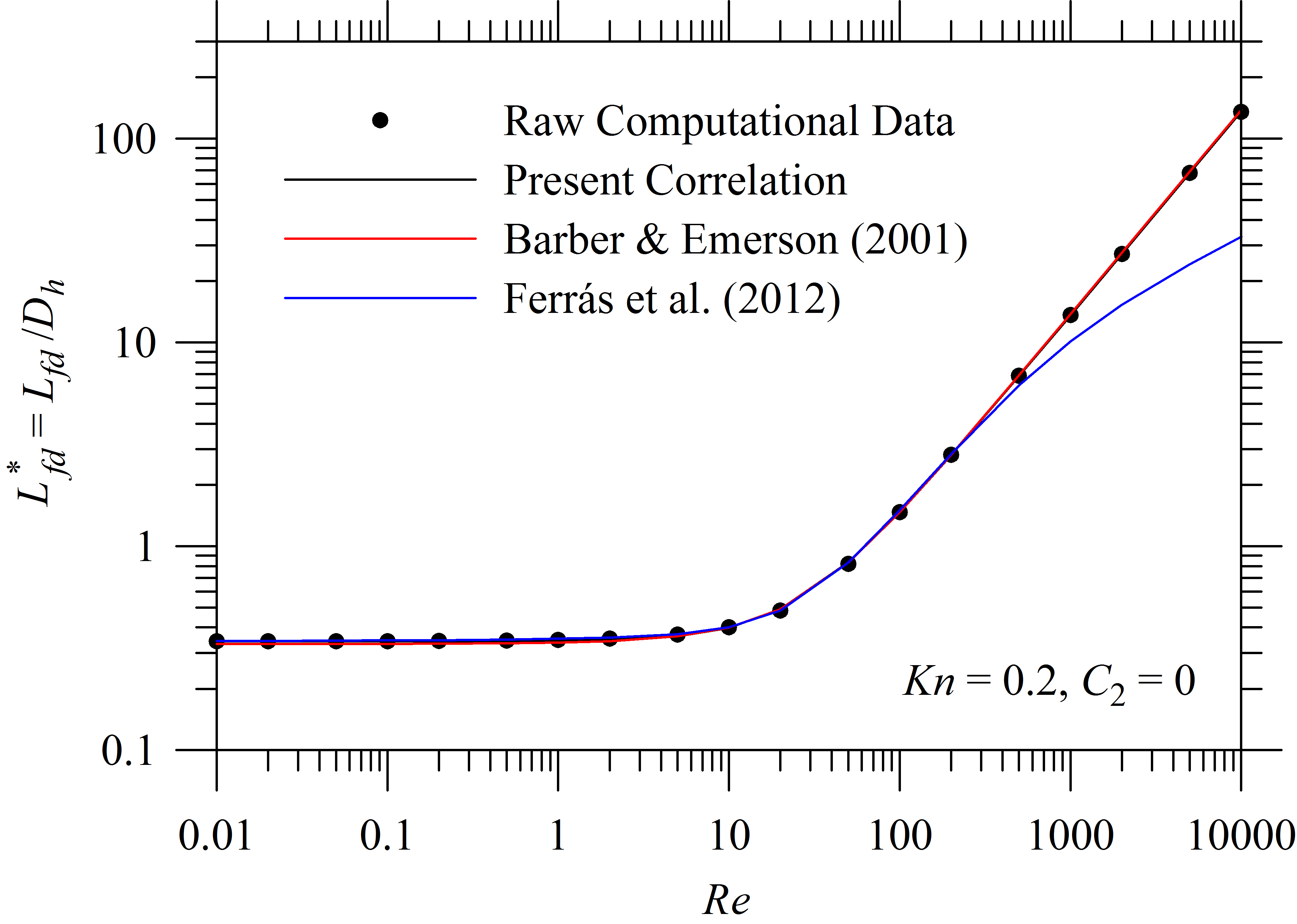}
\end{center}
\caption{Comparison of different correlations for $L^{*}_{fd}$ for channel flows with $Kn = 0.2$ and $C_{2} = 0$.}
\label{fig:LbyDCompFirstOrder}
\end{figure}

For completeness, a comparison of the present correlation for channel flows with the first order velocity slip condition at the wall\footnote{obtained by setting $C_{2} = 0$} with those proposed by \citet{BarberAndEmerson_2001} and \citet{Ferrasetal2012} is presented in Fig.~\ref{fig:LbyDCompFirstOrder} for $Kn = 0.2$. The figure clearly shows that although the high $Re$ asymptote is well represented by both Eqs.~(\ref{Eq:BarberAndEmerson_2001}) and (\ref{eq:corr_general}), the correlation of \citet{Ferrasetal2012} in Eq.~(\ref{Eq:Ferrasetal2012}) deviates considerably from the simulated data for $Re > 200$. Nevertheless, up to $Re = 200$, all correlations perform almost equally well and hence they are hardly distinguishable from each other in Fig.~\ref{fig:LbyDCompFirstOrder}. The maximum absolute errors for $Kn = 0.2$, however, have been obtained as $1.76$ \%, $3.06$ \% and $2.21$ \% for the present correlation (for $10^{-2} \le Re \le 10^{4}$), Eq.~(\ref{Eq:BarberAndEmerson_2001}) (for $10^{-2} \le Re \le 400$) and Eq.~(\ref{Eq:Ferrasetal2012}) (for $10^{-2} \le Re \le 200$), respectively. For $Kn = 0$, on the other hand, these errors are found to be $3.86$ \%, $8.59$ \% and $6.98$ \%, respectively, for $10^{-2} \le Re \le 10^{4}$, although these variations are not shown in Fig.~\ref{fig:LbyDCompFirstOrder} for brevity. It may, therefore, be safely concluded that the present correlation for channel flows is not only more general\footnote{Since it accounts for the second order velocity slip condition at the wall.} as compared to the recommendations from earlier studies, but also it has been proved to be the most accurate over the entire ranges of investigated parameters. \\

For pipe flows, however, no realistic comparison could be made since the only investigation from \citet{BarberAndEmerson_2001} recommended using the earlier correlations from \citet{Chen_1973} and \citet{Dombrowski_etal_1993} that were specifically developed for $Kn = 0$ in the continuum regime, even for $Kn \ne 0$. Therefore, the present correlations in Eq.~(\ref{eq:corr_general}) are recommended for evaluating $L_{fd}$ for both pipe and channel flows.

\subsection{Incremental Pressure Drop Number} \label{sub:pressure_drop}

The investigation on pressure drop in the entrance region of micro-channels has been carried out by analysing the variations in $K (x)$ as functions of $x / D_{h}$ for different operating conditions and their asymptotic behaviour in the fully-developed section has been analysed from the variations in $K_{fd}$ as functions or $Re$, $Kn$ and $C_{2}$. While $L^{*}_{fd}$ for $Kn \ne 0$ qualitatively remains similar to that for the no-slip case with $Kn = 0$ and hence could still be functionally represented by the similar form in Eq.~(\ref{eq:corr_general}), substantial qualitative differences have been observed in the pressure drop data. In this section, the variations in $K (x)$ and $K_{fd}$ are critically examined and the correlations for the latter, that are extremely important for the evaluation of pressure drop for $L \geq L_{fd}$, are proposed. \\

\begin{figure}[htbp]
	\begin{center}
		\subfigure[$Re = 0.1$]{
			\includegraphics[width=0.45\textwidth]{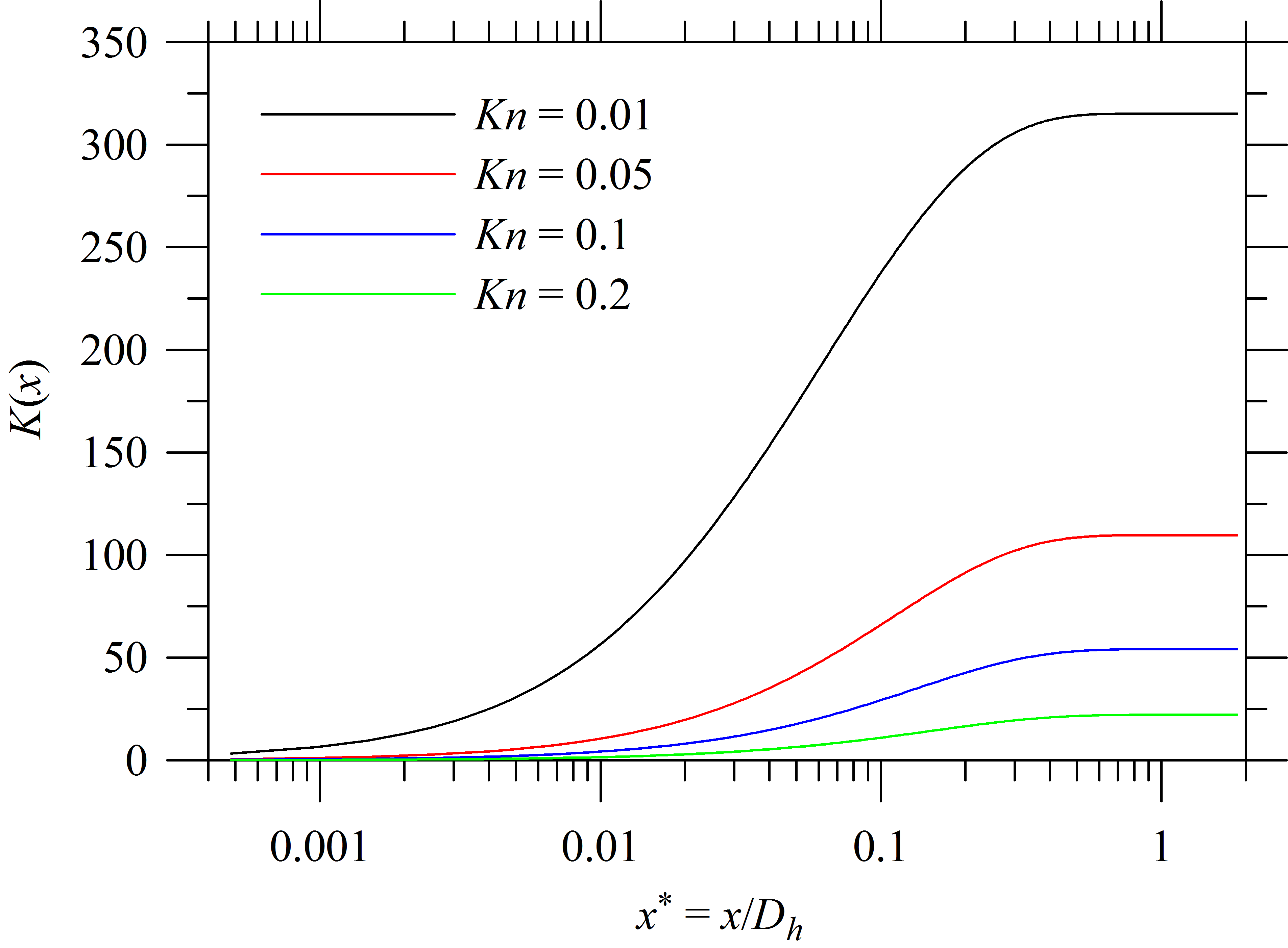}
		}\label{fig:K_Re01_C2_0}
		\subfigure[$Re = 1$]{
			\includegraphics[width=0.45\textwidth]{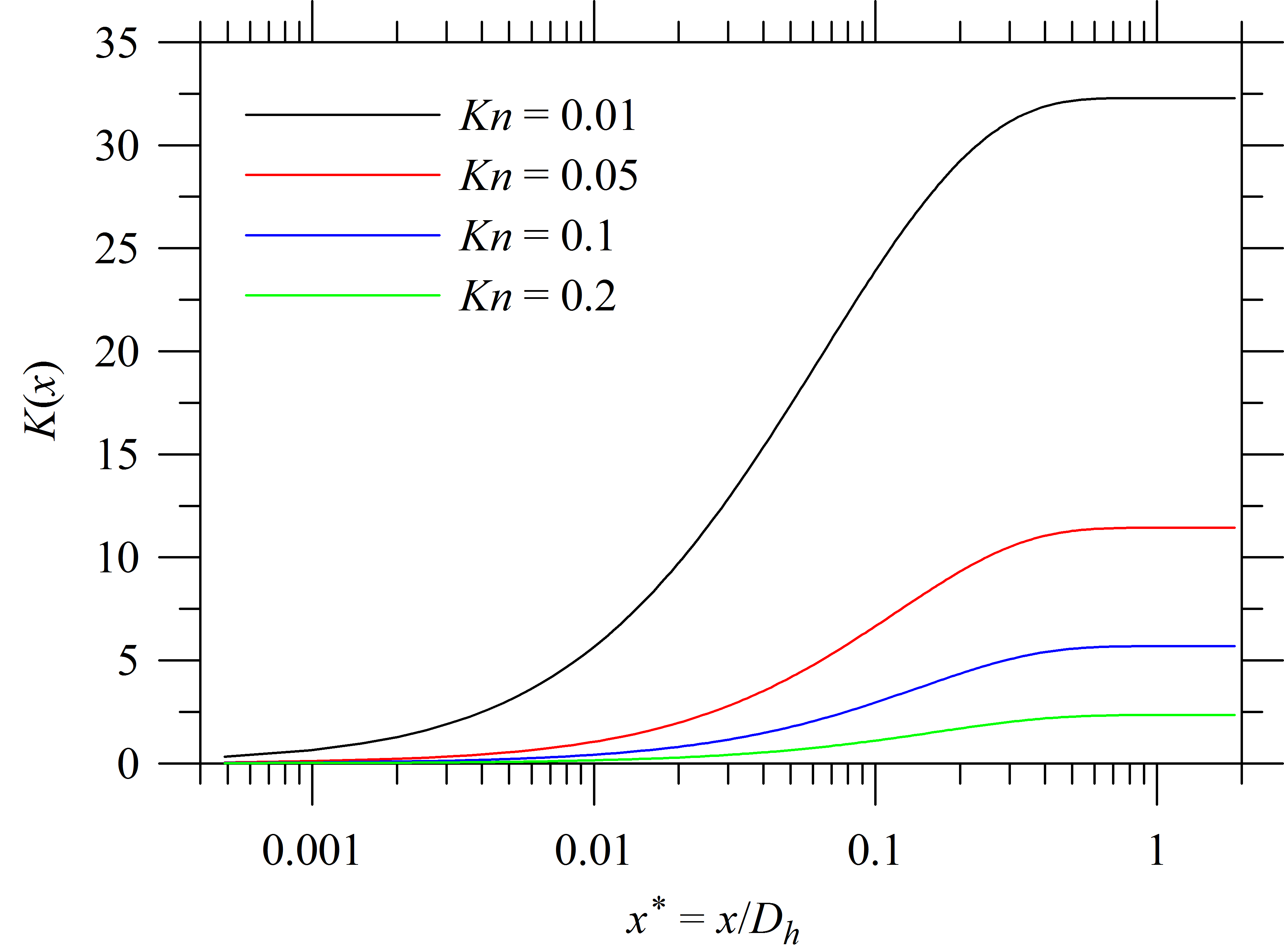}
		}\label{fig:K_Re1_C2_0}
		\subfigure[$Re = 10$]{
			\includegraphics[width=0.45\textwidth]{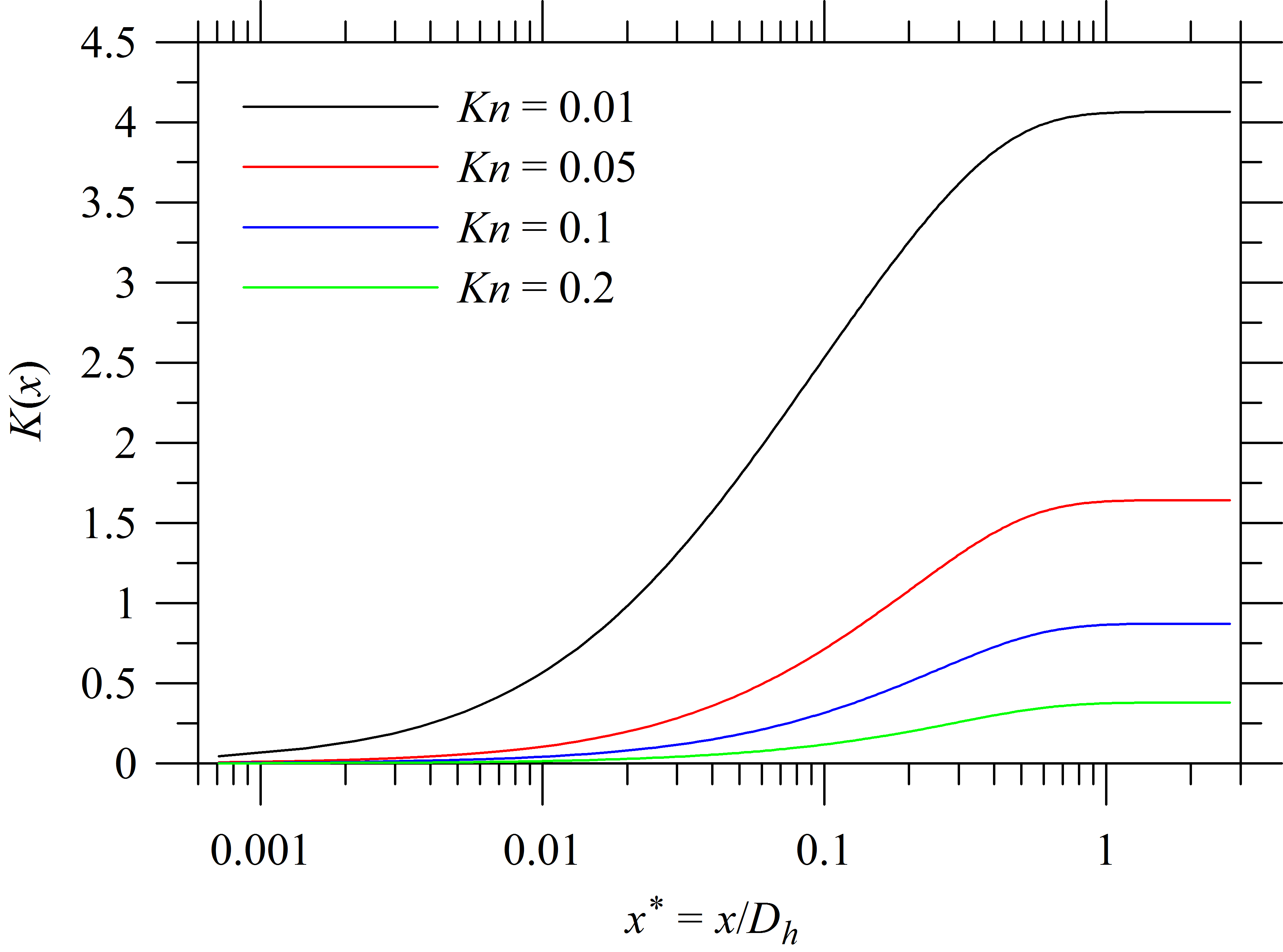}
		}\label{fig:K_Re10_C2_0}
		\subfigure[$Re = 100$]{
			\includegraphics[width=0.45\textwidth]{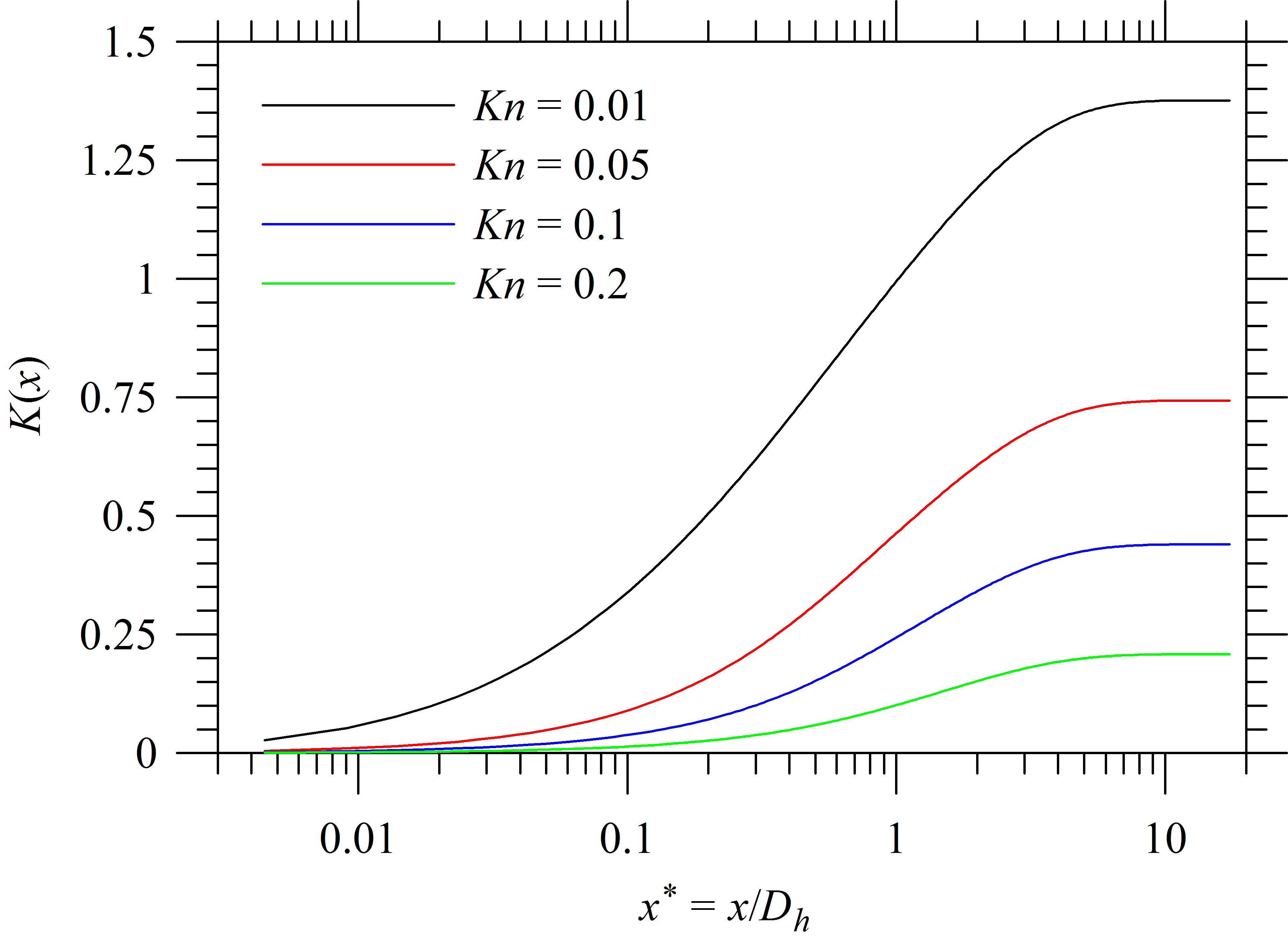}
		}\label{fig:K_Re100_C2_0}
	\end{center}
	\caption{Effects of $Kn$ on variations in $K(x)$ as functions of $x/D_{h}$ for pipe flows with $C_{2} = 0$.}
	\label{fig:K_Re_C2_0}
\end{figure}

For pipe flows with first order velocity slip condition at the wall, i.e., with $C_{2} = 0$, the variations in $K(x)$ for different $Re$ and $Kn$ are presented in Fig.~\ref{fig:K_Re_C2_0}, from which, it is evident that irrespective of $Kn$ and $Re$, $K(x)$ always increases from $0$ at $x^{*} = 0$ to its asymptotic value $K_{fd}$ for the fully-developed flow as $x^{*} \rightarrow \infty$, or, to be precise, for $x^{*} \ge L^{*}_{fd}$. These variations are similar to that observed for $Kn = 0$ with no-slip condition at the wall, although the magnitude of $K_{fd}$, as well as $K(x)$ for a particular $x^{*}$, reduces considerably with the increase in $Kn$. The reduction in $K_{fd}$ is due to the decrease in both $\Delta p^{*}_{m,\infty}$, which, as shown in Eq.~(\ref{Eq:DelPmInf}) and Fig.~\ref{fig:DelPmInf}, is independent of $Re$ and depends only on $Kn$ for a given $C_{2}$, and the velocity gradient and hence the shear stress at the wall, which, other than $Kn$ and $C_{2}$, also depends on $Re$, as demonstrated in Fig.~\ref{fig:overshoots}. These effects are more prominent as $Kn$ increases beyond $0.01$.\\

\begin{figure}[htbp]
	\begin{center}
		\subfigure[$Re = 0.1$]{
			\includegraphics[width=0.45\textwidth]{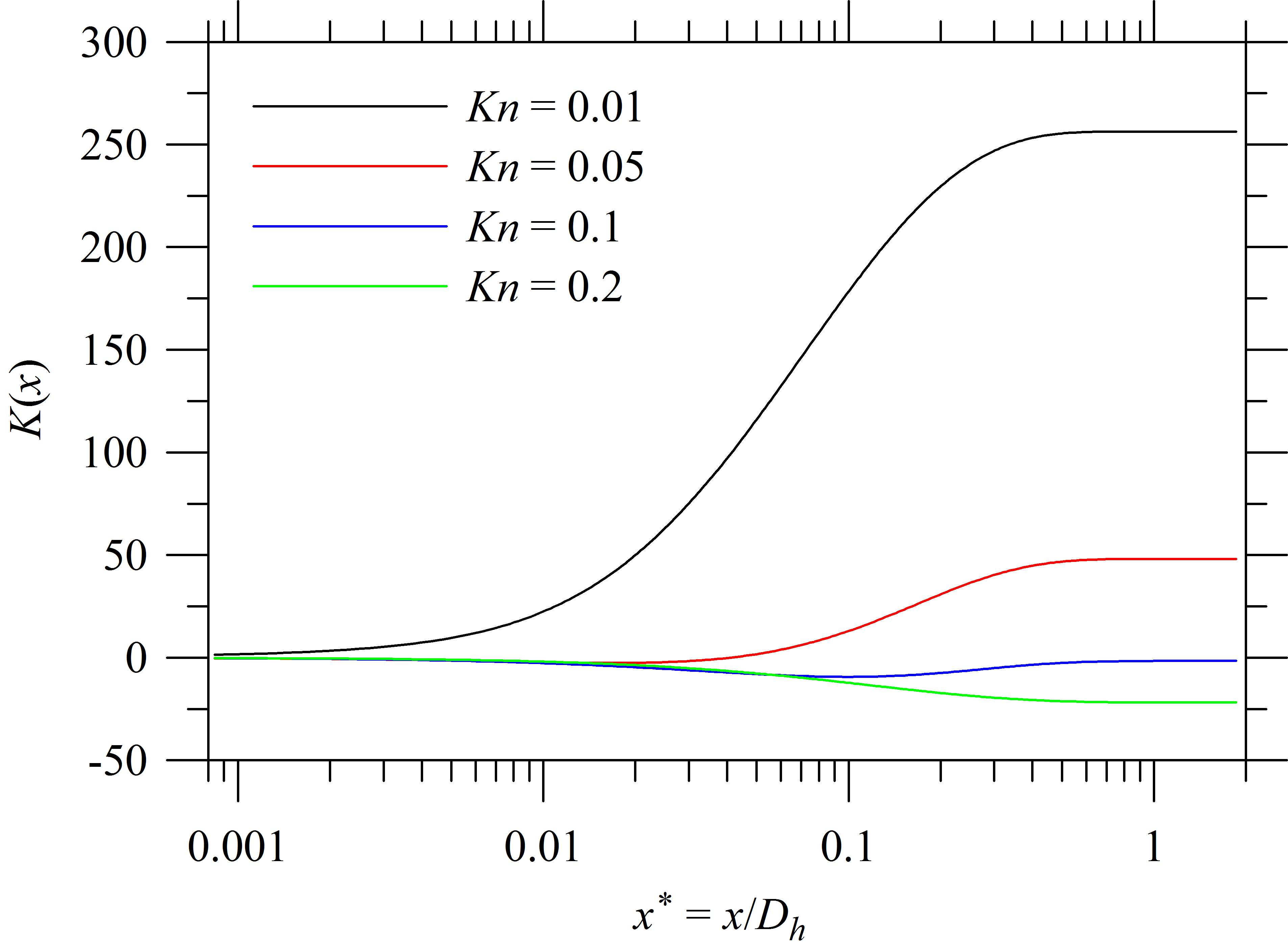}
		}\label{fig:K_Re01_C2_05}
		\subfigure[$Re = 1$]{
			\includegraphics[width=0.45\textwidth]{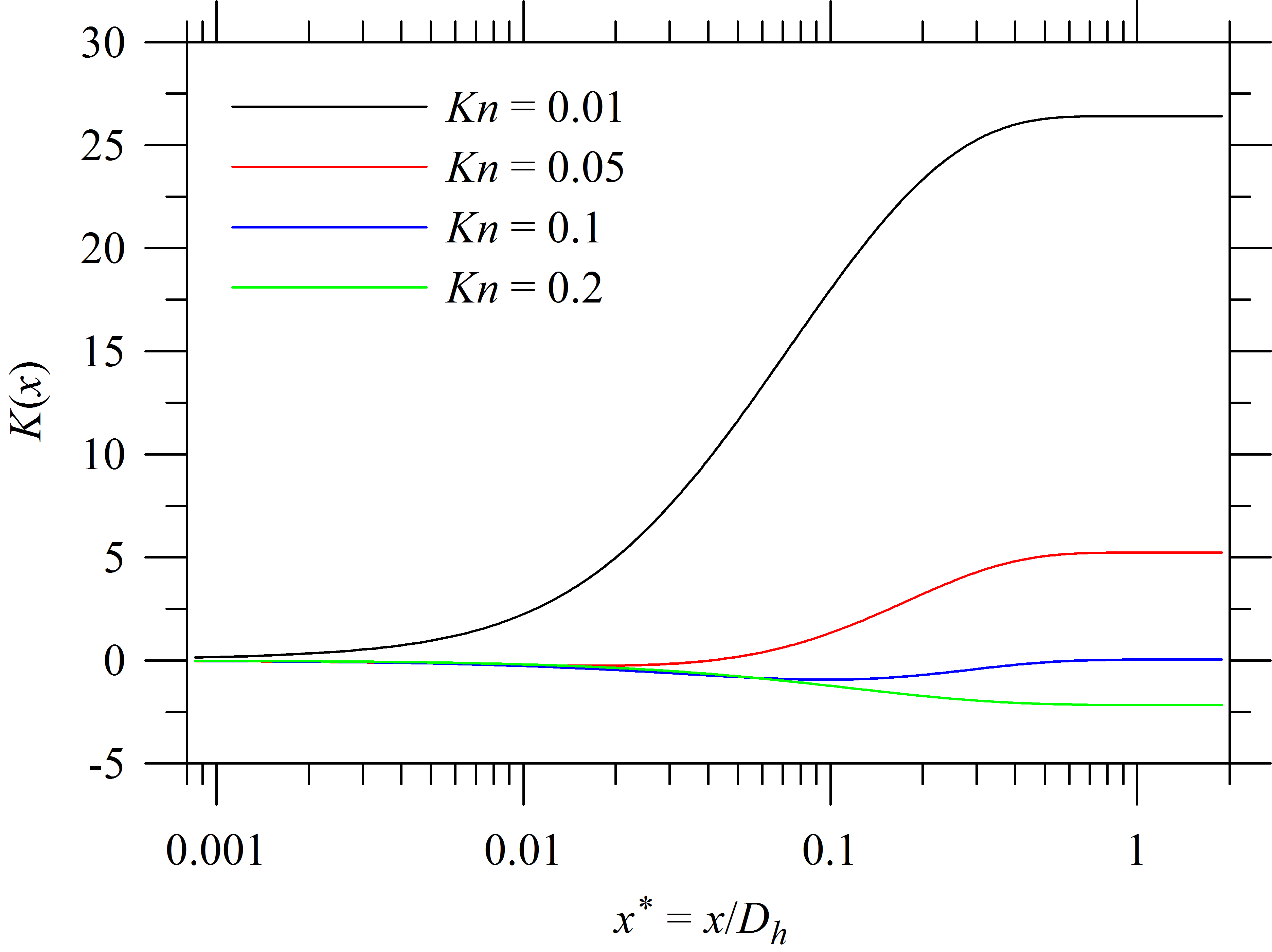}
		}\label{fig:K_Re1_C2_05}
		\subfigure[$Re = 10$]{
			\includegraphics[width=0.45\textwidth]{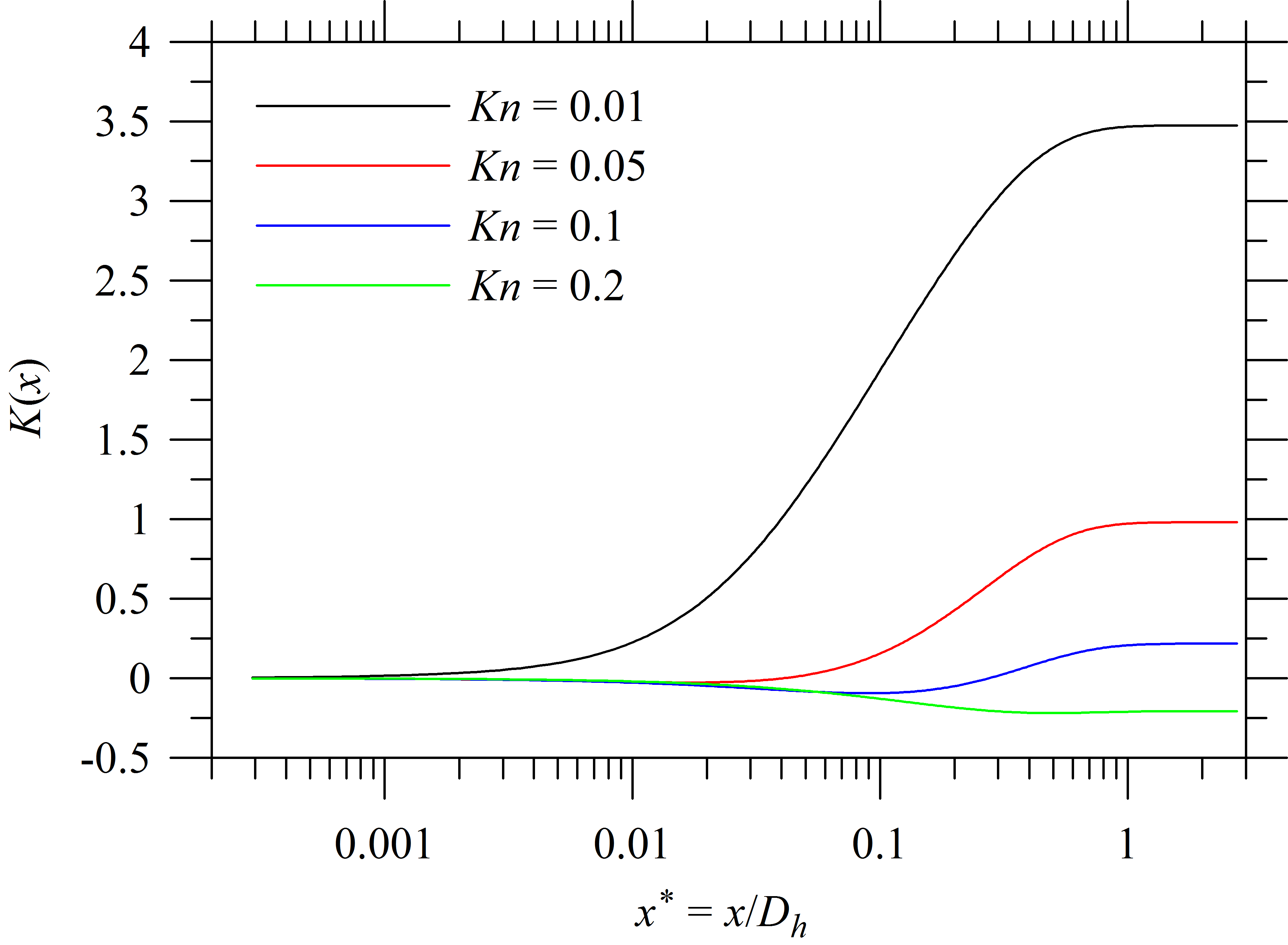}
		}\label{fig:K_Re10_C2_05}
		\subfigure[$Re = 100$]{
			\includegraphics[width=0.45\textwidth]{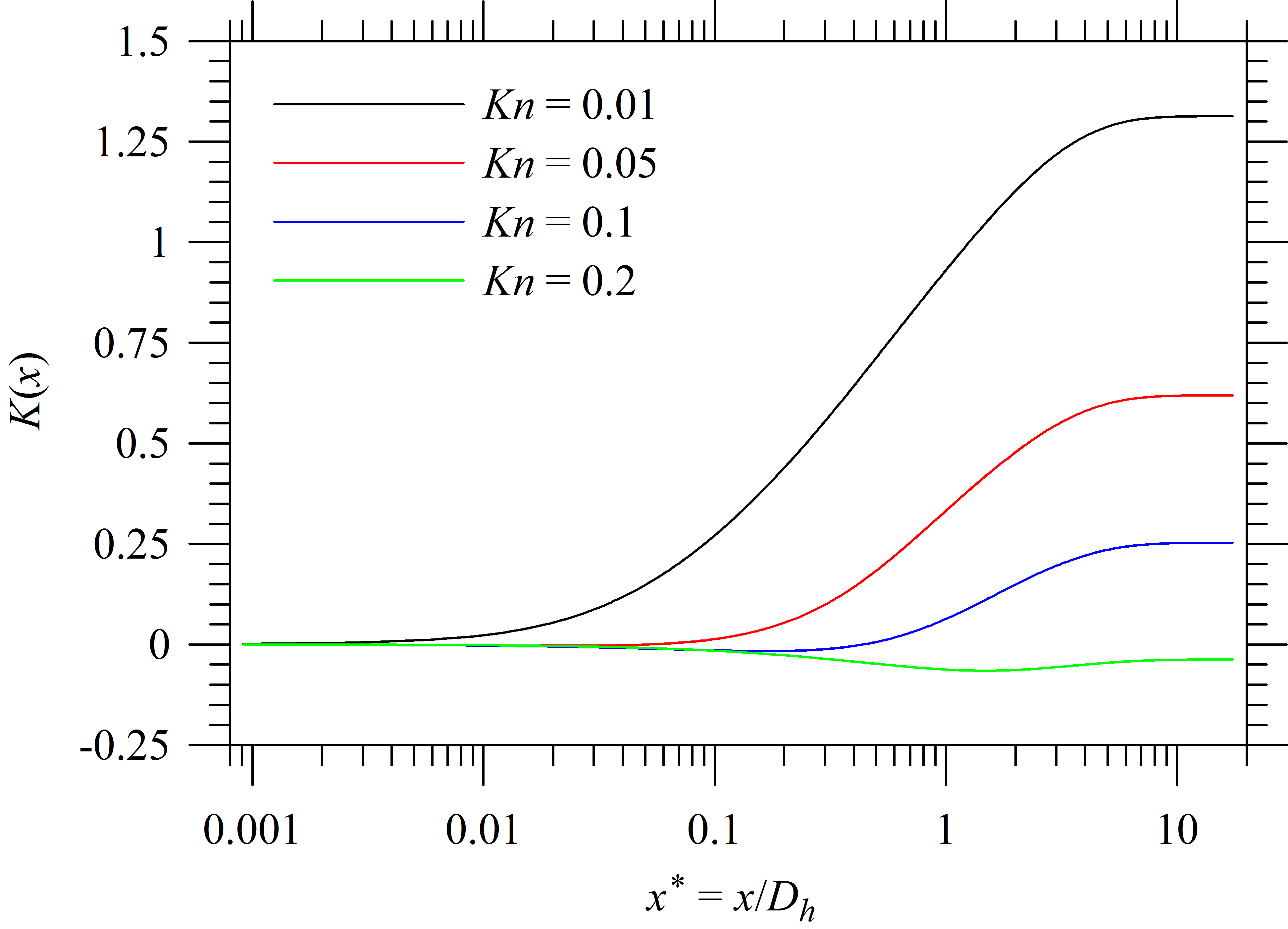}
		}\label{fig:K_Re100_C2_05}
	\end{center}
	\caption{Effects of $Kn$ on variations in $K(x)$ as functions of $x/D_{h}$ for pipe flows with $C_{2} = 0.5$.}
	\label{fig:K_Re_C2_05}
\end{figure}

Similar variations in Fig.~\ref{fig:K_Re_C2_05} for the second order velocity slip condition at the wall with $C_{2} = 0.5$, however, show some never reported interesting features. It is observed that $K(x)$ as well as $K_{fd}$, particularly for higher $Kn$, could even be negative, which, other than the already reported decrease in $\Delta p^{*}_{m,x}$ in Eq.~(\ref{Eq:Deltapx}), could be attributed to the reduction in wall shear stresses that occurs with the increase in both $Kn$ and $C_{2}$ (see Fig.~\ref{fig:overshoots}). It may be recognised that a positive $K(x)$, which is most often the case in the continuum regime for $Kn \le 0.01$ and with the first order velocity slip condition ($C_{2} = 0$), implies that if one calculates the true pressure drop using $f_{fd}$ rather than $f_{app,x}$ given in Eq.~(\ref{Eq:f_app}), one would underestimate $\Delta p$, while a negative $K(x)$ would lead to an overestimation of $\Delta p$ using the same method, which in certain cases, could be more acceptable than any underestimation. \\

\begin{figure}[htbp]
	\begin{center}
		\subfigure[$Re = 0.1$]{
			\includegraphics[width=0.45\textwidth]{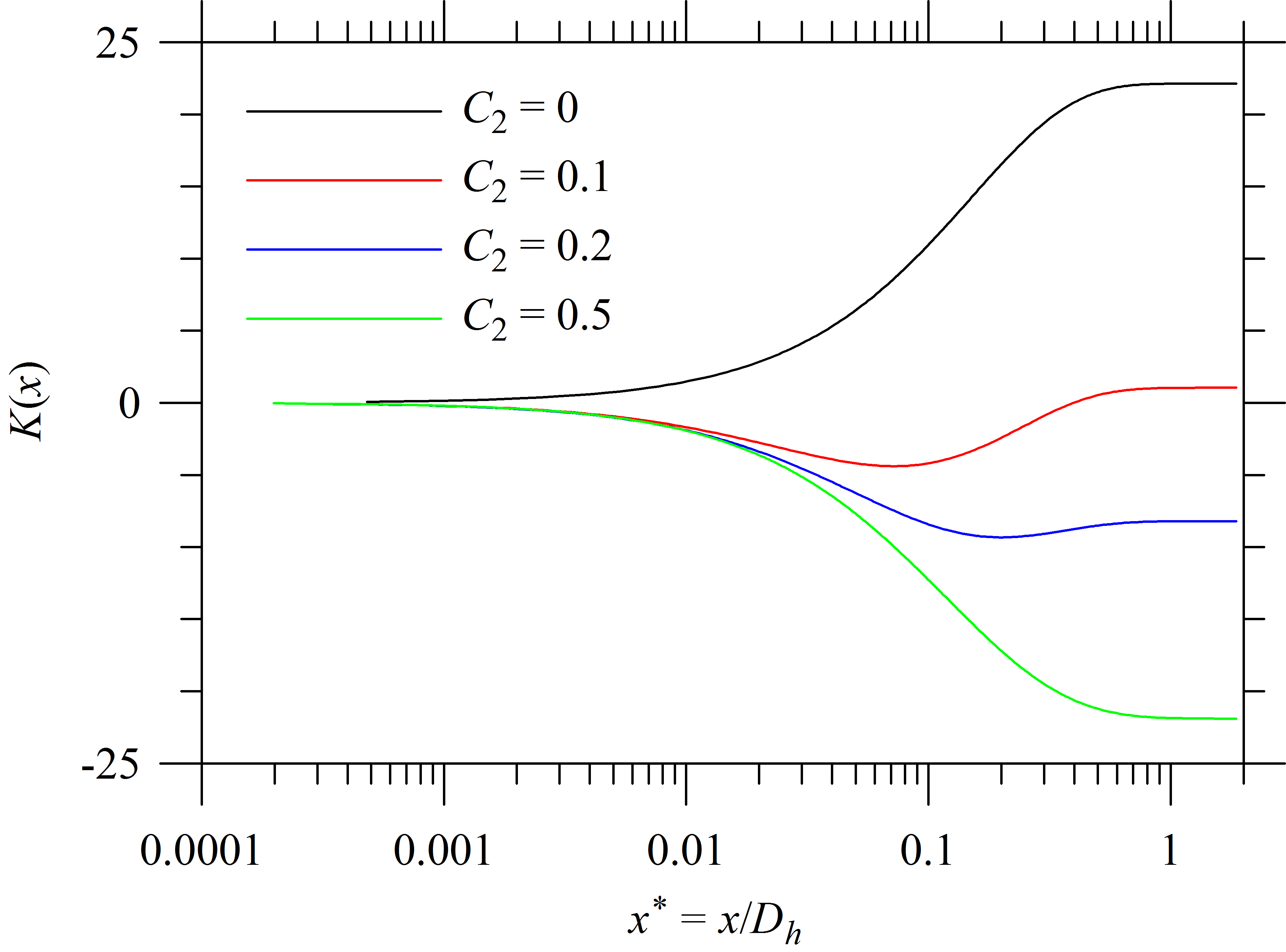}
		}\label{fig:K_Re01_Kn02}
		\subfigure[$Re = 1$]{
			\includegraphics[width=0.45\textwidth]{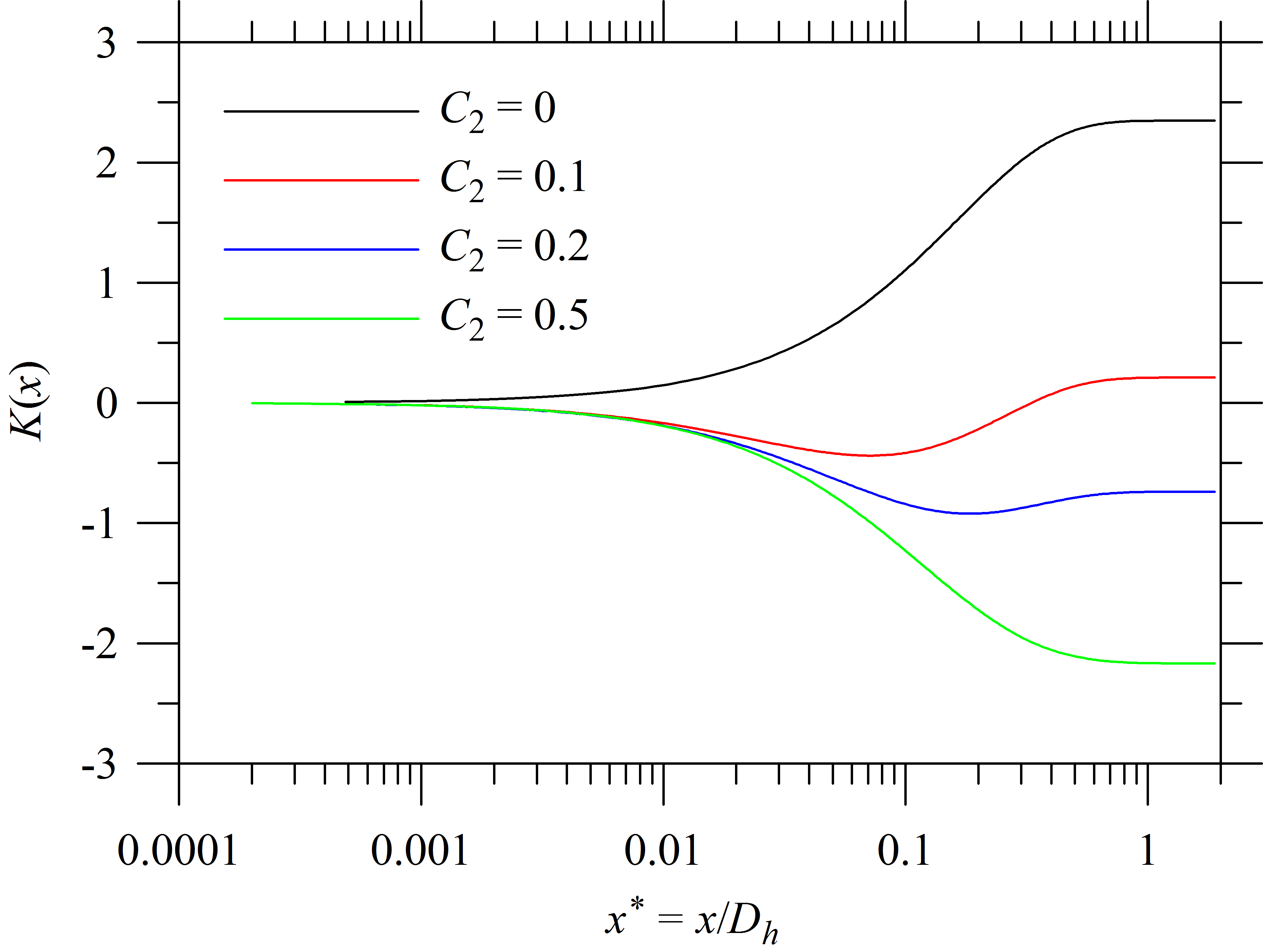}
		}\label{fig:K_Re1_Kn02}
		\subfigure[$Re = 10$]{
			\includegraphics[width=0.45\textwidth]{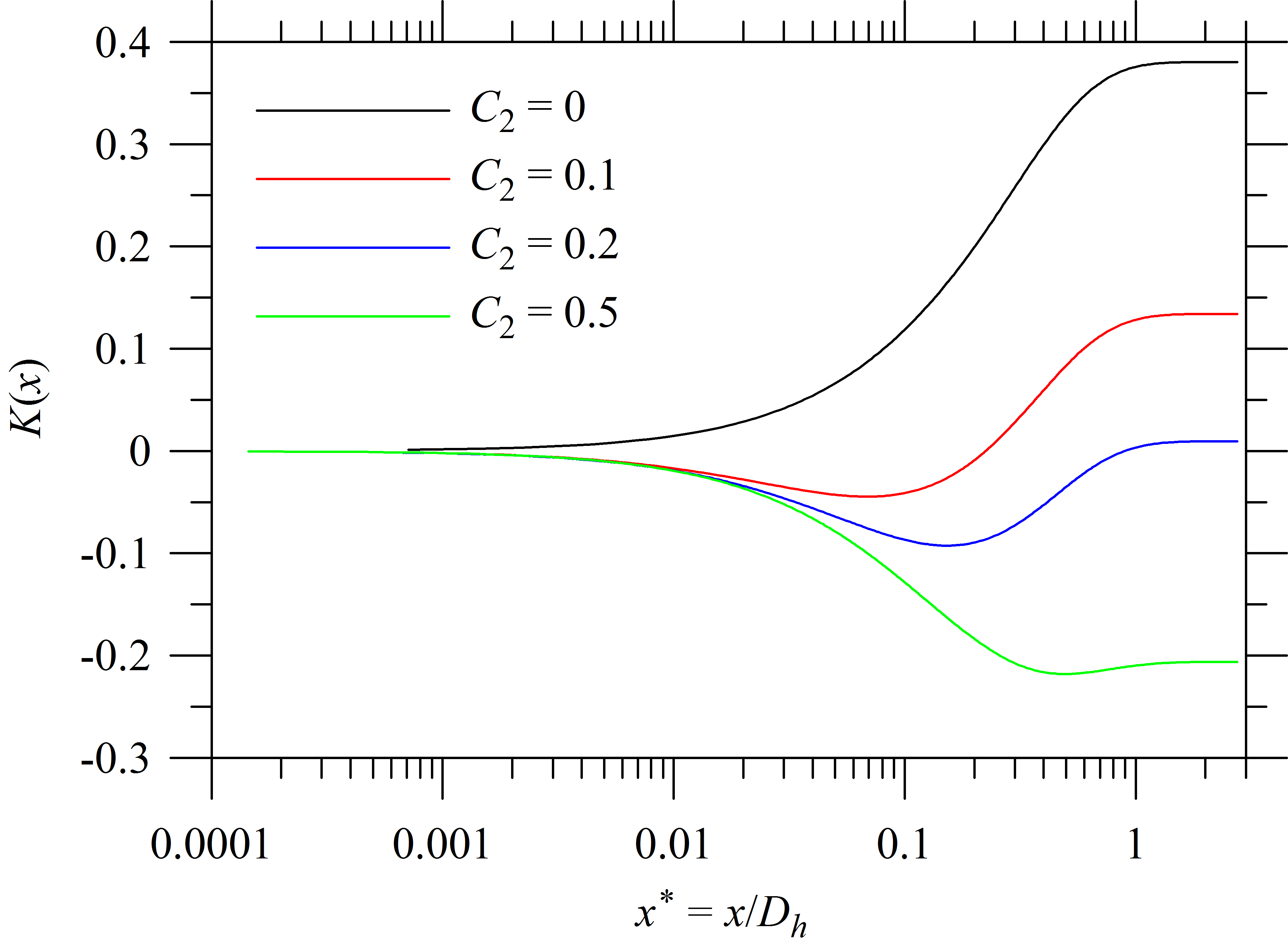}
		}\label{fig:K_Re10_Kn02}
		\subfigure[$Re = 100$]{
			\includegraphics[width=0.45\textwidth]{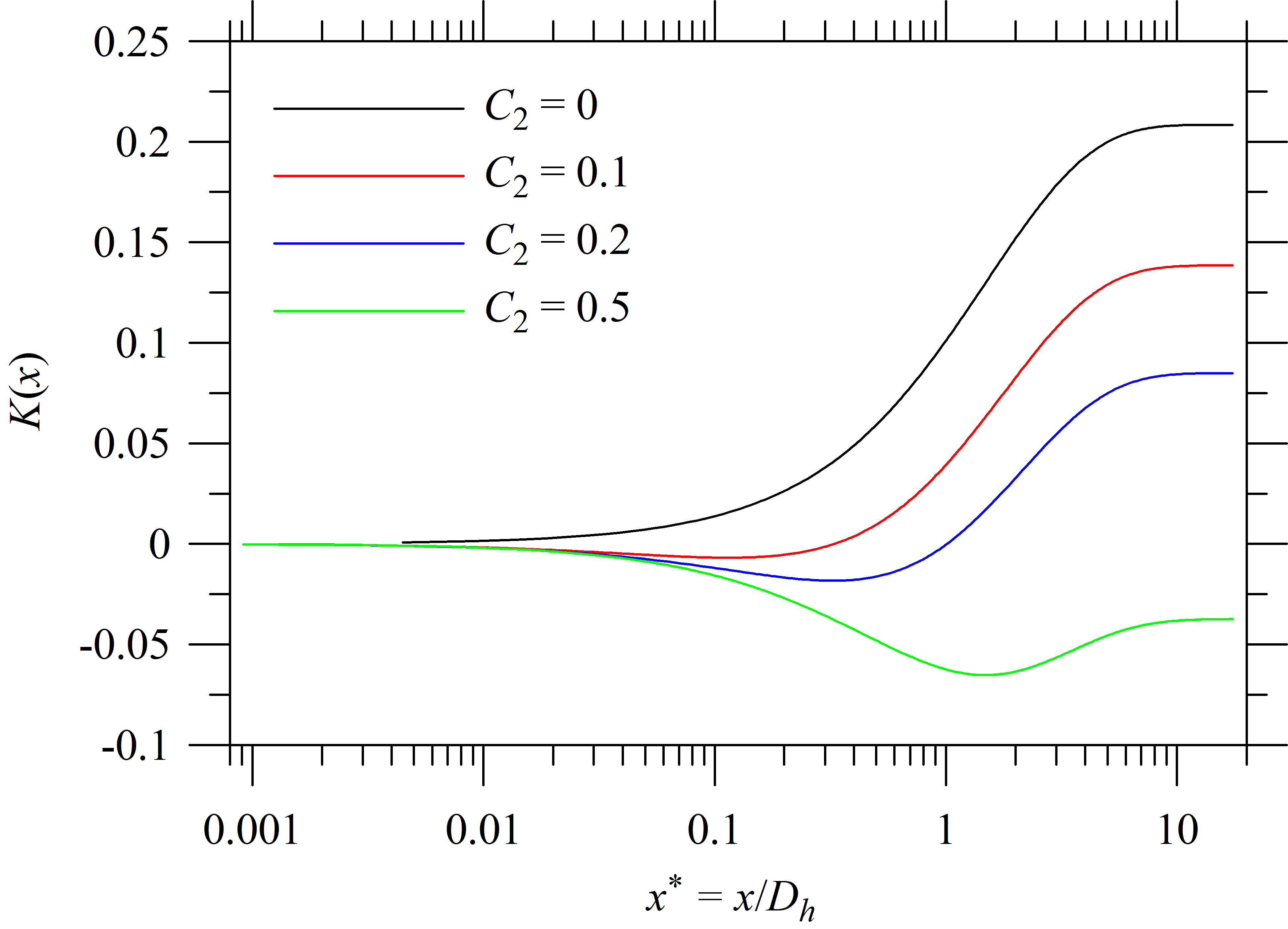}
		}\label{fig:K_Re100_Kn02}
	\end{center}
	\caption{Effects of $C_{2}$ on variations in $K(x)$ as functions of $x/D_{h}$ for pipe flows with $Kn = 0.2$.}
	\label{fig:K_Re_Kn02}
\end{figure}

Since negative values of $K(x)$ are observed only for higher $Kn$ with the second order velocity slip condition at the wall, the variations in $K(x)$ with $Kn = 0.2$ for different $Re$ and $C_{2}$ are presented in Fig.~\ref{fig:K_Re_Kn02}. The figure clearly shows that irrespective of $Re$, $K_{fd}$ is always negative for $C_{2} = 0.5$, whereas for lower values of $C_{2}$ (specifically for $C_{2} = 0$ and $0.1$ in Fig.~\ref{fig:K_Re_Kn02}), $K_{fd}$ is always positive, although $K(x)$ for $C_{2} = 0.1$ always remains negative in the entrance region. Similar variations in $K(x)$ and $K_{fd}$ for $C_{2} = 0.2$, on the other hand, show that although $K(x)$ is always negative at least in the developing region of the pipe, $K_{fd}$ remains negative for lower $Re$ (see cases with $Re = 0.1$ and $1$ in Fig.~\ref{fig:K_Re_Kn02}), while it increases with the increase in $Re$ and eventually becomes positive for higher $Re$ (see results for $Re = 10$ and $100$ in Fig.~\ref{fig:K_Re_Kn02}). \\

\begin{figure}[htbp]
	\begin{center}
		\subfigure[$Re = 0.1$]{
			\includegraphics[width=0.47\textwidth]{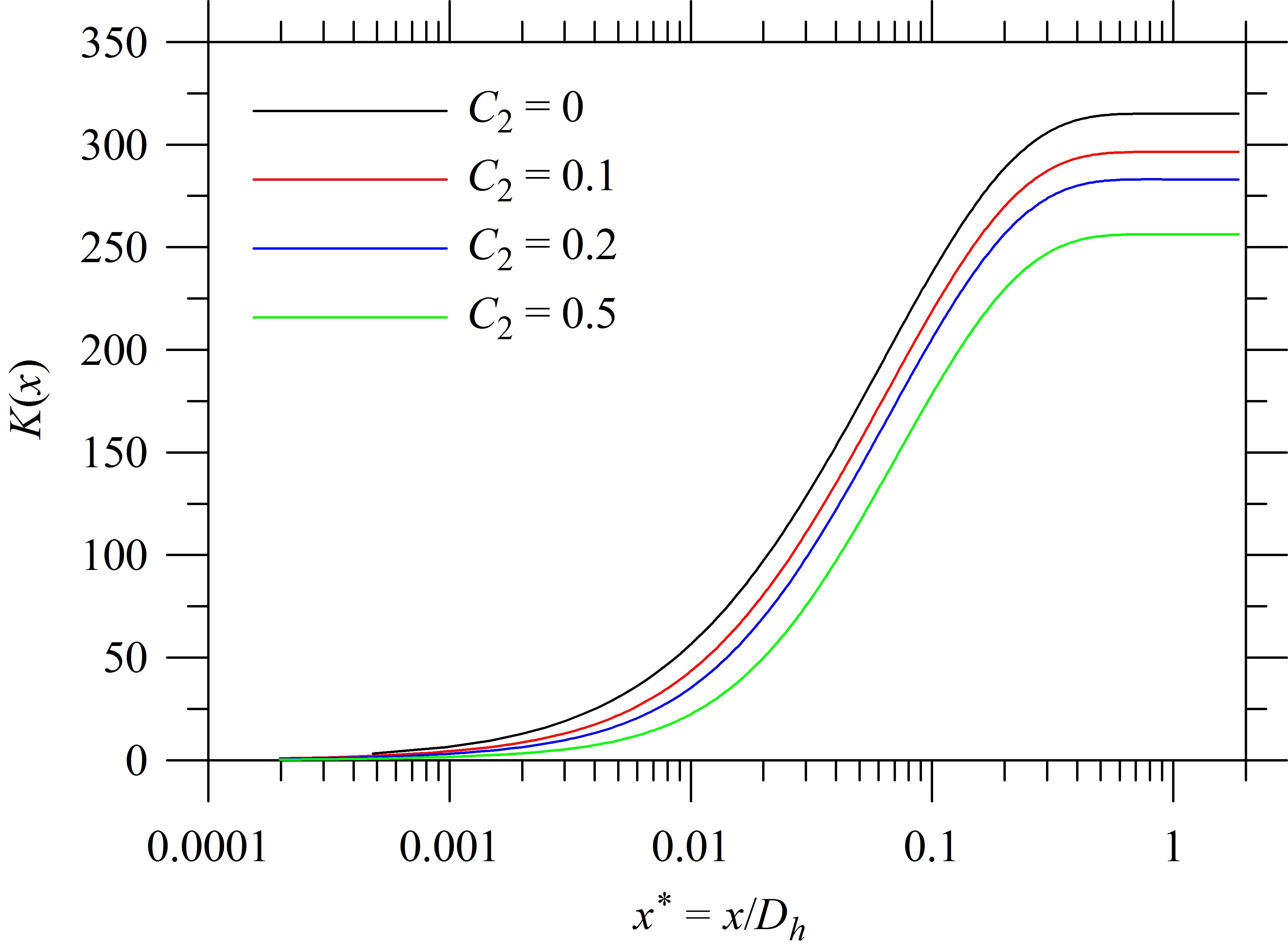}
		}\label{fig:K_Re01_Kn001}
		\subfigure[$Re = 100$]{
			\includegraphics[width=0.47\textwidth]{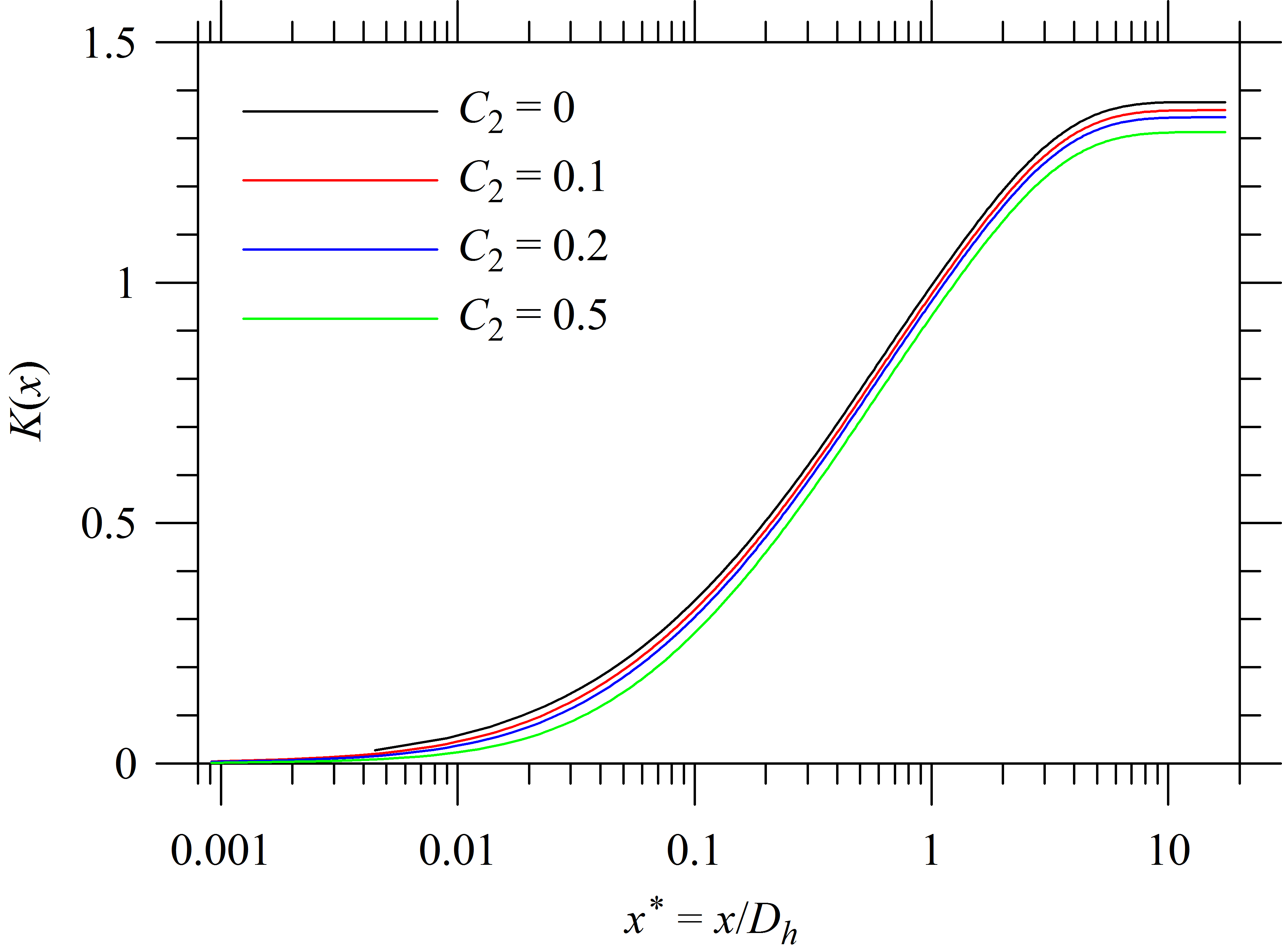}
		}\label{fig:K_Re100_Kn001}
	\end{center}
	\caption{Effects of $C_{2}$ on variations in $K(x)$ as functions of $x/D_{h}$ for pipe flows with $Kn = 0.01$.}
	\label{fig:K_Re_Kn001}
\end{figure}

Similar variations in $K(x)$ for $Kn = 0.01$, which lies in the border-line between the continuum and the slip flow regimes \citep{SchaafAndChambre_1961}, with $Re = 0.1$ and $100$ are presented in Fig.~\ref{fig:K_Re_Kn001} for different $C_{2}$. For brevity, the intermediate results for $Re = 1$ and $10$ are not presented. The comparison of results in Figs.~\ref{fig:K_Re_C2_0} -- \ref{fig:K_Re_Kn001} clearly shows that although $K(x)$ could be negative for higher $Kn$, irrespective of $Re$ and $C_{2}$, it always remains positive as $Kn$ is reduced to $0.01$. Most importantly, no negative $K(x)$, and hence $K_{fd}$, could be detected for the first order velocity slip condition at the wall, irrespective of $Re$ and $Kn$. \\

Another apparently inconsequential general observation from Figs.~\ref{fig:K_Re_C2_0} -- \ref{fig:K_Re_Kn001}, which has not been mentioned so far, is that for a given combination of $Kn$ and $C_{2}$, both $K(x)$ for a particular $x^{*}$ and $K_{fd}$ decrease substantially with the increase in $Re$. Such variations are, however, expected since according to Eq.~(\ref{Eq:Kx}), $K(x)$ is defined as the difference between the true and the expected fully-developed pressure drops that is normalised with respect to $ \rho u_{av}^{2} /2$, where by definition, for a given $D_{h}$ and the working fluid, $u_{av}$ is a linearly increasing function of $Re$. Therefore, any reduction in $K(x)$ or $K_{fd}$ does not necessarily imply a corresponding decrease in $\Delta p_{x} - \Delta p_{fd}$ and as such, $\Delta p_{x}$ should always be calculated according to Eq.~(\ref{Eq:Deltapx_From_Kx}). \\

The aforementioned observations may be summarised as follows. The generalised slip boundary condition allows a positive tangential velocity component at the wall that, other than the operating conditions, depends on the axial location. As a consequence, with the increase in both $Kn$ and $C_2$, the normal gradient of axial velocity at the wall decreases for a given combination of $Re$ and $x^{*}$. Therefore, the corresponding wall shear stress $\tau_{w,x}$ and hence as a consequence of Eq.~(\ref{Eq:fxRe}), the local friction factor $f_x$ are reduced. Finally, certain operating conditions can produce sufficient velocity slip at the wall that, owing to the complex interaction of three terms in Eq.~(\ref{Eq:Kx}), may eventually lead to not only negative $K(x)$ in the entrance region, but also a negative $K_{fd}$. \\

\begin{figure}[htbp]
	\begin{center}
		\subfigure[$Re = 0.1$]{
			\includegraphics[width=0.45\textwidth]{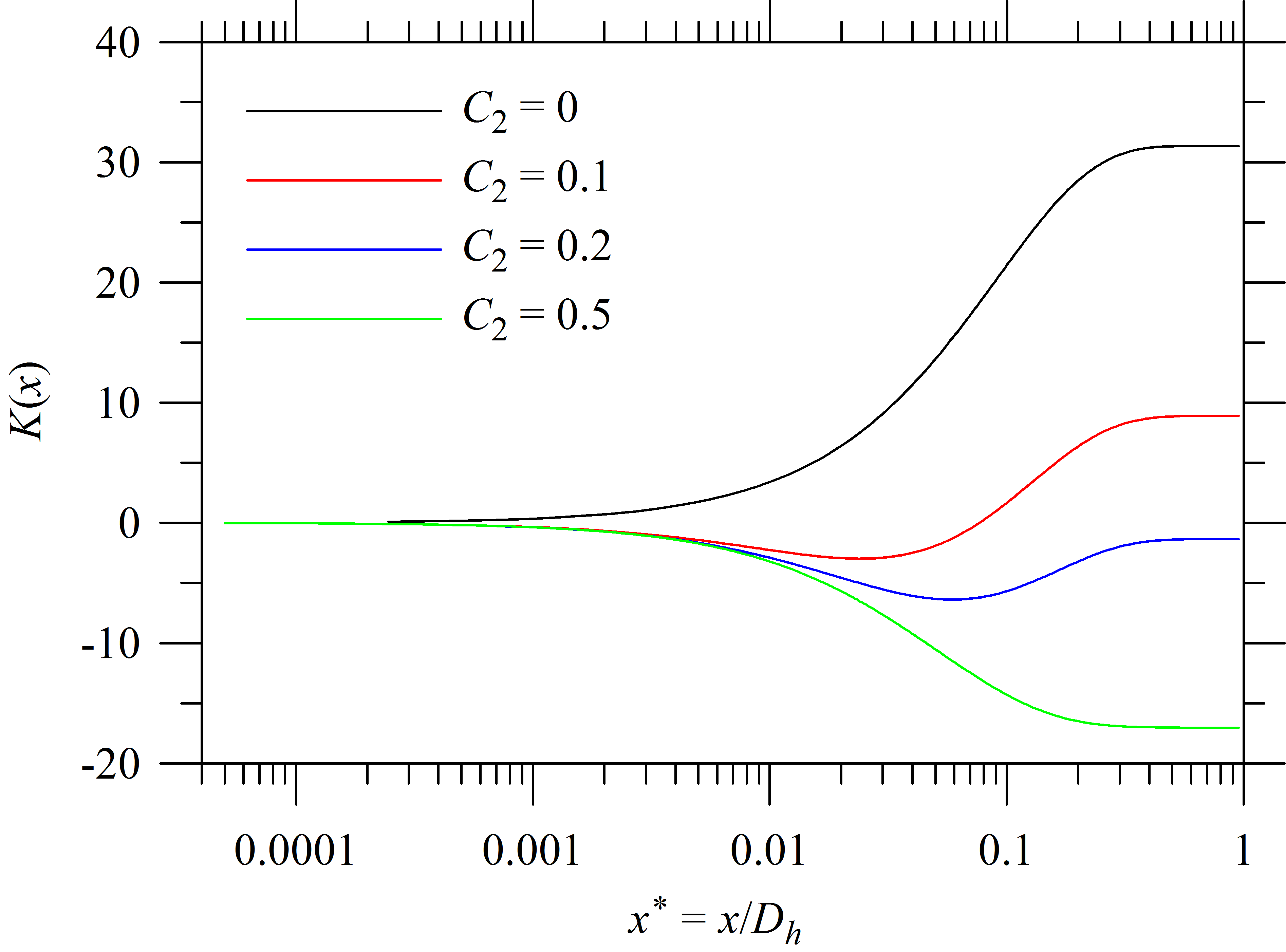}
		}\label{fig:K_Re01_Kn02_PP}
		\subfigure[$Re = 1$]{
			\includegraphics[width=0.45\textwidth]{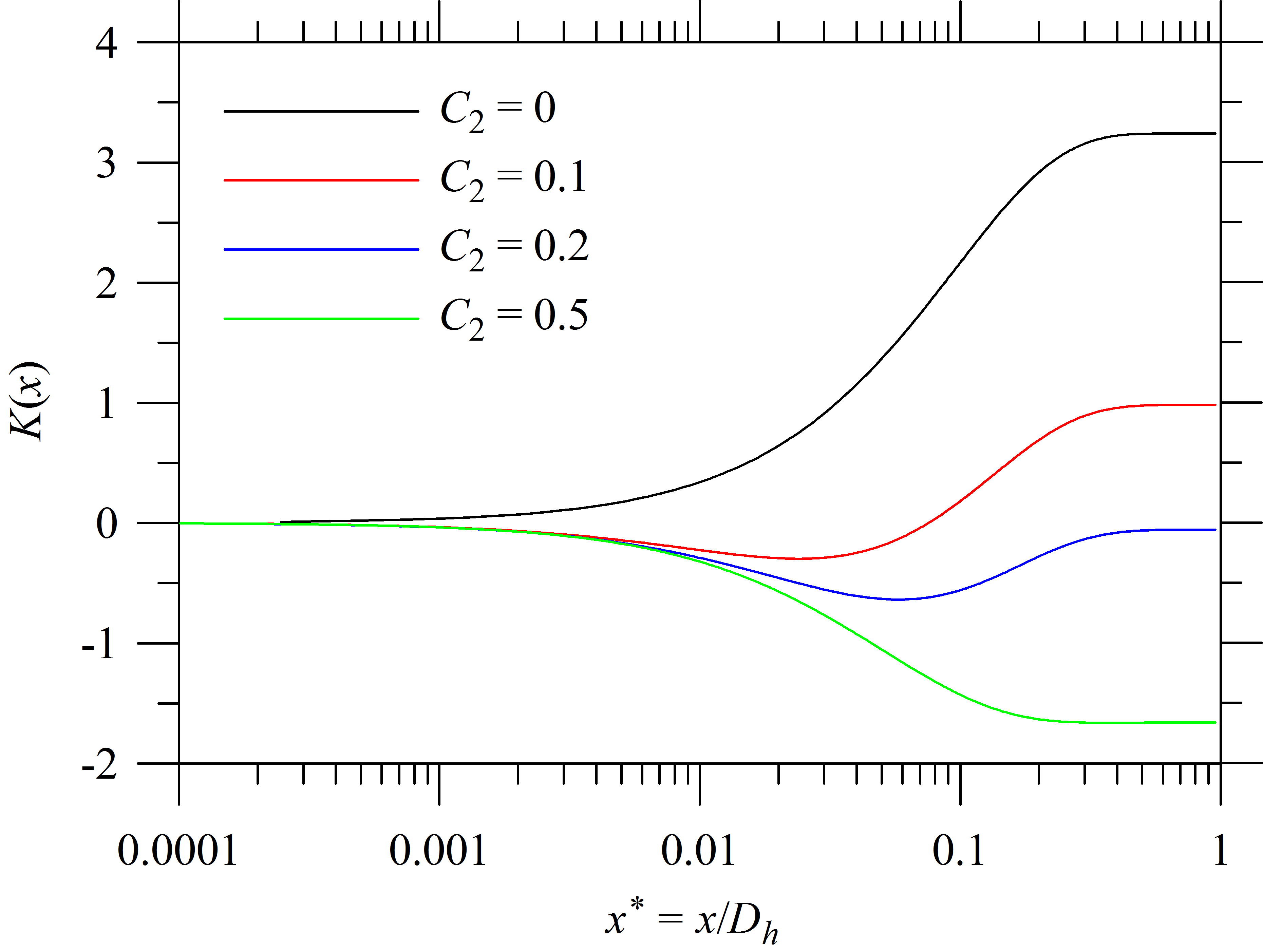}
		}\label{fig:K_Re1_Kn02_PP}
		\subfigure[$Re = 10$]{
			\includegraphics[width=0.45\textwidth]{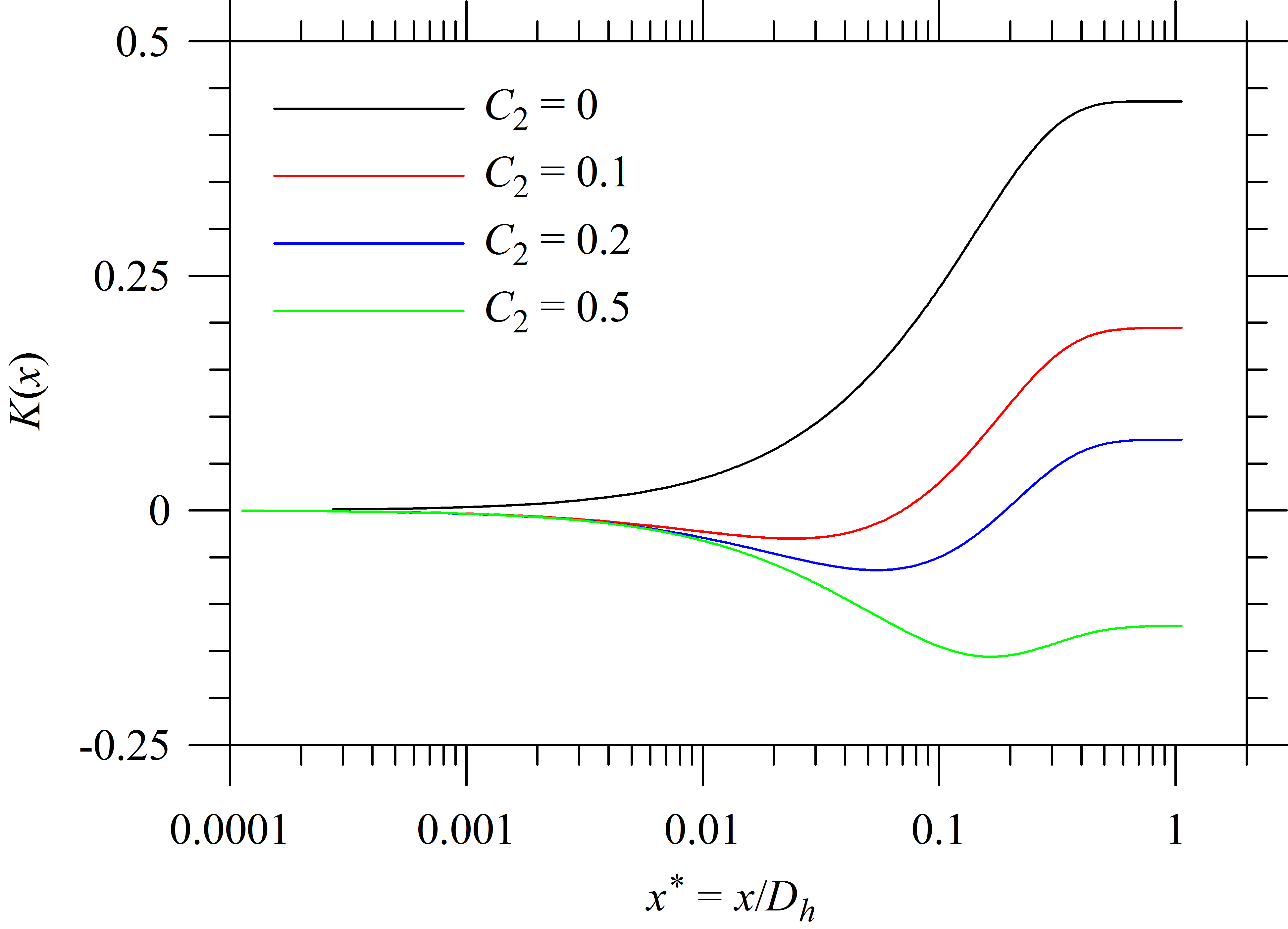}
		}\label{fig:K_Re10_Kn02_PP}
		\subfigure[$Re = 100$]{
			\includegraphics[width=0.45\textwidth]{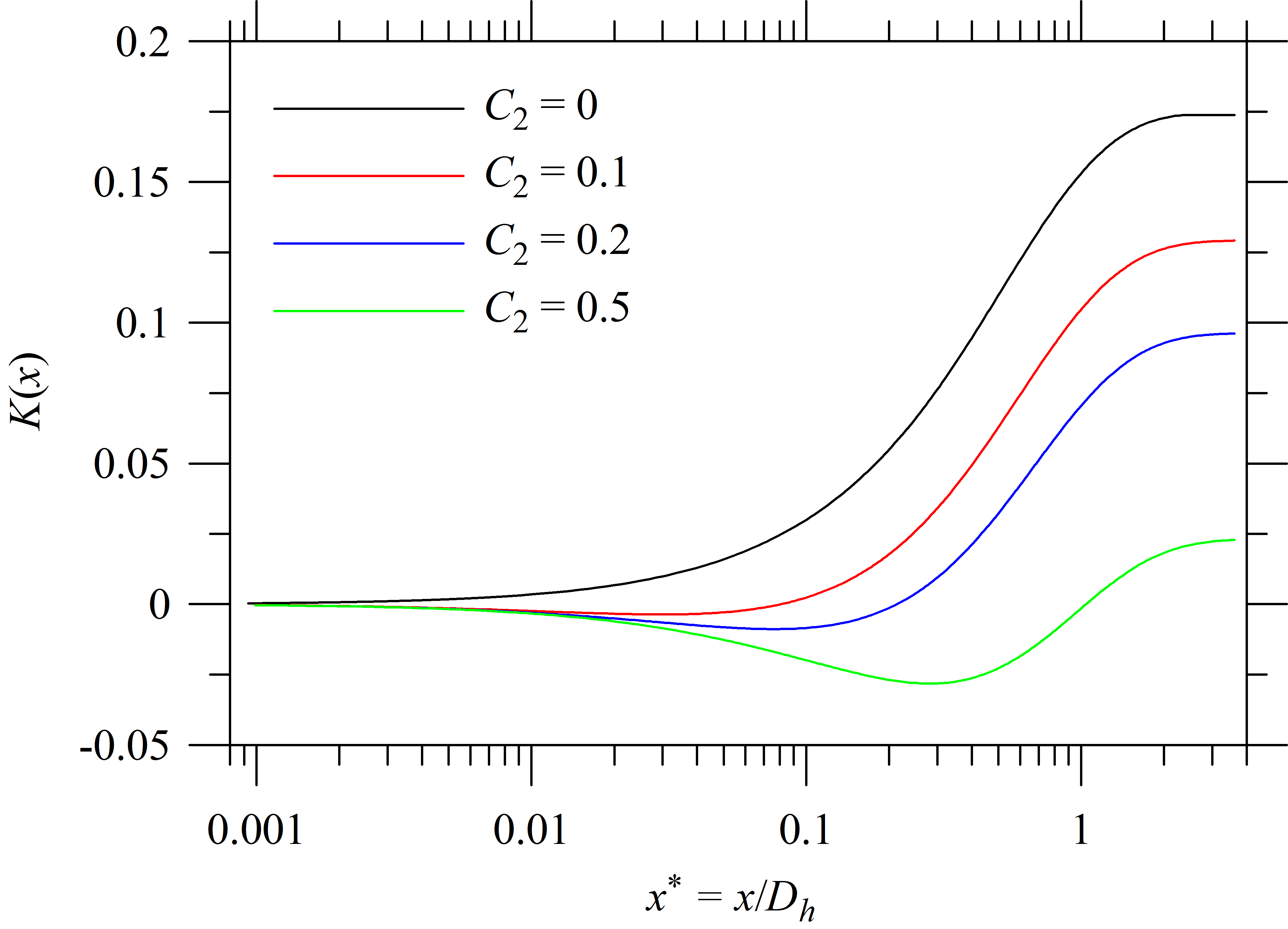}
		}\label{fig:K_Re100_Kn02_PP}
	\end{center}
	\caption{Effects of $C_{2}$ on variations in $K(x)$ as functions of $x/D_{h}$ for channel flows with $Kn = 0.2$.}
	\label{fig:K_Re_Kn02_PP}
\end{figure}

In order to demonstrate that the pressure drop characteristics for channel flows qualitatively remain nearly the same as that for pipe flows, the variations in $K(x)$ as functions of $x^{*}$ are presented in Fig.~\ref{fig:K_Re_Kn02_PP} with $Kn = 0.2$ for different $Re$ and $C_{2}$. Comparison of results in Figs.~\ref{fig:K_Re_Kn02} and \ref{fig:K_Re_Kn02_PP} adequately substantiates the claim and for brevity, no further variations in $K(x)$ for channel flows is presented in this article. Nevertheless, it is also obvious from Figs.~\ref{fig:K_Re_C2_0} -- \ref{fig:K_Re_Kn02_PP} that for both pipe and channel flows, the variations in $K(x)$ are quite complex functions of $x^{*}$ and their nature strongly depends on the operating condition. Therefore, at this point, no general functional form could be proposed that would adequately describe the observed variations in $K(x)$ and hence this issue has been left beyond the scope of the present investigation. \\

\begin{figure}[htbp]
	\begin{center}
		\subfigure[$Kn = 0.01$]{
			\includegraphics[width=0.45\textwidth]{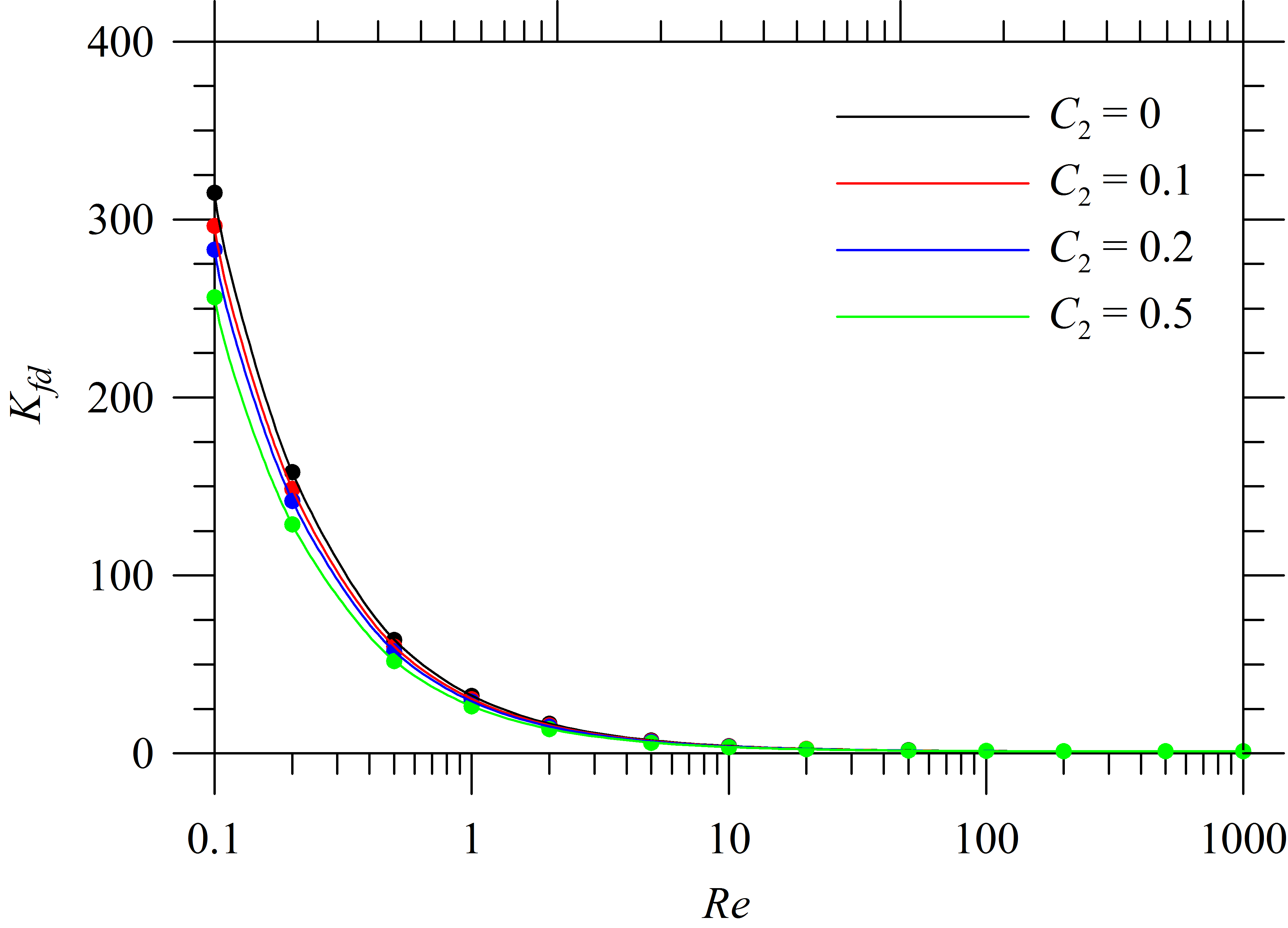}
		}\label{fig:KfdVsReKn001_Pipe}
		\subfigure[$Kn = 0.05$]{
			\includegraphics[width=0.45\textwidth]{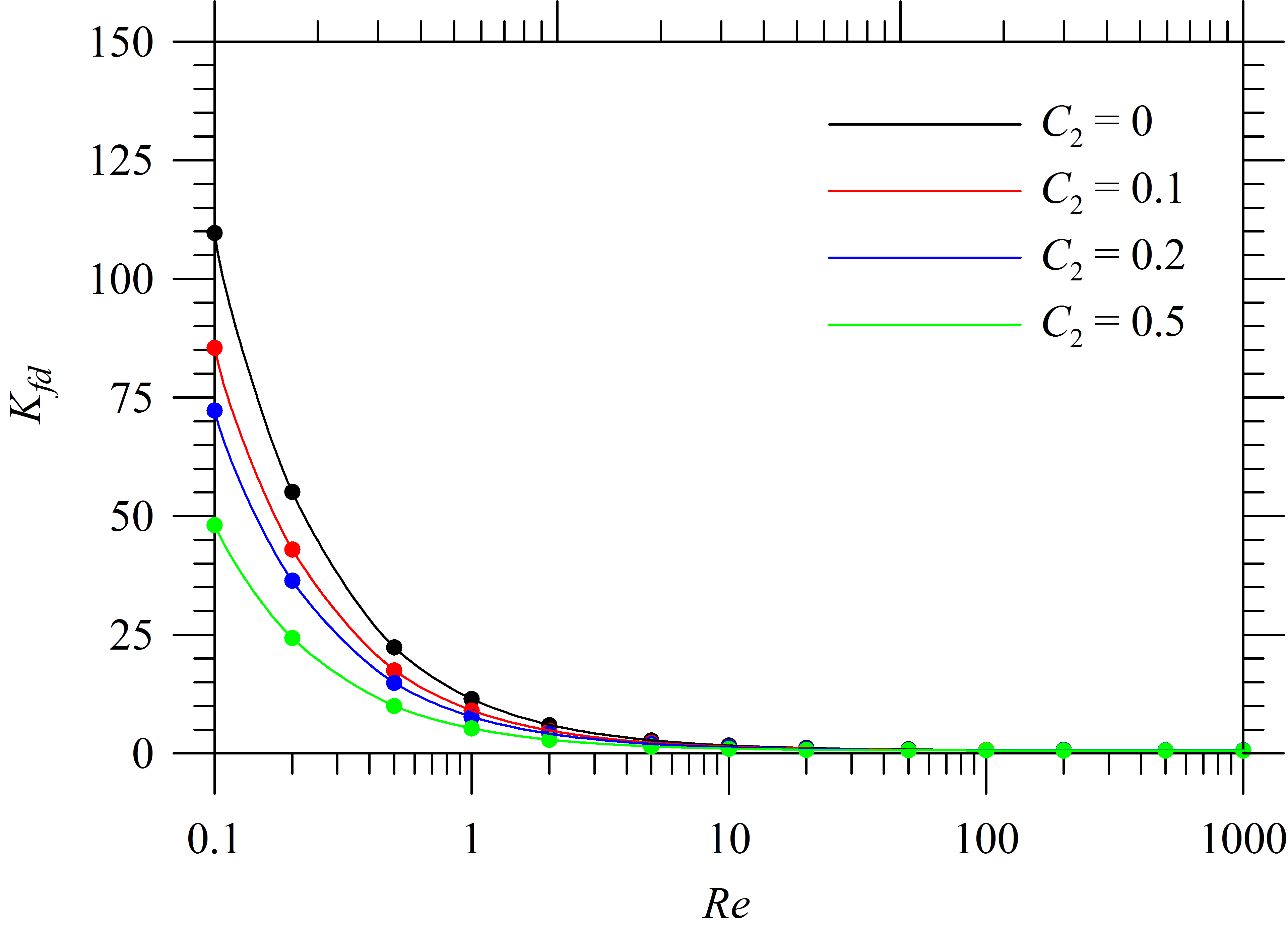}
		}\label{fig:KfdVsReKn005_Pipe}
		\subfigure[$Kn = 0.1$]{
			\includegraphics[width=0.45\textwidth]{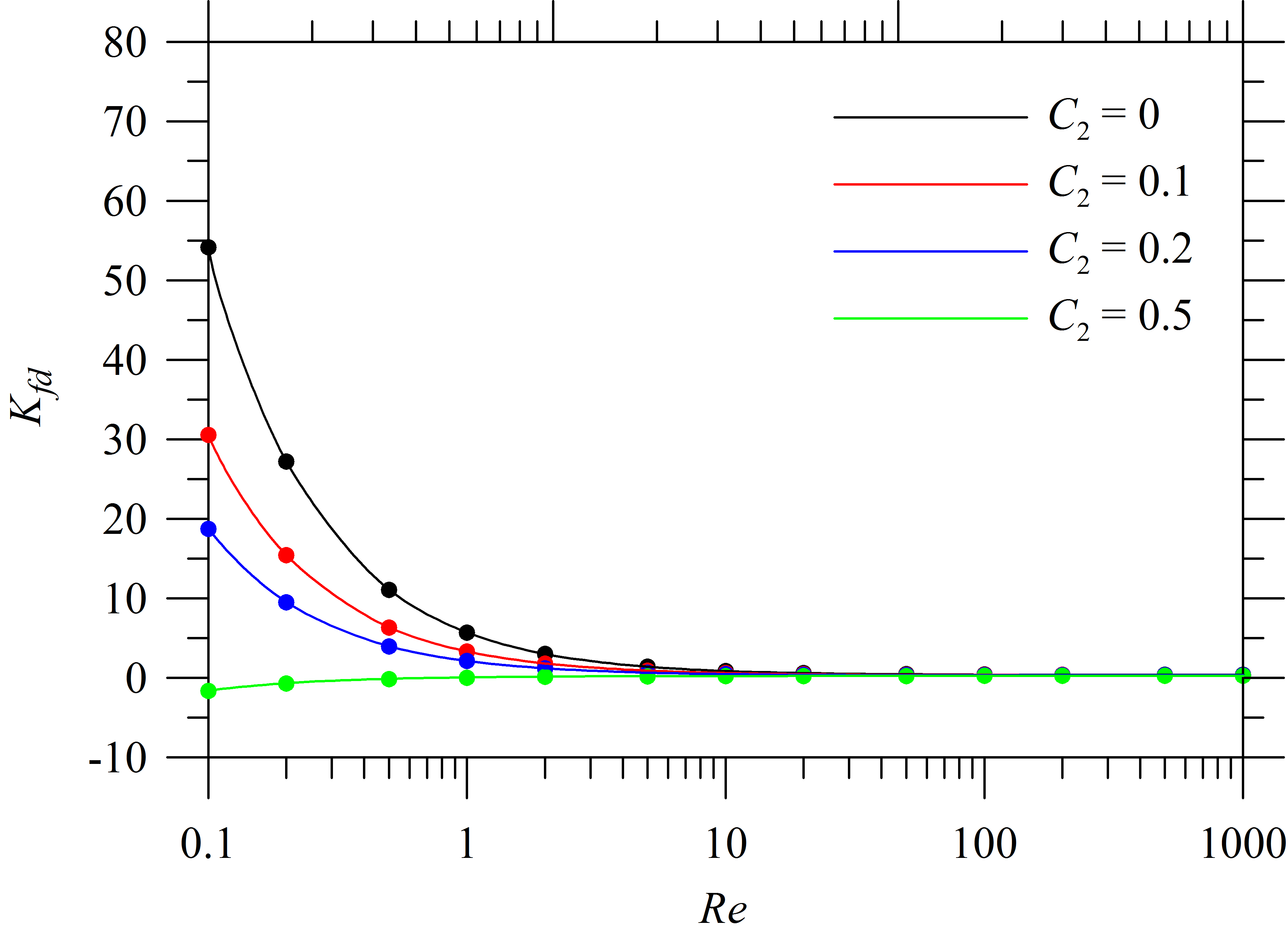}
		}\label{fig:KfdVsReKn01_Pipe}
		\subfigure[$Kn = 0.2$]{
			\includegraphics[width=0.45\textwidth]{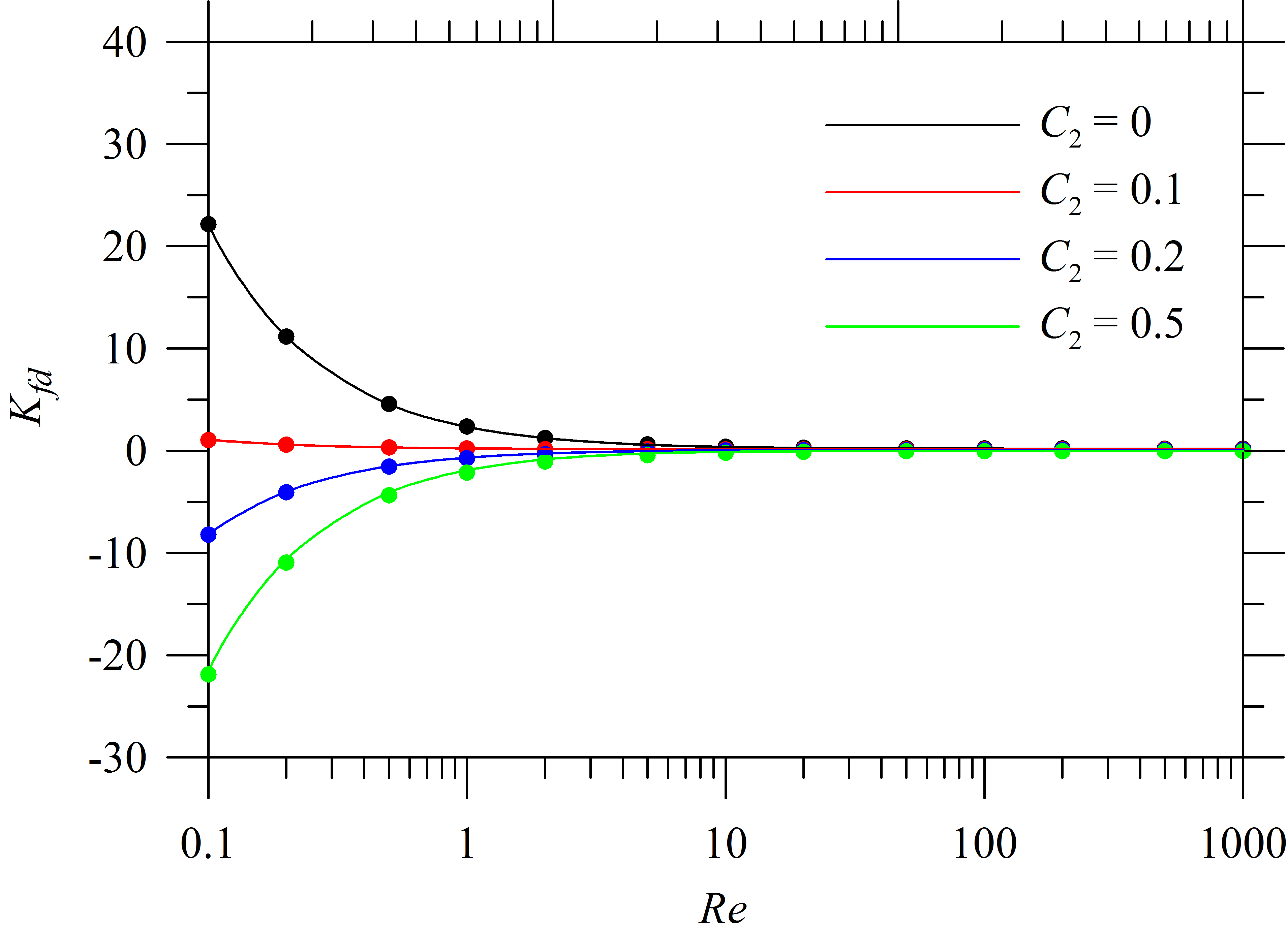}
		}\label{fig:KfdVsReKn02_Pipe}
	\end{center}
	\caption{Effects of $C_{2}$ on variations in $K_{fd}$ as functions of $Re$ for Pipe flows with different $Kn$. The symbols represent the simulated data and the lines are obtained according to Eq.~(\ref{Eq:Corr_Kfd_general}).}
	\label{fig:KfdVsRe_Pipe}
\end{figure}

\begin{figure}[htbp]
	\begin{center}
		\subfigure[$Kn = 0.01$]{
			\includegraphics[width=0.45\textwidth]{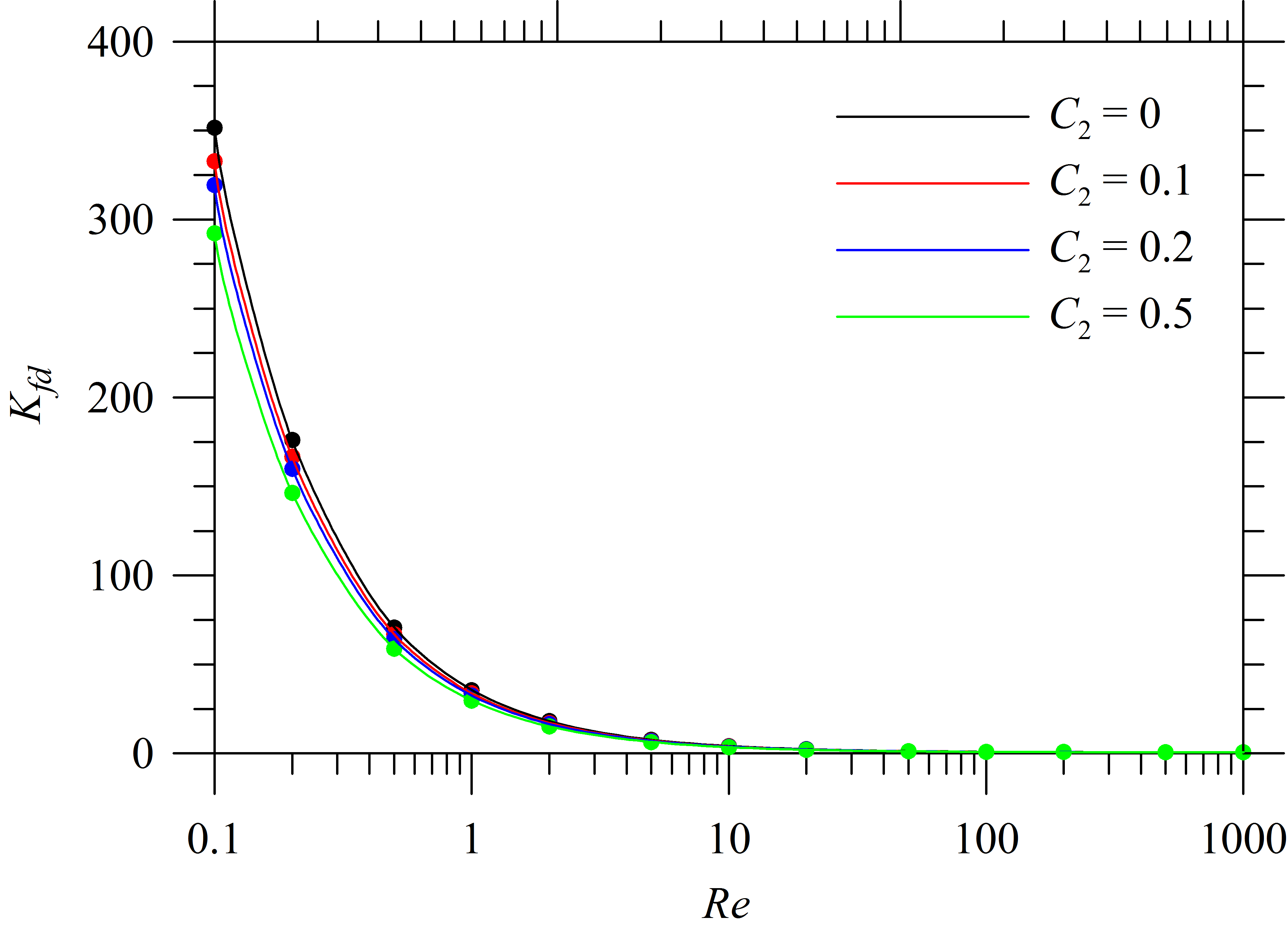}
		}\label{fig:KfdVsReKn001_Channel}
		\subfigure[$Kn = 0.05$]{
			\includegraphics[width=0.45\textwidth]{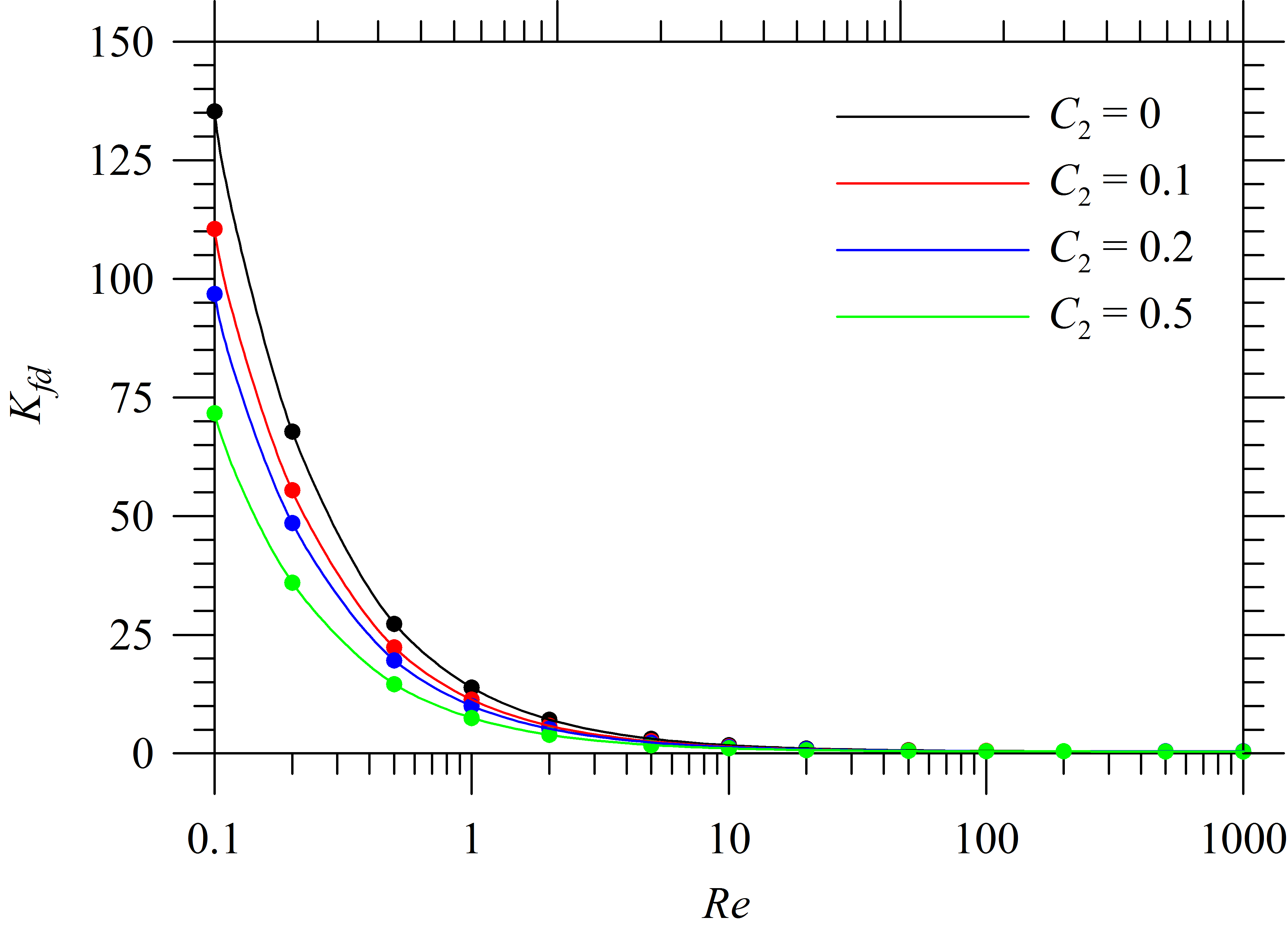}
		}\label{fig:KfdVsReKn005_Channel}
		\subfigure[$Kn = 0.1$]{
			\includegraphics[width=0.45\textwidth]{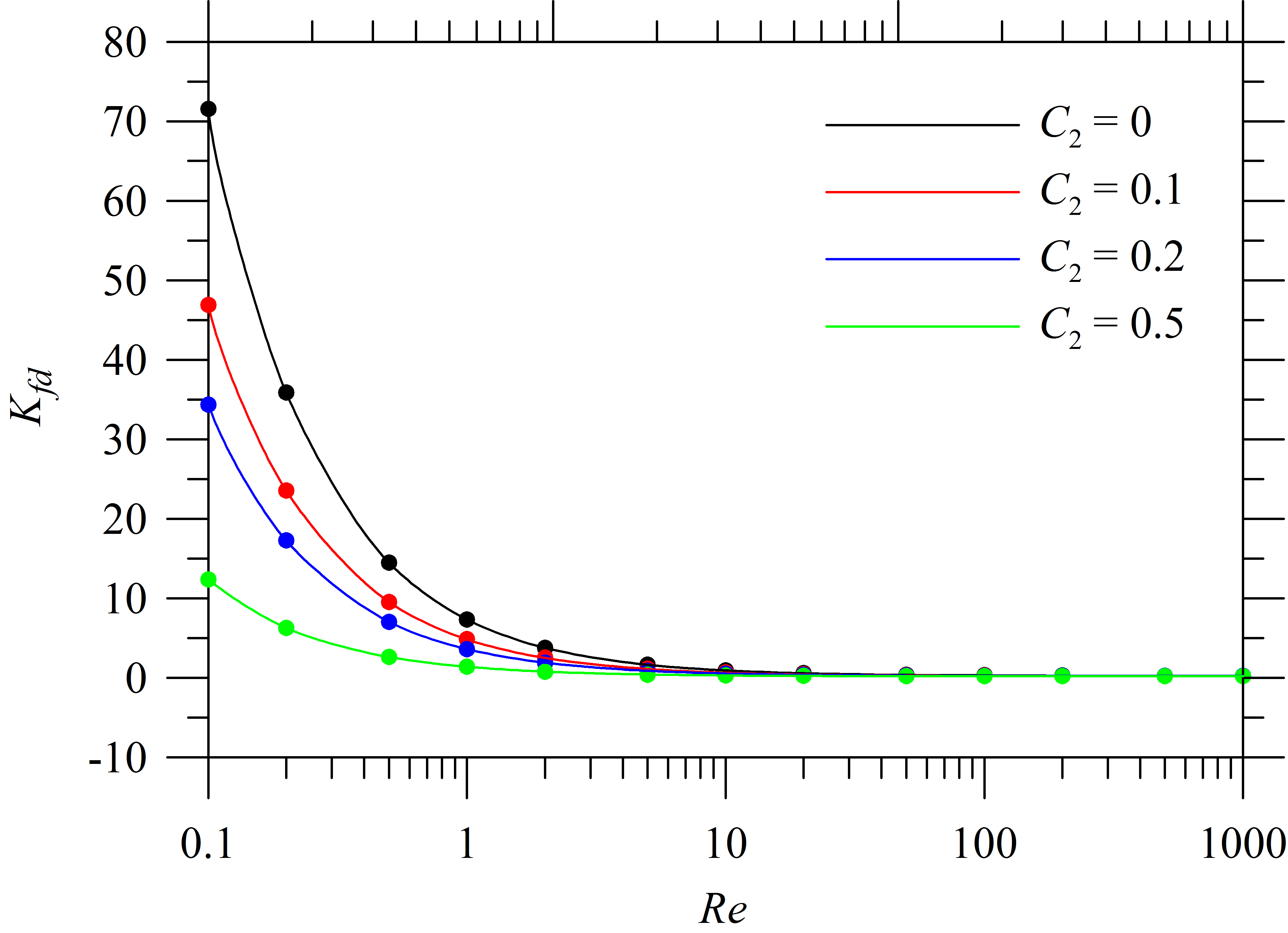}
		}\label{fig:KfdVsReKn01_Channel}
		\subfigure[$Kn = 0.2$]{
			\includegraphics[width=0.45\textwidth]{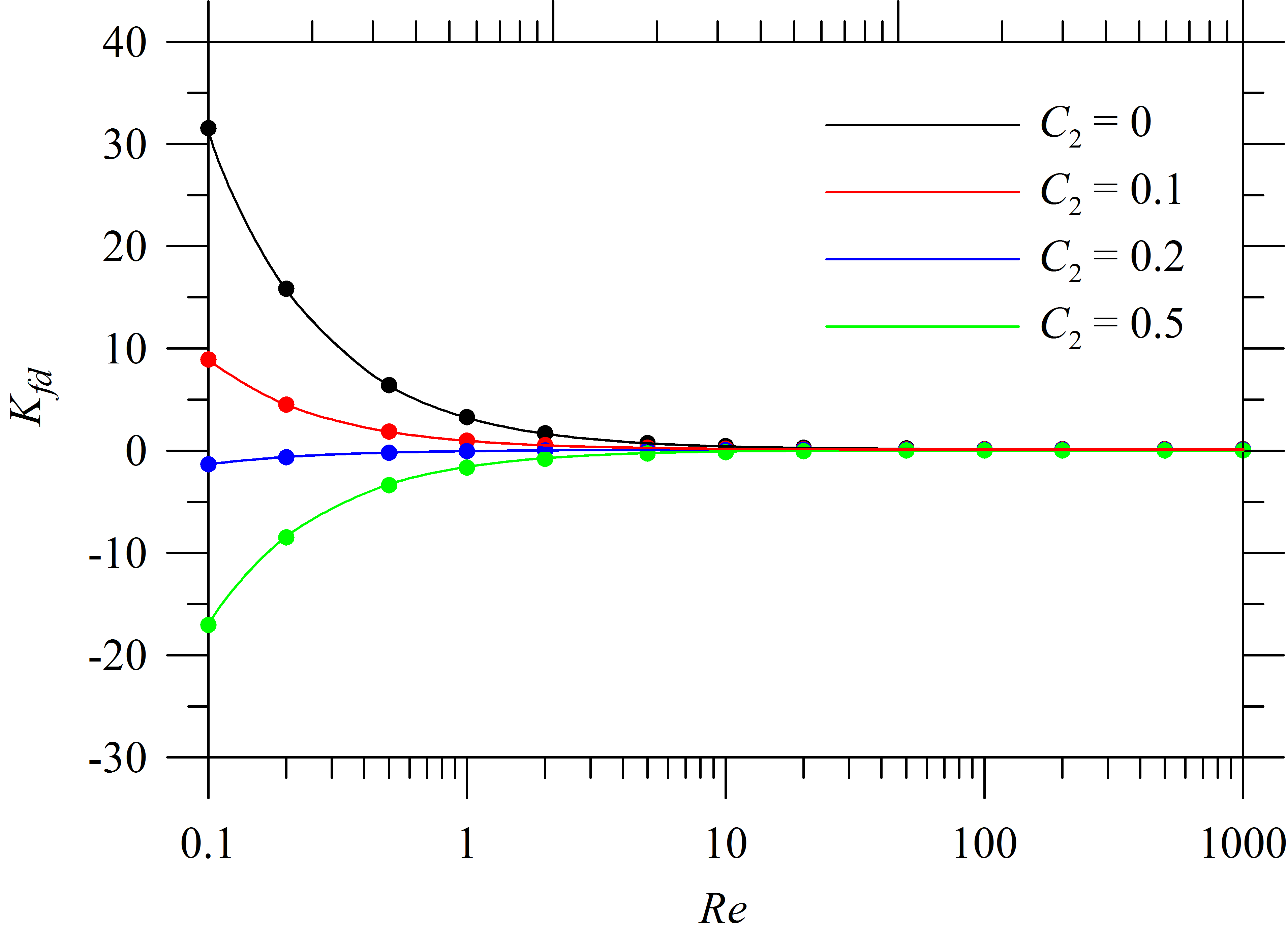}
		}\label{fig:KfdVsReKn02_Channel}
	\end{center}
	\caption{Effects of $C_{2}$ on variations in $K_{fd}$ as functions of $Re$ for Channel flows with different $Kn$. The symbols represent the simulated data and the lines are obtained according to Eq.~(\ref{Eq:Corr_Kfd_general}).}
	\label{fig:KfdVsRe_Channel}
\end{figure}

As mentioned earlier in section~\ref{mathform}, for a duct length $L \ge L_{fd}$, the actual pressure drop $\Delta p_{L}$ may be evaluated according to Eq.~(\ref{Eq:Deltapx_From_Kx}) by substituting $K(L) = K_{fd}$ and $x^{*} = L^{*}$. Therefore, similar to $L^{*}_{fd}$, $K_{fd}$ is also considered as another important characteristic of the developing flow through ducts that takes into account the pressure drop in the entrance region. The variations in $K_{fd}$ as functions of $Re$ for different $Kn$ and $C_{2}$ are presented in Figs.~\ref{fig:KfdVsRe_Pipe} and \ref{fig:KfdVsRe_Channel} for pipe and channel flows, respectively. It is evident from the figures that the magnitude of $K_{fd}$ decreases considerably with the increase in $Re$, $kn$ and $C_{2}$, although in the high $Re$ regime, $K_{fd} \rightarrow K_{1}$ and appears to be a weak function of $Kn$ and $C_{2}$. In addition, for lower $Kn$, $K_{fd}$ is less sensitive to the change in $C_{2}$ and the sensitivity increases with the increase in $Kn$. Since these observations may be explained in view of the foregoing discussions, they are not repeated here for brevity. \\

Figures~\ref{fig:KfdVsRe_Pipe} and \ref{fig:KfdVsRe_Channel} also clearly show that for lower $Kn$, irrespective of $Re$ and $C_{2}$, $K_{fd}$ is always positive, whereas for $Kn = 0.2$, depending upon $C_{2}$ and $Re$, $K_{fd}$ could assume either positive or negative values. However, as mentioned earlier, for the first order velocity slip condition at the wall ($C_{2} = 0$), or, even for lower $C_{2}$, $K_{fd}$ remains always positive for both pipe and channel flows, irrespective of $Re$ and $Kn$. \\

Nevertheless, irrespective of $Kn$ and $C_{2}$, the variations in $K_{fd}$ as functions of $Re$ are still described by two asymptotes, where the low and the high $Re$ asymptotes are given by $K_{0}/Re$ and $K_{1}$, respectively. Similar to $L_{0}$ and $L_{1}$, $K_{0}$ and $K_{1}$ have been obtained directly from the simulated data for $K_{fd} Re$ at $Re = 10^{-2}$ and $K_{fd}$ at $Re = 10^{4}$, respectively. The obtained data for different $C_{2}$ and $Kn$ are presented in Table~\ref{tab:K0K1_PipeCHannel}, from which it is evident that $K_{0}$ could assume either positive or negative values for both pipe and channel flows. On the other hand, $K_{1}$ for pipe flows is marginally negative only for higher $C_{2}$ and $Kn$, while it remains always positive for channel flows for the investigated ranges of $Kn$ and $C_{2}$. As a result, $K_{fd}$ in Figs.~\ref{fig:KfdVsRe_Pipe} and \ref{fig:KfdVsRe_Channel} could not be consistently presented using the logarithmic scale and in order to highlight that $K_{fd}$ assumes considerably smaller but non-zero values as $Re \rightarrow \infty$ on a linear scale, the variations of $K_{fd}$ in these figures are presented only for the range $1 \le Re \le 10^{3}$.\\

Depending upon $Kn$ and $C_{2}$, since $K_{0}$ and $K_{1}$ assume both positive and negative values, $K_{fd}$ could not be represented by the form similar to that for $L^{*}_{fd}$ in Eq.~(\ref{eq:corr_general}). Therefore, an alternative relation, that respects both low and high $Re$ asymptotes, has been adopted in order to correlate $K_{fd}$:
\begin{equation}
K_{fd} = \frac  {K_{0} / Re} {1 + K_{2}~Re} + K_{1}
\label{Eq:Corr_Kfd_general} 
\end{equation}
where $K_{2}$ is a function of $Kn$ and $C_{2}$. At this point, however, it is important to note that as $x \rightarrow 0$, the wall shear stress $\tau_{w,x} \rightarrow \infty$ and it has been found to be proportional to $x^{-m}$. Therefore, $f_{x}$ in this region is proportional to ${\left( x^{*} \right)}^{-m}$, according to the definitions of dimensionless variables. Similar consequence, although only in the convection dominated regime, could also be derived from the conventional boundary layer theory. Extensive numerical simulations show that for $Kn = 0$, $m$ varies from $1$ for $Re \rightarrow 0$, where the convective effects are insignificant, to $1/2$ as $Re \rightarrow \infty$, where the axial diffusion is negligible. Therefore, for $Kn = 0$, the first term on the right hand side of Eq.~(\ref{Eq:Deltapx}) $\Delta p^{*}_{f,x} \rightarrow \infty$ as $Re \rightarrow 0$ and hence no grid-independent solution either for $K (x)$ or $K_{fd}$ could be obtained in this limit, although $L^{*}_{fd}$ has been found to attain a grid-independent value \citep{Durstetal2005}. Further critical appreciation of this issue, arising out of the singularity in boundary conditions at the inlet, however, has been left beyond the scope of the present article. \\

Nevertheless, with the increase in $Kn$, substantial velocity slip occurs at the wall for a given $C_{2}$ that not only reduces the wall shear stress and hence $f_{x}$ close to the inlet, but also it modifies the exponent $m$ to a great extent, from unity to reasonable fractional values, such that grid independent solutions for $K (x)$ and hence $K_{fd}$ could be obtained even in the limit as $Re \rightarrow 0$. Therefore, correlations for $K_{fd}$ have been obtained only for $10^{-3} \le Kn \le 0.2$, while accommodating the complete range of investigated $Re$. \\

In order to develop these correlations, however, $K_{0}$ and $K_{1}$ in Table~\ref{tab:K0K1_PipeCHannel} have been directly used, since owing to their large variations with $kn$ and changing dependence on both $Kn$ and $C_{2}$, no reliable correlation for them could be derived. Subsequently, $K_{2}$ has been determined for each combination of $Kn$ and $C_{2}$, either by minimising the maximum absolute relative error in $K_{fd}$ when $\left| K_{fd} \right| >> 0$ for the entire range of $Re$, or using the least square method (i.e., by maximising the coefficient of determination $R^{2}$), when $K_{fd} \approx 0$ is expected for certain $Re$. The latter condition could be easily identified by checking i) if $\left| K_{1} \right| \leq \epsilon$, where $\epsilon$ has been taken as $0.05$, or ii) if the product of $K_{0}$ and $K_{1}$ is negative. Finally, for a given $C_{2}$, $K_{2}$ as a function of $Kn$ has been expressed as:
\begin{equation}
K_{2} = \sum_{j=0}^{2}~k_{2j}\ Kn^{j} = k_{20} + k_{21}\ Kn + k_{22}\ Kn^{2} 
\label{Eq:Corr_K2} 
\end{equation}
where $k_{2j}$ are functions of $C_{2}$. For pipe flows, they have been correlated as:
\begin{subequations}
\begin{eqnarray}
k_{20} & = & 1.778\times 10^{-4} - 3.5265\times 10^{-4}~C_{2} + 2.7782\times 10^{-4}~C^{2}_{2} \label{Eq:k20_Pipe} \\
k_{21} & = & 1.4482\times 10^{-1} + 1.9054\times 10^{-1}~C_{2} - 1.4346\times 10^{-1}~C^{2}_{2} \label{Eq:k21_Pipe} \\
k_{22} & = & 2.9901~C_{2} + 6.2006~C^{2}_{2} \label{Eq:k22_Pipe} 
\end{eqnarray}
\label{Eq:k2j_Pipe}
\end{subequations}
Similarly for channel flows, $k_{2j}$ could be expressed as: 
\begin{subequations}
\begin{eqnarray}
k_{00} & = & 1.3194\times 10^{-4} - 1.767\times 10^{-4}~C_{2} + 1.0306\times 10^{-4}~C^{2}_{2} \label{Eq:k20_Channel} \\
k_{01} & = & 5.3809\times 10^{-2} + 6.1618\times 10^{-2}~C_{2} - 2.412\times 10^{-2}~C^{2}_{2} \label{Eq:k21_Channel} \\
k_{02} & = & 1.6632\times 10^{-1} + 1.4324~C_{2} + 1.0509~C^{2}_{2} \label{Eq:k22_Channel}
\end{eqnarray}
\label{Eq:k2j_Channel}
\end{subequations}

\begin{figure}[htbp]
	\begin{center}
		\subfigure[Pipe]{
			\includegraphics[width=0.47\textwidth]{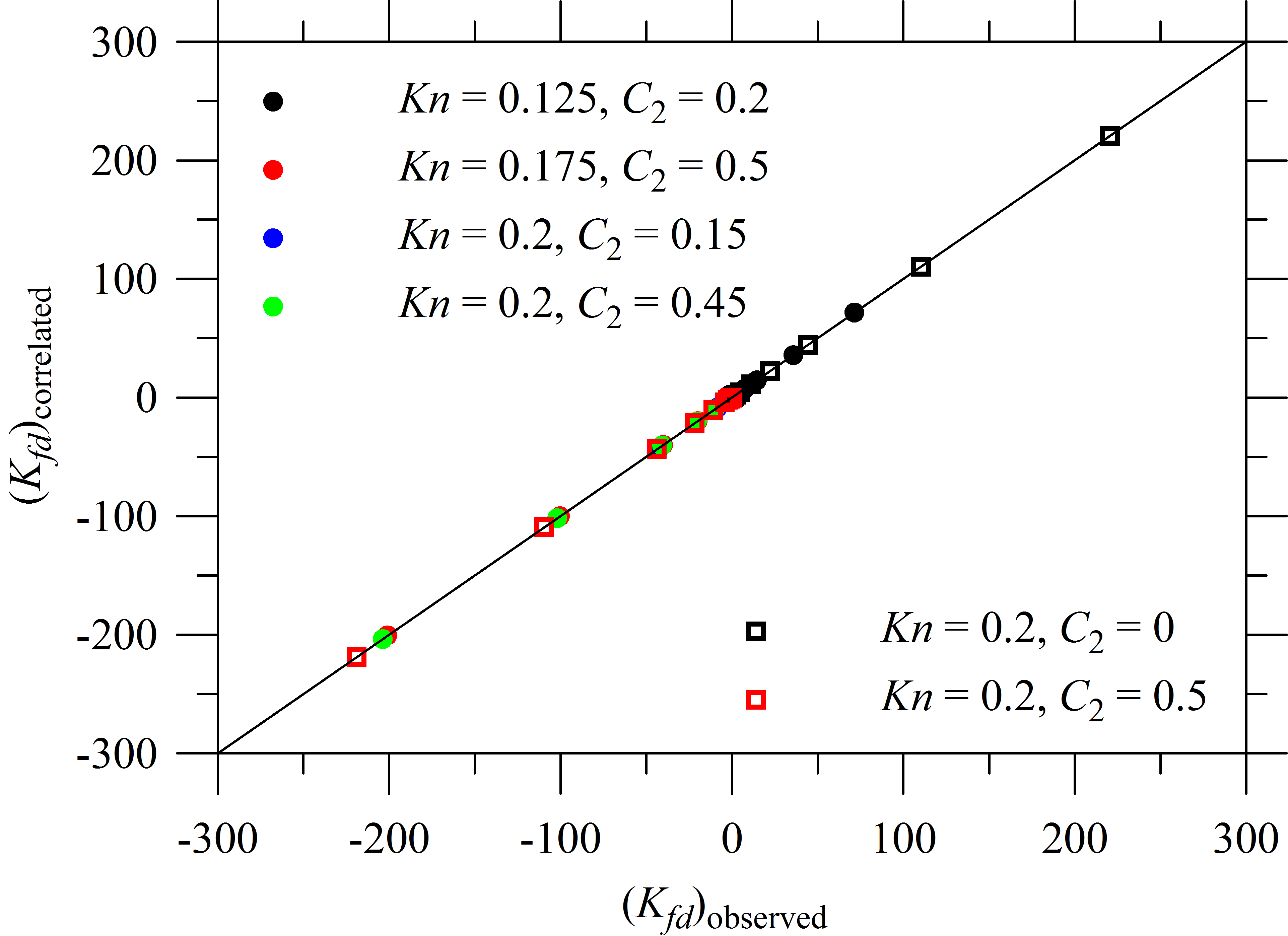}
		}\label{fig:KfdCorr_Pipe}
		\subfigure[Parallel Plate Channel]{
			\includegraphics[width=0.47\textwidth]{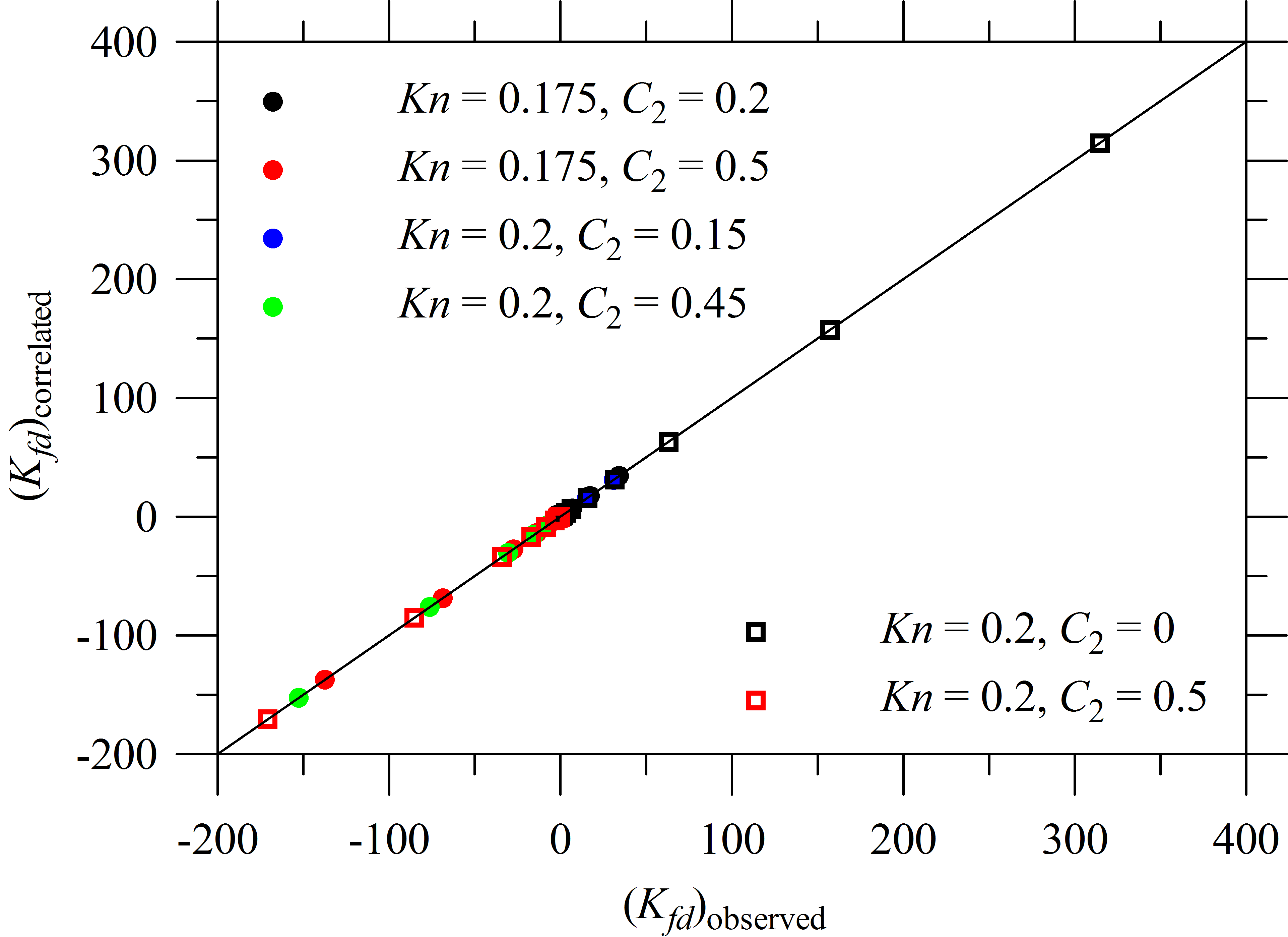}
		}\label{fig:KfdCorr_PP}
	\end{center}
	\caption{Performance of correlations for $K_{fd}$. The solid lines represent $100$ \% accuracy.}
	\label{fig:KfdCorr}
\end{figure}

Since the tabulated values of $K_{0}$ and $K_{1}$ have been used in order to evaluate $K_{2}$, the correlations for $K_{fd}$ in Eq.~(\ref{Eq:Corr_Kfd_general}) are expected to perform extremely well for all simulated cases. Therefore, in order to demonstrate that these correlations are still accurate for the intermediate values of $Kn$ and $C_{2}$, where $K_{0}$ and $K_{1}$ are required to be interpolated, four additional cases for both geometries have been carefully chosen between two tabulated results, where either $K_{0}$ or $K_{1}$ changes its sign. Comparison between the observed and the correlated values of $K_{fd}$ for these additional cases are presented in Fig.~\ref{fig:KfdCorr}. In addition, the results for two tabulated cases with $C_{2} = 0$ and $C_{2} = 0.5$ for $Kn = 0.2$ are also shown in the same figure in order to demonstrate that the proposed correlations indeed perform extremely well for the already reported cases, where $K_{0}$ and $K_{1}$ could be directly obtained from Table~\ref{tab:K0K1_PipeCHannel} without any interpolation. While quantifying the performance of the correlations, however, it must be noted that since $K_{fd} \approx 0$ has been recorded for some of the cases, the maximum absolute relative error cannot be consistently considered an appropriate measure of deviation and hence the coefficient of determination $R^{2}$ has been reported in order to demonstrate the accuracy of the correlations. \\

At this point, a few comments regarding the interpolation of $K_{0}$ and $K_{1}$ for intermediate $Kn$ and $C_{2}$ are essential. It has been observed that the linearly interpolated $K_{0}$ and $K_{1}$ do not produce satisfactory agreement, owing to the non-linear nature of their variations as functions of both $Kn$ and $C_{2}$. The minimum values of coefficient of determination have been obtained as $R^{2}_{\min} \approx 0.9$ for both pipe and channel flows,\footnote{Among four additional cases, $R^{2}_{\min}$ has been observed for $C_{2} = 0.2$ with $Kn = 0.125$ and $Kn = 0.175$ for pipe and channel flows, respectively.} when linear interpolation is used while considering $Kn$ and $C_{2}$ as independent variables. In order to improve the prediction, $Kn^{1/4}$ and $C_{2}^{1/2}$ have been subsequently considered as independent variables for interpolation. Even then, the use of linear interpolation has resulted in $R^{2}_{\min} \approx 0.96$ for both geometries, although these data are not shown in Fig.~\ref{fig:KfdCorr} for brevity. Finally, cubic functions\footnote{In order to fit cubic functions, if possible, two points on either sides of the desired $Kn$ have been selected (e.g., for $Kn = 0.125$ for pipe flow). Otherwise, depending upon the availability of data, three points on one side and one point on the other side of the desired $Kn$ have been chosen (e.g., for $Kn = 0.175$ for both geometries).} of $Kn^{1/4}$ and quadratic functions\footnote{Quadratic functions have been fitted preferably by selecting two points on the higher side and one point on the lower side of the desired $C_{2}$ (e.g., for $C_{2} = 0.15$). In the absence of such data, one point on the higher side and two points on the lower side have been chosen (e.g., for $C_{2} = 0.45$).} of $C_{2}^{1/2}$ have been employed for interpolating $K_{0}$ and $K_{1}$, that, rounded up to the fourth place of decimal, produce $R^{2} = 1$ for all additional cases involving both geometries. \\

From the foregoing discussions, it is obvious that for a duct length $L < L_{fd}$, evaluation of the actual pressure drop $\Delta p_{L}$ from Eq.~(\ref{Eq:Deltapx_From_Kx}) requires accurate information about $K (L)$ in the developing region. However, since no correlation for $K (x)$ could be derived from the present investigation owing to its complex dependence on the operating conditions, few comments on the estimation of $\Delta p_{L}$ for $L < L_{fd}$ would be useful. While selecting the pumping device and determining its power requirement for a particular micro-channel application, an overestimation is often considered more acceptable than any underestimation and hence in order to be on the safer side, $\Delta p_{L}$ could be estimated as:
\begin{subequations}
\begin{eqnarray}
\Delta p_{L} & = & \frac {1} {2} \rho u^{2}_{av} \left( K_{fd} + 4 f_{fd} L^{*} \right) ~~~~~ \mbox{for}~K_{fd} > 0,~\mbox{set}~K(L) = K_{fd} \label{Eq:Delta_P_L_forKfdGE0} \\
& = & 2 \rho u^{2}_{av} f_{fd} L^{*}  ~~~~~ \mbox{for} ~ K_{fd} \le 0,~\mbox{set}~K(L) = 0 \label{Eq:Delta_P_L_forKfdLE0} 
\end{eqnarray}
\label{Eq:Delta_P_L}
\end{subequations}
It must be emphasised that $\Delta p_{L}$, calculated from Eq.~(\ref{Eq:Delta_P_L}) for $L < L_{fd}$, should never be used for optimisation since the estimated value would always be greater than the true pressure drop and may be grossly erroneous. For $L \ge L_{fd}$, however, the correlation for $K_{fd}$ in Eq.~(\ref{Eq:Corr_Kfd_general}) may be used for accurately evaluating $\Delta p_{L}$.

\section{Summary, Conclusions and Final Remarks}\label{concls}
In the present investigation, developing laminar flows through micro-capillaries and micro-channels have been numerically analysed by solving the conventional NS equations (\ref{eq:mass}) -- (\ref{eq:y-mom}) along with the second-order velocity slip condition at the wall in Eq.~(\ref{Eq:General-Second-Order-Slip-Condition-Dimless}). For both pipe and channel flows, thorough investigations have been conducted for $10^{-2} \le Re \le 10^{4}$ and $10^{-4} \le Kn \le 0.2$, while specifying $C_{1} = 1$ and varying $C_{2}$ from $0$ to $0.5$. The results for the first order velocity slip condition, similar to that presented by \citet{BarberAndEmerson_2001} and \citet{Ferrasetal2012} for parallel plate channels, have been obtained by setting $C_{2} = 0$, whereas that of \citet{Durstetal2005} in the continuum regime have been generated with $Kn = 0$. \\

Before carrying out the extensive parametric study, the maximum length $x^{*}_{\max}$ of the computational domain has been first ascertained. The development of the axial velocity profile shows that analytical solutions for the fully-developed flow have always been obtained at the exit of the duct, irrespective of the operating condition. Since the obtained numerical solutions are also grid independent, this observation confirms the authenticity of the predicted results. The converged solutions for each combination of $Re$, $Kn$ and $C_2$ have been further post-processed in order to evaluate the local friction factor $f_{x}$, the dimensionless development length $L^{*}_{fd}$, the local and the fully-developed incremental pressure drop number $K(x)$ and $K_{fd}$, respectively. The obtained data have been carefully analysed and correlations for $L^{*}_{fd}$ and $K_{fd}$ have been proposed for both geometries. The major conclusions from the present investigation may be summarised as follows: \\
\begin{enumerate}
\item The development of the axial velocity profile shows that although irrespective of $Re$, $Kn$ and $C_{2}$, $u^{*}_{s}$ for any $x^{*} \le L^{*}_{fd}$ is always higher than $u^{*}_{s,fd}$, the velocity gradients at the wall and hence $\tau_{w,x}$ could be considerably less than that in the fully-developed section, particularly for higher $Kn$ and $C_{2}$. These effects are more prominent for lower $Re$ and significantly affect the variations in $K(x)$ and hence $K_{fd}$. The velocity overshoots, on the other hand, are observed to be more pronounced for higher $Re$ with lower velocity slip at the wall and are considerably reduced with the decrease in $Re$ as well as the increase in both $Kn$ and $C_{2}$.

\item The development length $L^{*}_{fd}$ increases with the increase in both $Kn$ and $C_{2}$, except for pipe flows in the high $Re$ regime and for both geometries in the low $Re$ regime, but only with $C_{2} = 0$. For the first exception, $L^{*}_{fd}$ initially decreases and subsequently increases with the increase in $Kn$, except for $C_{2} = 0$, for which, although the variation is marginal, it decreases once again. For the second exception, on the other hand, $L^{*}_{fd}$ decreases marginally for higher $Kn$, which could be detected for $Kn = 0.2$ for both geometries.

\item Similar to the continuum regime, $L^{*}_{fd}$ in the presence of ssecond order velocity slip at the wall is also characterised by the low and the high $Re$ asymptotes and hence it has been correlated according to Eq.~(\ref{eq:corr_general}), where the expressions for $L_{0}$, $L_{1}$ and $q$ are provided in Eqs.~(\ref{Eq:Corr_Ai}) -- (\ref{Eq:n_PipeChannel}). The proposed correlations for $L^{*}_{fd}$ produce $2.47$ \% and $3.86$ \% errors for pipe and channel flows, respectively. On the other hand, with a constant $q = 1.6$, as proposed by \citet{Durstetal2005}, these errors are obtained as $3.70$ \% and $3.87$ \%, respectively. Even for flows through parallel plate micro-channels with the first order velocity slip condition at the wall, the present correlation has been found to be more accurate than those proposed by \citet{BarberAndEmerson_2001} and \citet{Ferrasetal2012} for the entire range of Reynolds number.

\item Pressure drop characteristics in the entrance region as well as in the fully-developed section of micro-channels have been thoroughly analysed. It has been observed that depending upon the operating condition, particularly for higher $Kn$ and $C_{2}$, both $K(x)$ and $K_{fd}$ could assume negative values. This important as well as interesting feature, which is more prominent in the low $Re$ regime, has never been reported in the literature. These results imply that in the presence of substantial velocity slip at the wall, the pressure gradient in the developing section could be even less than that in the fully-developed region.

\item No reliable correlation for $K(x)$ could be proposed from the present investigation since for both geometries, $K(x)$ is a quite complicated function of the axial location and the functional form depends strongly on the operating condition. Although the calculation of true pressure drop for a duct of length $L < L_{fd}$ requires precise knowledge about $K(x)$, a method has been proposed in order to obtain an conservative estimate of the pressure drop that, other than the fully-developed friction factor $f_{fd}$, also depends on $K_{fd}$ for $K_{fd} > 0$.

\item The magnitude of $K_{fd}$ decreases consistently with the increase in $Re$, $Kn$ and $C_{2}$, although it is a weak function of $Kn$ and $C_{2}$ in the high $Re$ regime where $K_{fd} \rightarrow K_{1}$. Irrespective of $C_{2}$, it has been found to be a strong function of $Kn$, whereas it appears to be weak function of $C_{2}$ for lower $Kn$ and its sensitivity to the change in $C_{2}$ increases with the increase in $Kn$.

\item Similar to $L^{*}_{fd}$, the variations in $K_{fd}$ as functions of $Re$ are also represented by the low and the high $Re$ asymptotes, $K_{0}/Re$ and $K_{1}$, respectively for any combination of $Kn$ and $C_{2}$. However, since these asymptotic values vary over a large range and their functional dependence on both $Kn$ and $C_{2}$ changes considerably over the investigated ranges, no reliable correlation for either $K_{0}$ or $K_{1}$ could be proposed. Moreover, since $K_{0}$ for both geometries and $K_{1}$ for pipe flows assume both positive and negative values, $K_{fd}$ could not be expressed by the form similar to that for $L^{*}_{fd}$ and hence an alternative expression in Eq.~(\ref{Eq:Corr_Kfd_general}) has been adopted that satisfies both low and high $Re$ limits. Using the tabulated values of $K_{0}$ and $K_{1}$, $K_{2}$ has been determined for each $C_{2}$, while allowing $Kn$ to vary from $10^{-3}$ to $0.2$, from which, correlations for $K_{2}$ has been proposed that are presented in Eqs.~(\ref{Eq:Corr_K2}) -- (\ref{Eq:k2j_Channel}).

\item Four additional cases for both geometries have been specifically simulated for intermediate values of $Kn$ and $C_{2}$, where $K_{0}$ and $K_{1}$ are required to be carefully interpolated using cubic functions of $Kn^{1/4}$ and quadratic functions of $C_{2}^{1/2}$. The performance of the proposed correlations show excellent agreement with the simulated data for all additional cases, with $R^{2} = 1$ for both geometries.
\end{enumerate}

As a final remark, it may be mentioned that in the present investigation, although $Re$, $Kn$ and $C_{2}$ have been varied over wide ranges, $C_{1}$ has been kept fixed to unity. Previous investigations and the present analysis based on the dimensionless parameters clearly show that the effects of $Kn$ and $C_{1}$ could not be separated if the first order velocity slip condition at the duct surface is applied. Therefore, in the future, $C_{1}$ in the second order velocity slip boundary condition at the wall should be varied over a realistic range in order to quantify its effects on both development length and pressure drop in the entrance region for pipe and channel flows.

\addcontentsline{toc}{section}{References}
\bibliography{DevLenSlip}

\begin{thebibliography}{75}
\providecommand{\natexlab}[1]{#1}
\providecommand{\url}[1]{\texttt{#1}}
\expandafter\ifx\csname urlstyle\endcsname\relax
  \providecommand{\doi}[1]{doi: #1}\else
  \providecommand{\doi}{doi: \begingroup \urlstyle{rm}\Url}\fi

\bibitem[Aidun and Clausen(2010)]{AidunAndClausen_2010}
C.~K. Aidun and J.~R. Clausen.
\newblock Lattice-{B}oltzmann method for complex flows.
\newblock \emph{Annual Review of Fluid Mechanics}, 42:\penalty0 439–--472,
  2010.

\bibitem[Ansumali et~al.(2007)Ansumali, Karlin, Arcidiacono, Abbas, and
  Prasianakis]{Ansumali_etal_2007}
S.~Ansumali, I.~V. Karlin, S.~Arcidiacono, A.~Abbas, and N.~Prasianakis.
\newblock Hydrodynamics beyond navier–stokes: exact solution to the lattice
  {B}oltzmann hierarchy.
\newblock \emph{Physical Review Letters}, 98:\penalty0 124502, 2007.

\bibitem[Arkilic et~al.(1997)Arkilic, Schmidt, and Breuer]{Akrilic_etal_1997}
E.~B. Arkilic, M.~A. Schmidt, and K.~S. Breuer.
\newblock Gaseous slip flow in microchannels.
\newblock \emph{Journal of Microelectromechanical Systems}, 6:\penalty0
  167–--174, 1997.

\bibitem[Barber and Emerson(2001)]{BarberAndEmerson_2001}
R.~W. Barber and D.~R. Emerson.
\newblock A numerical investigation of low reynolds number gaseous slip flow at
  the entrance of circular and parallel plate micro-channels.
\newblock In \emph{Proceedings of ECCOMAS Computational Fluid Dynamics
  Conference}, 2001.

\bibitem[Barber and Emerson(2006)]{BarberAndEmerson_2006}
R.~W. Barber and D.~R. Emerson.
\newblock Challenges in modeling gas-phase flow in microchannels: from slip to
  transition.
\newblock \emph{Heat Transfer Engineering}, 27:\penalty0 3–--12, 2006.

\bibitem[Beskok and Karniadakis(1996)]{BeskokAndKarniadakis_1996}
A.~Beskok and G.~E. Karniadakis.
\newblock Rarefaction and compressibility effects in gas microflows.
\newblock \emph{ASME J. Fluids Eng.}, 118:\penalty0 448–--456, 1996.

\bibitem[Beskok and Karniadakis(1999)]{BeskokAndKarniadakis_1999}
A.~Beskok and G.~E. Karniadakis.
\newblock A model for flows in channels, pipes, and ducts at micro and nano
  scales.
\newblock \emph{Microscale Thermophysical Engineering}, 3:\penalty0 43–--77,
  1999.

\bibitem[Bird(1994)]{Bird_1994}
G.~A. Bird.
\newblock \emph{Molecular gas dynamics and the direct simulation of gas flows}.
\newblock Clarendon Press, Oxford, 1994.

\bibitem[Bird et~al.(2002)Bird, Stewart, and Lightfoot]{TransportPhenomena}
R.~B. Bird, W.~E. Stewart, and E.~N. Lightfoot.
\newblock \emph{Transport Phenomena}.
\newblock John Wiley \& Sons, New York, 2002.

\bibitem[Cao et~al.(2009)Cao, Sun, Chen, and Guo]{Cao_etal_2009}
B.~Y. Cao, J.~Sun, M.~Chen, and Z.~Y. Guo.
\newblock Molecular momentum transport at fluid-solid interfaces in mems/nems:
  a review.
\newblock \emph{International Journal of Molecular Sciences}, 10:\penalty0
  4638–--4706, 2009.

\bibitem[Cercignani(1975)]{Cercignani_1975}
C.~Cercignani.
\newblock \emph{Theory and application of the {B}oltzmann equation}.
\newblock Scottish Academic Press, Edinburgh, 1975.

\bibitem[Cercignani(1988)]{Cercignani_1988}
C.~Cercignani.
\newblock \emph{The {B}oltzmann equation and its applications}.
\newblock Springer-Verlag, New York, 1988.

\bibitem[Cercignani(1990)]{Cercignani_1990}
C.~Cercignani.
\newblock \emph{Mathematical methods in kinetic theory}.
\newblock Plenum, New York, 1990.

\bibitem[Cercignani(2000)]{Cercignani_2000}
C.~Cercignani.
\newblock \emph{Rarefied gas dynamics}.
\newblock Cambridge University Press, Cambridge, 2000.

\bibitem[Chakraborty and Durst(2007)]{SumanAndDurst_2007}
S.~Chakraborty and F.~Durst.
\newblock Derivations of extended navier-stokes equations from upscaled
  molecular transport considerations for compressible ideal gas flows: Towards
  extended constitutive forms.
\newblock \emph{Phys. Fluids}, 19:\penalty0 088104, 2007.

\bibitem[Chen and Bogy(2010)]{ChenAndBogy_2010}
D.~Chen and D.~B. Bogy.
\newblock Comparisons of slip-corrected reynolds lubrication equations for the
  air bearing film in the head-disk interface of hard disk drives.
\newblock \emph{Tribology Letters}, 37:\penalty0 191–--201, 2010.

\bibitem[Chen(1973)]{Chen_1973}
R.~Y. Chen.
\newblock Flow in the entrance region at low reynolds numbers.
\newblock \emph{ASME J. Fluids Eng.}, 95:\penalty0 153--158, 1973.

\bibitem[Colin(2005)]{Colin_2005}
S.~Colin.
\newblock Rarefaction and compressibility effects on steady and transient gas
  flows in microchannels.
\newblock \emph{Microfluidics and Nanofluidics}, 1:\penalty0 268–--279, 2005.

\bibitem[Colin(2012)]{Colin_2012}
S.~Colin.
\newblock Gas microflows in the slip flow regime: a critical review on
  convective heat transfer.
\newblock \emph{ASME Journal of Heat Transfer}, 134:\penalty0 020908, 2012.

\bibitem[Cornubert et~al.(1991)Cornubert, d{'}Humieres, and
  Levermore]{Cornubert_etal_1991}
R.~Cornubert, D.~d{'}Humieres, and D.~Levermore.
\newblock A knudsen layer theory for lattice gases.
\newblock \emph{Physica D}, 47:\penalty0 241–--259, 1991.

\bibitem[Dombrowski et~al.(1993)Dombrowski, Foumeny, Ookawara, and
  Riza]{Dombrowski_etal_1993}
N.~Dombrowski, E.~A. Foumeny, S.~Ookawara, and A.~Riza.
\newblock The influence of reynolds number on the entry length and pressure
  drop for laminar pipe flow.
\newblock \emph{The Canadian Journal of Chemical Engineering}, 71:\penalty0
  472--476, 1993.

\bibitem[Dongari et~al.(2007)Dongari, Agrawal, and Agrawal]{Dongari2007}
N.~Dongari, A.~Agrawal, and A.~Agrawal.
\newblock Analytical solution of gaseous slip flow in long microchannels.
\newblock \emph{Int. J. Heat Mass Transfer}, 50:\penalty0 3411--3421, 2007.

\bibitem[Dongari et~al.(2009)Dongari, Sambasivam, and Durst]{Dongari_etal_2009}
N.~Dongari, R.~Sambasivam, and F.~Durst.
\newblock Extended navier–stokes equations and treatments of micro-channel gas
  flows.
\newblock \emph{Journal of Fluid Science and Technology}, 4:\penalty0
  454–--467, 2009.

\bibitem[Dongari et~al.(2010)Dongari, Durst, and
  Chakraborty]{Dongari_etal_2010}
N.~Dongari, F.~Durst, and S.~Chakraborty.
\newblock Predicting microscale gas flows and rarefaction effects through
  extended navier–stokes–fourier equations from phoretic transport
  considerations.
\newblock \emph{Microfluidics and Nanofluidics}, 9:\penalty0 831–--846, 2010.

\bibitem[Dongari et~al.(2011{\natexlab{a}})Dongari, Zhang, and
  Reese]{Dongari_etal_2011_1}
N.~Dongari, Y.~H. Zhang, and J.~M. Reese.
\newblock Molecular free path distribution in rarefied gases.
\newblock \emph{Journal of Physics D, Applied Physics}, 44:\penalty0 125502,
  2011{\natexlab{a}}.

\bibitem[Dongari et~al.(2011{\natexlab{b}})Dongari, Zhang, and
  Reese]{Dongari_etal_2011_2}
N.~Dongari, Y.~H. Zhang, and J.~M. Reese.
\newblock Modeling of knudsen layer effects in micro/nanoscale gas flows.
\newblock \emph{ASME J. Fluids Eng.}, 133:\penalty0 071101, 2011{\natexlab{b}}.

\bibitem[Durst et~al.(2005)Durst, \"Unsal, Ray, and Saleh]{Durstetal2005}
F.~Durst, B.~\"Unsal, S.~Ray, and O.~Saleh.
\newblock The development lengths of laminar pipe and channel flows.
\newblock \emph{ASME J. Fluids Eng.}, 127:\penalty0 1154--1160, 2005.

\bibitem[Ferr\'as et~al.(2012)Ferr\'as, Alfonso, Alves, N\'obrega, and
  Pinho]{Ferrasetal2012}
L.~Ferr\'as, A.~Alfonso, M.~Alves, J.~N\'obrega, and F.~Pinho.
\newblock Development length in planar channel flows of newtonian fluids under
  the influence of wall slip.
\newblock \emph{ASME J. Fluids Eng.}, 134:\penalty0 104503--1--104503--5, 2012.

\bibitem[Ferziger and Peri\'c(1999)]{peric}
J.~H. Ferziger and M.~Peri\'c.
\newblock \emph{Computational Methods for Fluid Dynamics}.
\newblock Springer, Berlin, 1999.

\bibitem[Fox et~al.(2010)Fox, McDonald, and Pritchard]{FoxMcDonald}
R.~W. Fox, A.~T. McDonald, and P.~J. Pritchard.
\newblock \emph{Introduction to Fluid Mechanics}.
\newblock John Wiley \& Sons, New York, 2010.

\bibitem[Gad{-}el{-}Hak(1999)]{Gad-el-Hak_1999}
M.~Gad{-}el{-}Hak.
\newblock The fluid mechanics of microdevices —- the freeman scholar lecture.
\newblock \emph{JFE}, 121:\penalty0 5–--33, 1999.

\bibitem[Gad{-}el{-}Hak(2001)]{Gad-el-Hak_2001}
M.~Gad{-}el{-}Hak.
\newblock Flow physics in mems.
\newblock \emph{M\'ecanique \& Industries}, 2:\penalty0 313–--341, 2001.

\bibitem[Gad{-}el{-}Hak(2006)]{Gad-el-Hak_2006}
M.~Gad{-}el{-}Hak.
\newblock Gas and liquid transport at the microscale.
\newblock \emph{Heat Transfer Engineering}, 27:\penalty0 13–--29, 2006.

\bibitem[Hadjiconstantinou(2000)]{Hadjiconstantinou_2000}
N.~G. Hadjiconstantinou.
\newblock Analysis of discretization in the direct simulation monte carlo.
\newblock \emph{Phys. Fluids}, 12:\penalty0 2634–--2638, 2000.

\bibitem[Hadjiconstantinou(2006)]{Hadjiconstantinou_2006}
N.~G. Hadjiconstantinou.
\newblock The limits of navier–stokes theory and kinetic extensions for
  describing small-scale gaseous hydrodynamics.
\newblock \emph{Phys. Fluids}, 18:\penalty0 111301, 2006.

\bibitem[Ho and Tai(1998)]{HoAndTai_1998}
C.~M. Ho and Y.~C. Tai.
\newblock Micro-electro-mechanical-systems (mems) and fluid flows.
\newblock \emph{Annual Review of Fluid Mechanics}, 30:\penalty0 579–--612,
  1998.

\bibitem[Jin and Slemrod(2001)]{JinAndSlemrod_2001}
S.~Jin and M.~Slemrod.
\newblock Regularization of the burnett equations via relaxation.
\newblock \emph{Journal of Statistical Physics}, 103:\penalty0 1009–--1033,
  2001.

\bibitem[Kandlikar et~al.(2013)Kandlikar, Colin, Garimella, Pease, Brandner,
  and Tuckerman]{Kandlikar_etal_2013}
S.~G. Kandlikar, S.~Colin, Y.~P.~S. Garimella, R.~F. Pease, J.~J. Brandner, and
  D.~B. Tuckerman.
\newblock Heat transfer in microchannels —- 2012 status and research needs.
\newblock \emph{ASME Journal of Heat Transfer}, 135:\penalty0 091001, 2013.

\bibitem[Karniadakis and Beskok(2002)]{KarniadakisAndBeskok_2002}
G.~E. Karniadakis and A.~Beskok.
\newblock \emph{Micro flows: fundamentals and simulation}.
\newblock Springer-Verlag, New York, 2002.

\bibitem[Karniadakis et~al.(2005)Karniadakis, Beskok, and
  Aluru]{MicroflowsNanoflows}
G.~E. Karniadakis, A.~Beskok, and N.~Aluru.
\newblock \emph{Microflows and Nanoflows: Fundamentals and Simulation}.
\newblock Springer-Verlag, New York, 2005.

\bibitem[Khosla and Rubin(1974)]{khosla-rubin}
P.~K. Khosla and S.~G. Rubin.
\newblock A diagonally dominant second order accurate implicit scheme.
\newblock \emph{Comput. Fluids}, 2:\penalty0 207--209, 1974.

\bibitem[Knudsen(1909)]{Knudsen1909}
M.~Knudsen.
\newblock Die gesetze der molekularstr\"omung und der inneren
  reibungsstr\"omung der gase durch r\"ohren.
\newblock \emph{Annalen der Physik}, 333\penalty0 (1):\penalty0 75--130, 1909.

\bibitem[Li and Kwok(2003)]{LiAndKwok_2003}
B.~Li and D.~Y. Kwok.
\newblock Discrete {B}oltzmann equation for microfluidics.
\newblock \emph{Physical Review Letters}, 90:\penalty0 124502, 2003.

\bibitem[Lilley and Sader(2008)]{LilleyAndSader_2008}
C.~R. Lilley and J.~E. Sader.
\newblock Velocity profile in the knudsen layer according to the {B}oltzmann
  equation.
\newblock \emph{Proceedings of Royal Society A}, 464:\penalty0 2015–--2035,
  2008.

\bibitem[Lockerby and Reese(2008)]{LockerbyAndReese_2008}
D.~A.~. Lockerby and J.~M. Reese.
\newblock On the modelling of isothermal gas flows at the microscale.
\newblock \emph{J. Fluid Mechanics}, 604:\penalty0 235–--261, 2008.

\bibitem[Lockerby et~al.(2004)Lockerby, Reese, Emerson, and
  Barber]{Lockerby_etal_2004}
D.~A. Lockerby, J.~M. Reese, D.~R. Emerson, and R.~W. Barber.
\newblock Velocity boundary condition at solid walls in rarefied gas
  calculations.
\newblock \emph{Physical Review E}, 70:\penalty0 017303, 2004.

\bibitem[Loyalka(1971)]{Loyalka_1971}
S.~K. Loyalka.
\newblock Approximate method in kinetic theory.
\newblock \emph{Phys. Fluids}, 14:\penalty0 2291–--2294, 1971.

\bibitem[Pan et~al.(1999)Pan, Liu, and Lam]{Pan_etal_1999}
L.~S. Pan, G.~R. Liu, and K.~Y. Lam.
\newblock Determination of slip coefficient for rarefied gas flows using direct
  simulation monte carlo.
\newblock \emph{Journal of Micromechanics and Microengineering}, 9:\penalty0
  89–--96, 1999.

\bibitem[Patankar(1980)]{patankar}
S.~V. Patankar.
\newblock \emph{Numerical Heat Transfer and Fluid Flow}.
\newblock Hemisphere, Washington, DC, 1980.

\bibitem[Patankar and Spalding(1972)]{patankar-spalding}
S.~V. Patankar and D.~B. Spalding.
\newblock A calculation procedure for heat, mass and momentum transfer in
  three-dimensional parabolic flows.
\newblock \emph{Int. J. Heat Mass Transfer}, 15\penalty0 (10):\penalty0
  1787--1806, 1972.

\bibitem[Rhie and Chow(1983)]{RhieChow1983}
C.~Rhie and W.~Chow.
\newblock Numerical study of the turbulent flow past an airfoil with trailing
  edge separation.
\newblock \emph{AIAA J.}, 21\penalty0 (11):\penalty0 1525--1532, 1983.

\bibitem[Roohi and Darbandi(2009)]{RoohiAndDarbandi_2009}
E.~Roohi and M.~Darbandi.
\newblock Extending the navier–stokes solutions to transition regime in
  two-dimensional micro- and nanochannel flows using information preservation
  scheme.
\newblock \emph{Phys. Fluids}, 21:\penalty0 082001, 2009.

\bibitem[Sambasivam(2013)]{Sambasivam_PhD_2013}
R.~Sambasivam.
\newblock \emph{Extended Navier-Stokes Equations: Derivations and Applications
  to Fluid Flow Problems}.
\newblock PhD thesis, Friedrich Alexander Universit\"at, Erlangen-N\"urenberg,
  Germany, 2013.

\bibitem[Sbragaglia and Succi(2005)]{SbragagliaAndSucci_2005}
M.~Sbragaglia and Succi.
\newblock Analytical calculation of slip flow in lattice {B}oltzmann models
  with kinetic boundary conditions.
\newblock \emph{Phys. Fluids}, 17:\penalty0 093602, 2005.

\bibitem[Schaaf and Chambre(1961)]{SchaafAndChambre_1961}
S.~A. Schaaf and P.~L. Chambre.
\newblock \emph{Rarefied Gas Dynamics}.
\newblock Princeton University Press, Princeton, 1961.

\bibitem[Shah and London(1978)]{ShahandLondon1978}
R.~K. Shah and A.~L. London.
\newblock \emph{Advances in Heat Transfer, Supplement I: Laminar Flow Forced
  Convection in Ducts}.
\newblock Academic Press, New York, San Francisco, London, 1978.

\bibitem[Shan et~al.(2006)Shan, Yuan, and Chen]{Shan_etal_2006}
X.~Shan, X.~Yuan, and H.~Chen.
\newblock Kinetic theory representation of hydrodynamics: a way beyond the
  navier–stokes equation.
\newblock \emph{J. Fluid Mechanics}, 550:\penalty0 413–--441, 2006.

\bibitem[Sharipov(2003)]{Sharipov_2003}
F.~Sharipov.
\newblock Application of the cercignani-lampis scattering kernel to
  calculations of rarefied gas flows. ii. slip and jump coefficients.
\newblock \emph{European Journal of Mechanics B/Fluid}, 22:\penalty0 133–--143,
  2003.

\bibitem[Sharipov and Seleznev(1998)]{SharipovAndSeleznev_1998}
F.~Sharipov and V.~Seleznev.
\newblock Data on internal rarefied gas flows.
\newblock \emph{J Phys Chem Ref Data}, 27:\penalty0 657–--706, 1998.

\bibitem[Siewert and Sharipov(2002)]{SiewertAndSharipov_2002}
C.~E. Siewert and F.~Sharipov.
\newblock Model equations in rarefied gas dynamics: Viscous-slip and
  thermal-slip coefficients.
\newblock \emph{Phys. Fluids}, 14:\penalty0 4123–--4129, 2002.

\bibitem[Stone(1968)]{stone}
H.~L. Stone.
\newblock Iterative solution of implicit approximations of multidimensional
  partial differential equations.
\newblock \emph{SIAM. J. Num. Anal.}, 5:\penalty0 530--541, 1968.

\bibitem[Struchtrup and Torrilhon(2003)]{StruchtrupAndTorrilhon_2003}
H.~Struchtrup and M.~Torrilhon.
\newblock Regularization of grid's 13 moment equations: derivation and linear
  analysis.
\newblock \emph{Phys. Fluids}, 15:\penalty0 2668–--2680, 2003.

\bibitem[Struchtrup and Torrilhon(2008)]{StruchtrupAndTorrilhon_2008}
H.~Struchtrup and M.~Torrilhon.
\newblock Higher-order effects in rarefied channel flows.
\newblock \emph{Physical Review E}, 78:\penalty0 046301, 2008.

\bibitem[Tang et~al.(2007{\natexlab{a}})Tang, He, and Tao]{Tang_etal_2007a}
G.~H. Tang, Y.~L. He, and W.~Q. Tao.
\newblock Comparison of gas slip models with solutions of linearized
  {B}oltzmann equation and direct simulation of monte carlo method.
\newblock \emph{International Journal of Modern Physics C}, 18:\penalty0
  203–--216, 2007{\natexlab{a}}.

\bibitem[Tang et~al.(2007{\natexlab{b}})Tang, Zhuo, He, and
  Tao]{Tang_etal_2007b}
G.~H. Tang, L.~Zhuo, Y.~L. He, and W.~Q. Tao.
\newblock Experimental study of compressibility, roughness and rarefaction
  influences on microchannel flow.
\newblock \emph{Int. J. Heat Mass Transfer}, 50:\penalty0 2282–--2295,
  2007{\natexlab{b}}.

\bibitem[Tang et~al.(2008)Tang, Zhang, and Emerson]{Tang_etal_2008}
G.~H. Tang, Y.~H. Zhang, and D.~R. Emerson.
\newblock Lattice {B}oltzmann models for nonequilibrium gas flows.
\newblock \emph{Physical Review E}, 77:\penalty0 046701, 2008.

\bibitem[Weng and Chen(2008)]{WengAndChen_2008}
H.~C. Weng and C.~K. Chen.
\newblock A challenge in navier–stokes-based continuum modeling:
  Maxwell-burnett slip law.
\newblock \emph{Phys. Fluids}, 20:\penalty0 106101, 2008.

\bibitem[White(2003)]{White}
F.~M. White.
\newblock \emph{Fluid Mechanics}.
\newblock McGraw-Hill, New York, 2003.

\bibitem[Wu and Bogy(2001)]{WuAndBogy_2001}
L.~Wu and D.~B. Bogy.
\newblock A generalized compressible reynolds lubrication equation with bounded
  contact pressure.
\newblock \emph{Phys. Fluids}, 13:\penalty0 2237–--2244, 2001.

\bibitem[Zhang et~al.(2012{\natexlab{a}})Zhang, Zhang, Zheng, and
  Ye]{ZhangHW_etal_2012}
H.~W. Zhang, Z.~Q. Zhang, Y.~G. Zheng, and H.~F. Ye.
\newblock Molecular dynamics-based prediction of boundary slip of fluids in
  nanochannels.
\newblock \emph{Microfluidics and Nanofluidics}, 12:\penalty0 107–--115,
  2012{\natexlab{a}}.

\bibitem[Zhang(2011)]{Zhang_2011}
J.~Zhang.
\newblock Lattice {B}oltzmann method for microfluidics: models and
  applications.
\newblock \emph{Microfluidics and Nanofluidics}, 10:\penalty0 1–--28, 2011.

\bibitem[Zhang et~al.(2006{\natexlab{a}})Zhang, Shan, and
  Chen]{ZhangR_etal_2006}
R.~Zhang, X.~Shan, and H.~Chen.
\newblock Efficient kinetic method for fluid simulation beyond the
  navier–stokes equation.
\newblock \emph{Physical Review E}, 74:\penalty0 046703, 2006{\natexlab{a}}.

\bibitem[Zhang et~al.(2012{\natexlab{b}})Zhang, Meng, and Wei]{Zhang_etal_2012}
W.~Zhang, G.~Meng, and X.~Wei.
\newblock A review on slip models for gas microflows.
\newblock \emph{Microfluidics and Nanofluidics}, 13:\penalty0 845--882,
  2012{\natexlab{b}}.

\bibitem[Zhang et~al.(2006{\natexlab{b}})Zhang, Gu, Barber, and
  Emerson]{ZhangYH_etal_2006}
Y.~H. Zhang, X.~J. Gu, R.~W. Barber, and D.~R. Emerson.
\newblock Capturing knudsen layer phenomena using a lattice {B}oltzmann model.
\newblock \emph{Physical Review E}, 74:\penalty0 046704, 2006{\natexlab{b}}.

\bibitem[Zheng et~al.(2006)Zheng, Reese, Scanlon, and
  Lockerby]{Zheng_etal_2006}
Y.~Zheng, J.~M. Reese, T.~J. Scanlon, and D.~A. Lockerby.
\newblock Scaled navier-stokes-fourier equations for gas flow and heat transfer
  phenomena in micro- and nanosystems.
\newblock In \emph{Proceedings of ASME ICNMM2006}, Limerick, Ireland 96066,
  June 19--21 2006.

\end{thebibliography}
\bibliographystyle{abbrvnat}
\begin{landscape}
\begin{table}
\small
	\centering
	\caption{$L_{0}$ and $L_{1}$ for different $C_{2}$ and $Kn$.}
		\begin{tabular}{c r c c c c c c c c c c c c}
			\toprule 
		& & \multicolumn{2}{c}{$C_2 = 0$} & \multicolumn{2}{c}{$C_2 = 0.1$} & \multicolumn{2}{c}{$C_2 = 0.2$} & \multicolumn{2}{c}{$C_2 = 0.3$} & \multicolumn{2}{c}{$C_2 = 0.4$} & \multicolumn{2}{c}{$C_2 = 0.5$} \\
		& \multicolumn{1}{c}{$Kn$} & \multicolumn{1}{c}{$L_{0}$} & \multicolumn{1}{c}{$L_{1} / 10^{-2}$} & \multicolumn{1}{c}{$L_{0}$} & \multicolumn{1}{c}{$L_{1} / 10^{-2}$} & \multicolumn{1}{c}{$L_{0}$} & \multicolumn{1}{c}{$L_{1} / 10^{-2}$} & \multicolumn{1}{c}{$L_{0}$} & \multicolumn{1}{c}{$L_{1} / 10^{-2}$} & \multicolumn{1}{c}{$L_{0}$} & \multicolumn{1}{c}{$L_{1} / 10^{-2}$} & \multicolumn{1}{c}{$L_{0}$} & \multicolumn{1}{c}{$L_{1} / 10^{-2}$}\\
		\midrule
		\parbox[t]{2mm}{\multirow{9}{*}{\rotatebox[origin=c]{90}{Pipe}}}
		& 0      & 0.6044      & 5.5935      & 0.6044 & 5.5935      & 0.6044 & 5.5935      & 0.6044 & 5.5935      & 0.6044 & 5.5935      & 0.6044 & 5.5935 \\
		& 0.0001 & 0.6045      & 5.5931      & 0.6045 & 5.5931      & 0.6045 & 5.5931      & 0.6045 & 5.5931      & 0.6045 & 5.5931      & 0.6045 & 5.5931 \\
		& 0.001  & 0.6054      & 5.5882      & 0.6054 & 5.5884      & 0.6054 & 5.5884      & 0.6054 & \bf{5.5882} & 0.6054 & \bf{5.5882} & 0.6054 & \bf{5.5882} \\
		& 0.01   & 0.6136      & \bf{5.5757} & 0.6138 & \bf{5.5795} & 0.6141 & \bf{5.5840} & 0.6144 & 5.5888      & 0.6146 & 5.5934      & 0.6149 & 5.5981 \\
		& 0.02   & 0.6220      & 5.5795      & 0.6232 & 5.5975      & 0.6244 & 5.6159      & 0.6256 & 5.6344      & 0.6268 & 5.6543      & 0.6281 & 5.6743 \\
		& 0.05   & 0.6399      & \bf{5.5918} & 0.6471 & 5.6894      & 0.6544 & 5.7903      & 0.6623 & 5.8943      & 0.6703 & 6.0014      & 0.6786 & 6.1116 \\
		& 0.1    & 0.6531      & 5.5768      & 0.6756 & 5.8708      & 0.6999 & 6.1890      & 0.7263 & 6.5322      & 0.7550 & 6.9053      & 0.7866 & 7.3167 \\
		& 0.15   & 0.6554      & 5.5253      & 0.6951 & 6.0496      & 0.7398 & 6.6360      & 0.7907 & 7.3044      & 0.8501 & 8.0913      & 0.9240 & 9.0654 \\ 
		& 0.2    & \bf{0.6524} & 5.4554      & 0.7092 & 6.2142      & 0.7751 & 7.0958      & 0.8538 & 8.1542      & 0.9555 & 9.5227      & 1.1160 & 11.6313 \\
		\midrule
		\parbox[t]{2mm}{\multirow{9}{*}{\rotatebox[origin=c]{90}{Channel}}} 
		& 0      & 0.3152      & 1.0984 & 0.3152 & 1.0984 & 0.3152 & 1.0984 & 0.3152 & 1.0984 & 0.3152 & 1.0984 & 0.3152 & 1.0984 \\
		& 0.0001 & 0.3152      & 1.0985 & 0.3152 & 1.0985 & 0.3152 & 1.0985 & 0.3152 & 1.0985 & 0.3152 & 1.0985 & 0.3152 & 1.0985 \\
		& 0.001  & 0.3156      & 1.0992 & 0.3156 & 1.0992 & 0.3156 & 1.0992 & 0.3156 & 1.0992 & 0.3156 & 1.0992 & 0.3156 & 1.0992 \\
		& 0.01   & 0.3198      & 1.1130 & 0.3199 & 1.1133 & 0.3200 & 1.1136 & 0.3200 & 1.1140 & 0.3201 & 1.1144 & 0.3202 & 1.1148 \\
		& 0.02   & 0.3241      & 1.1337 & 0.3245 & 1.1359 & 0.3249 & 1.1383 & 0.3254 & 1.1409 & 0.3258 & 1.1434 & 0.3262 & 1.1461 \\
		& 0.05   & 0.3338      & 1.1950 & 0.3364 & 1.2105 & 0.3391 & 1.2265 & 0.3419 & 1.2429 & 0.3447 & 1.2597 & 0.3476 & 1.2766 \\
		& 0.1    & 0.3417      & 1.2740 & 0.3502 & 1.3277 & 0.3591 & 1.3839 & 0.3683 & 1.4422 & 0.3780 & 1.5027 & 0.3879 & 1.5655 \\
		& 0.15   & 0.3436      & 1.3248 & 0.3590 & 1.4279 & 0.3751 & 1.5378 & 0.3920 & 1.6537 & 0.4096 & 1.7763 & 0.4280 & 1.9059 \\
		& 0.2    & \bf{0.3425} & 1.3549 & 0.3647 & 1.5129 & 0.3880 & 1.6821 & 0.4123 & 1.8636 & 0.4377 & 2.0565 & 0.4647 & 2.2658 \\
		\bottomrule
	\end{tabular}
	\label{tab:L0L1_PipeCHannel}
\end{table}

\begin{table}
\small
	\centering
	\caption{$K_{0}$ and $K_{1}$ for different $C_{2}$ and $Kn$.}
		\begin{tabular}{c r r r r r r r r r r r r r}
			\toprule
		& & \multicolumn{2}{c}{$C_2 = 0$} & \multicolumn{2}{c}{$C_2 = 0.1$} & \multicolumn{2}{c}{$C_2 = 0.2$} & \multicolumn{2}{c}{$C_2 = 0.3$} & \multicolumn{2}{c}{$C_2 = 0.4$} & \multicolumn{2}{c}{$C_2 = 0.5$} \\
		& \multicolumn{1}{c}{$Kn$} & \multicolumn{1}{c}{$K_{0}$} & \multicolumn{1}{c}{$K_{1}$} & \multicolumn{1}{c}{$K_{0}$} & \multicolumn{1}{c}{$K_{1}$} & \multicolumn{1}{c}{$K_{0}$} & \multicolumn{1}{c}{$K_{1}$} & \multicolumn{1}{c}{$K_{0}$} & \multicolumn{1}{c}{$K_{1}$} & \multicolumn{1}{c}{$K_{0}$} & \multicolumn{1}{c}{$K_{1}$} & \multicolumn{1}{c}{$K_{0}$} & \multicolumn{1}{c}{$K_{1}$} \\
		\midrule
		\parbox[t]{2mm}{\multirow{8}{*}{\rotatebox[origin=c]{90}{Pipe}}} 
		& 0.0001 & 76.3599 & 1.2655 &	76.3526 & 1.2655 &  76.3453  & 1.2655 &  76.3380  & 1.2655 &  76.3307  & 1.2655 &  76.3234  &  1.2655 \\
		& 0.001  & 63.2900 & 1.2454 & 62.9725 & 1.2453 &  62.6644  & 1.2453 &  62.3652  & 1.2452 &  62.0746  & 1.2451 &  61.7920  &  1.2450 \\
		& 0.01   & 31.4299 & 1.0866 & 29.5644 & 1.0850 &  28.2273  & 1.0835 &  27.1745  & 1.0822 &  26.3005  & 1.0810 &  25.5503  &  1.0798 \\
		& 0.02   & 21.6288 & 0.9478 & 19.4039 & 0.9439 &  18.0081  & 0.9403 &  16.9569  & 0.9369 &  16.1016  & 0.9335 &  15.3752  &  0.9302 \\
		& 0.05   & 10.9205 & 0.6591 &  8.5012 & 0.6449 &   7.1769  & 0.6312 &   6.2058  & 0.6176 &   5.4263  & 0.6043 &   4.7707  &  0.5911   \\
		& 0.1    &  5.3864 & 0.4029 &  3.0287 & 0.3709 &   1.8488  & 0.3403 &   1.0098  & 0.3109 &   0.3544  & 0.2826 & $-0.1826$ &  0.2554  \\
		& 0.15   &  3.2671 & 0.2712 &  1.0311 & 0.2254 & $-0.0133$ & 0.1831 & $-0.7275$ & 0.1439 & $-1.2648$ & 0.1075 & $-1.6889$ &  0.0737 \\
		& 0.2    &  2.2047 & 0.1948 &  0.0940 & 0.1391 & $-0.8284$ & 0.0896 & $-1.4311$ & 0.0456 & $-1.8646$ & 0.0062 & $-2.1910$ & $-0.0291$ \\
		\midrule
		\parbox[t]{2mm}{\multirow{7}{*}{\rotatebox[origin=c]{90}{Channel}}}
		& 0.0001 & 80.3390 & 0.6802 & 80.3745 & 0.6912 &  80.3672  & 0.6912 &  80.3599  & 0.6912 &  80.3526  & 0.6912 &  80.3453  & 0.6912 \\
		& 0.001  & 67.2768 & 0.6726 & 66.9628 & 0.6814 &  66.6541  & 0.6813 &  66.3544  & 0.6813 &  66.0632  & 0.6812 &  65.7801  & 0.6812 \\
		& 0.01   & 35.1348 & 0.6086 & 33.2312 & 0.6110 &  31.8826  & 0.6101 &  30.8196  & 0.6093 &  29.9364  & 0.6086 &  29.1775  & 0.6079 \\
		& 0.02   & 25.0101 & 0.5505 & 22.7359 & 0.5513 &  21.3181  & 0.5496 &  20.2478  & 0.5480 &  19.3757  & 0.5466 &  18.6339  & 0.5452 \\
		& 0.05   & 13.5077 & 0.4184 & 11.0197 & 0.4147 &   9.6489  & 0.4088 &   8.6385  & 0.4030 &   7.8251  & 0.3974 &   7.1392  & 0.3918 \\
		& 0.1    &  7.1224 & 0.2825 &  4.6717 & 0.2700 &   3.4163  & 0.2555 &   2.5153  & 0.2415 &   1.8063  & 0.2279 &   1.2214  & 0.2147 \\
		& 0.15   &  4.5213 & 0.2055 &  2.1511 & 0.1823 &   1.0139  & 0.1606 &   0.2249  & 0.1401 & $-0.3764$ & 0.1209 & $-0.8570$ & 0.1028 \\
		& 0.2    &  3.1242 & 0.1522 &  0.8822 & 0.1252 & $-0.1416$ & 0.0984 & $-0.8240$ & 0.0740 & $-1.3239$ & 0.0519 & $-1.7076$ & 0.0317 \\
		\bottomrule
	\end{tabular}
	\label{tab:K0K1_PipeCHannel}
\end{table}
\end{landscape}

\addcontentsline{toc}{section}{List of Figures}
\listoffigures

\addcontentsline{toc}{section}{List of Tables}
\listoftables

\end{document}